\documentclass[a4paper,11pt]{article}
\pdfoutput=1
\usepackage{jheppub}
\usepackage{graphicx,color}
\usepackage{amsmath,autobreak}
\usepackage{amsbsy}
\usepackage{array}
\usepackage{ulem}
\newcolumntype{P}[1]{>{\centering\arraybackslash}p{#1}}
\allowdisplaybreaks
\RequirePackage{ifpdf}
\usepackage{amsmath}
\usepackage{mathtools}

\usepackage{jheppub}
\usepackage[final]{pdfpages}
\usepackage{ifpdf}
\usepackage{slashed}

\usepackage{hyperref}

\usepackage{color}
\usepackage{graphics}

\usepackage{etoolbox}
\usepackage{fixmath}
\usepackage{caption}
\usepackage{subcaption}
\usepackage{amsfonts}

\usepackage{multirow}
\usepackage{epstopdf}

\usepackage{relsize}
\usepackage{float}

\usepackage[table]{xcolor}
\usepackage{tabularx}

\newcommand{\EuGa}{\gamma_E}
\newcommand{\gt}{\overline{\mathcal{G}}}
\newcommand{\g}{g}
\newcommand{\Lt}{\overline L_w}
\def\arXiv{{\fontfamily{qcr}\selectfont arXiv}}

\usepackage{tikz}
\usetikzlibrary{positioning,arrows}
\usetikzlibrary{decorations.pathmorphing}
\usetikzlibrary{decorations.markings}
\usetikzlibrary{shapes.geometric}
\usepackage{endnotes}
\tikzset{
    vector/.style={decorate, decoration={snake}, draw},
    provector/.style={decorate, decoration={snake,amplitude=2.5pt}, draw},
    antivector/.style={decorate, decoration={snake,amplitude=-2.5pt}, draw},
    fermion/.style={draw=black,
      postaction={decorate},decoration={markings,mark=at position .55
        with {\arrow[draw=black]{>}}}},
    fermionbar/.style={draw=black, postaction={decorate},
                       decoration={markings,mark=at position .55 with {\arrow[draw=black]{<}}}},
    fermionnoarrow/.style={draw=black},
    gluon/.style={decorate, draw=black,decoration={coil,amplitude=4pt, segment length=6pt}},
    scalar/.style={dashed,draw=black,
      postaction={decorate},decoration={markings,mark=at position .55
        with {\arrow[draw=black]{>}}}},
    scalarbar/.style={dashed,draw=black,
      postaction={decorate},decoration={markings,mark=at position .55
        with {\arrow[draw=black]{<}}}},
    scalarnoarrow/.style={dashed,draw=black},
    electron/.style={draw=black,
      postaction={decorate},decoration={markings,mark=at position .55
        with {\arrow[draw=black]{>}}}},
    bigvector/.style={decorate, decoration={snake,amplitude=4pt}, draw},
}
\title{On next to soft corrections to Drell-Yan and Higgs Boson productions}

\author{A.H. Ajjath, Pooja Mukherjee and V. Ravindran}
\emailAdd{
ajjathah@imsc.res.in,
poojamukherjee@imsc.res.in,
ravindra@imsc.res.in}

\affiliation{The Institute of Mathematical Sciences, HBNI, Taramani, Chennai 600113, India}

\preprint{IMSc/2020/11/03}

\abstract{ \textcolor{black}{
We present a framework that resums threshold enhanced large logarithms to all orders in perturbation theory
for the production of a pair of leptons in Drell-Yan process and of Higgs boson in gluon fusion as well as in
bottom quark annihilation.  We restrict ourselves to contributions from diagonal partonic channels.  These logarithms include the distributions $((1-z)^{-1} \ln^i(1-z))_+$ resulting from
soft plus virtual (SV) and the logarithms $\ln^i(1-z)$ from next-to-SV (NSV) contributions.  
We use collinear factorisation and 
renormalisation group invariance to achieve this.  
The former allows one to define a Soft-Collinear (SC) function which encapsulates soft and collinear dynamics of the perturbative results to all orders in strong coupling constant.  The logarithmic structure of these results are governed by universal infrared anomalous dimensions and process dependent functions of Sudakov differential equation that the SC satisfies.  \textcolor{black}{The solution to the differential equation}  is obtained by proposing an all-order ansatz in dimensional regularisation, owing to  several state-of-the-art perturbative results available to third order.     The $z$ space solutions thus obtained provide
an integral representation to sum up large logarithms originating  from both soft and collinear configurations, 
conveniently in Mellin $N$ space.
We show that in $N$ space, tower of logarithms 
$a_s^n/N^\alpha \ln^{2n-\alpha} (N), 
a_s^n/N^\alpha \ln^{2n-1-\alpha}(N) \cdots $ etc for $\alpha =0,1$ are summed to all orders in $a_s$.}}

\begin{document}
\allowdisplaybreaks[4]
\unitlength1cm
\maketitle
\flushbottom
\let\footnote=\endnote
\renewcommand*{\thefootnote}{\fnsymbol{footnote}}
\section{Introduction}
Precision studies in the context of Large Hadron Collider (LHC) play an important role
to decipher the experimental data to understand the physics at extremely small length scales. 
The tests \cite{Brooijmans:2018xbu} of the Standard Model (SM) of high energy physics at the LHC with unprecedented 
accuracy can provide indirect clues to unravel physics beyond SM (BSM).  
Accurate measurements of SM observables such as the productions of lepton pairs, vector bosons such as photons, 
Zs and Ws, top quarks, Higgs bosons etc are underway.  From the theory side, the predictions for
these observables are available taking into account various higher order quantum effects.
Both in electroweak sector of SM and in quantum chromodynamics (QCD), the observables are computed 
in power series expansion of their coupling constants viz., 
$e,g_{EW}$ in SM and $g_s$ in QCD.
To name few, the inclusive cross sections for deep inelastic scattering (DIS) and Higgs boson production in hadron colliders
are known to third order in QCD,  see \cite{Vermaseren:2005qc,Soar:2009yh} and \cite{Anastasiou:2015ema,
Mistlberger:2018etf,Duhr:2019kwi} respectively and for invariant mass distribution up to third order in QCD 
see \cite{Hamberg:1990np,Harlander:2002wh,Duhr:2020seh}, for complete list see \cite{Georgi:1977gs,Graudenz:1992pv,Djouadi:1991tka,Spira:1995rr,Catani:2001ic,Harlander:2001is,Anastasiou:2002yz,Harlander:2002wh,Catani:2003zt,Ravindran:2003um,Moch:2005ky,Ravindran:2006cg,deFlorian:2012za,Bonvini:2014jma,deFlorian:2014vta,Anastasiou:2014vaa,Li:2014afw,Anastasiou:2015yha,Anastasiou:2015vya,Das:2020adl} for Higgs production in gluon fusion 
and \cite{Altarelli:1978id,Altarelli:1979ub,Matsuura:1987wt,Matsuura:1988nd,Matsuura:1988sm,Matsuura:1990ba,Hamberg:1990np,vanNeerven:1991gh,Harlander:2002wh,Moch:2005ky,Ravindran:2006cg,deFlorian:2012za,Ahmed:2014cla,Catani:2014uta,Li:2014afw,Duhr:2020seh} for Drell-Yan production.

The LHC being the hadronic machine, even electroweak induced processes do get large quantum corrections
resulting from strong interaction.  QCD being the theory of strong interactions
provide framework to compute these corrections.  
The measurements and predictions from QCD have reached the level that demand the inclusion of
electroweak effects (EW).  The EW corrections to hadronic observables are hard to compute at higher orders
due to the presence of heavy particles such as Ws,Zs and tops in the loops.  
The results of higher order quantum effects from QCD and EW theory provide theoretical laboratory to understand
both ultraviolet (UV) and infrared (IR) structure of the underlying quantum field theory (QFT) and 
also to demonstrate the universal structure.  For IR, see \cite{Catani:1998bh,Becher:2009cu,Becher:2009qa,
Gardi:2009qi,Catani:1998bh} (see \cite{Ajjath:2019vmf,H:2019nsw} for a QFT with mixed gauge groups).  
This is due to certain factorisation properties of scattering amplitudes in UV and IR regions.  
The consequence of the factorisation is the renormalisation group (RG) invariance which demonstrates the structure of
logarithms of the renormalisation scale $\mu_R$ from UV and of the factorisation scale $\mu_F$ from IR 
to all orders in perturbation theory.  
The renormalisation scale separates UV divergent part from the finite part of the Green's function or on-shell 
amplitudes, quantifying
the arbitrariness in the finite part. While the parameters of the renormalised version of the theory
are functions of the renormalisation scale, the physical observables are expected to be independent of this
scale.  This is the consequence of renormalisation group invariance.  
The anomalous dimensions of the RG equations govern the structure 
of the logarithms of renormalisation scale in the perturbation theory to all orders.  
Like UV sector, the infrared sectors of
both SM and QCD are also very rich.  Massless gauge fields such as photons in QED and gluons in QCD
and light matter particles at high energies give soft and collinear divergences, collectively called 
IR divergences, in scattering amplitudes.    The IR divergences
are shown to factorise from on-shell amplitudes and from certain cross sections respectively in a process 
independent way at an arbitrary factorisation scale.  
The resulting IR renormalisation group equations are governed by IR anomalous dimensions.  The
IR renormalisation group equations  are peculiar in the sense that the resulting evolution is
not only controlled by the factorisation scale 
but also by the energy scale(s) in the amplitude or in the scattering process.   
\textcolor{black}{Unlike the UV divergences which are removed by appropriate renormalisation constants, the IR divergences
do not require any such renormalisation procedure} as they add up to zero for infrared safe observables thanks to 
KLN theorem \cite{Kinoshita:1962ur,Lee:1964is}.  The structure of resulting IR logarithms at every order in the perturbation theory is governed 
by the IR anomalous dimensions.  Hence, most of the logarithms present at higher orders are due to UV and IR divergences
present at the intermediate stages of the computations.  The logarithms of renormalisation and factorisation scales
present in the perturbative expansions often play important role to estimate the error that results due to the truncation
of the perturbative series.   Lesser the dependence on these scales, more the reliability of the truncated results.  
Note that there are also logarithms that are functions of physical scales or the 
corresponding scaling variables in the observables.
In certain kinematical regions, these logarithms that are present at every order can be large enough to spoil
the reliability of the truncated perturbative series.  Since the structure of these logarithms at every order is 
controlled by anomalous dimensions of IR renormalisation group equations, they can be 
systematically summed up to all orders.  This procedure is called 
resummation.  There are classic examples in QCD.  For example, the threshold logarithms of the kind 
\begin{eqnarray}
{\cal D}_i(z) = \left({\ln^i(1-z) \over 1-z} \right)_+
\end{eqnarray}
are present in the perturbative results of inclusive cross section in deep inelastic scattering and of 
invariant mass distribution of pair of leptons in Drell-Yan process.  Here the subscript $+$  means 
that ${\cal D}_i(z)$ is a plus distributions.  For DIS, the scaling variable is $z=-q^2/2 p.q$ and 
$z=M^2_{l^+l^-}/\hat s$ for DY.
The momentum transfer from lepton to parton with momentum $p$ in DIS is denoted by $q$ and the invariants
$\hat s$ and $M^2_{l^+l^-}$ are center of mass energy
of incoming partons and invariant mass of final state leptons in DY.  The distributions ${\cal D}_i(z)$ are 
often called threshold logarithms as they dominate in the threshold region namely $z$ approaches $1$.  
In this limit, the entire energy of the incoming particles in the 
scattering event goes into producing a set of hard particles along with infinite number of 
soft gluons each carrying almost zero momentum.
In particular, the logarithms of the form $\ln^i(1-z)/(1-z)$ 
result from the processes involving real radiations of soft gluons and collinear particles.  
While these contributions are ill defined in 4 space-time dimensions in the limit $z \rightarrow 1$, the inclusion of  pure virtual contributions  
gives distributions ${\cal D}_i(z)$ and $\delta(1-z)$.  
The terms that constitute 
these distributions and $\delta(1-z)$ are called soft plus virtual (SV) contributions.   
The SV results in QCD are available for numerous observables in hadron
colliders.  For SV results up to third order, see \cite{Moch:2005ky,Ravindran:2005vv,Ravindran:2006cg,deFlorian:2012za,Ahmed:2014cha, Kumar:2014uwa,Ahmed:2014cla,Catani:2014uta,Li:2014bfa}.  
These logarithms in the perturbative results when convoluted 
with appropriate parton distribution functions to obtain hadronic cross section can not only dominate over 
other contributions but also give large contributions at every order.   Presence of these large 
corrections at every order spoil the reliability of the predictions from the truncated series.
The seminal works by Sterman \cite{Sterman:1986aj} and Catani and Trentedue \cite{Catani:1989ne} provide resolution to 
this problem through reorganisation of the perturbative series called threshold resummation, for its applications 
to various inclusive processes, see \cite{Catani:1996yz,Moch:2005ba,Kramer:1996iq,Bonvini:2012an,Bonvini:2014joa,Bonvini:2014tea,Bonvini:2016frm} for Higgs production in gluon fusion, 
\cite{Bonvini:2016fgf,H:2019dcl} for bottom quark annihilation and for DY \cite{Moch:2005ba,Bonvini:2010ny,Bonvini:2012sh,H.:2020ecd,Catani:2014uta}.   
Since $z$ space results involve convolutions of these distributions,
Mellin space approach using the conjugate variable $N$ is used for resummation. 
\textcolor{black}{In Mellin space, large logarithms of the kind ${\cal D}_i(z)$ become functions of  $\ln^{j+1}N, j\leq i$ with ${\cal O}(1/N)$} 
suppressed terms in the corresponding  $N$ space threshold limit, namely $N\rightarrow \infty$.  
Threshold resummation allows one to resum $\omega=2 a_s(\mu_R^2) \beta_0 \ln N$ terms to all orders in $\omega$ 
and then to organise the resulting perturbative result
in powers of coupling constant $a_s(\mu_R^2) = g_s^2(\mu_R^2)/16 \pi^2$, where $g_s$ is the strong coupling constant.  
Here, $\beta_0$ is the leading coefficient of QCD beta function.
If ${\cal O}_N$ is an observable in Mellin $N$ space, with $N$ being the
conjugate variable to $z$ of the observable ${\cal O}(z)$ in $z$ space,
then the resummation of threshold logarithms gives
\textcolor{black}{\begin{eqnarray}
\label{resumgen}
\ln {\cal O}_N = \ln N g^{\cal O}_1(\omega) + \sum_{i=0}^\infty a_s^i(\mu_R^2) g^{\cal O}_{i+2}(\omega) 
+ \ln g^{\cal O}_0(a_s(\mu_R^2))\,,
\end{eqnarray} }
where $g^{\cal O}_0(a_s(\mu_R^2))$ is $N$ independent and is given by 
\begin{eqnarray}
g^{\cal O}_0(a_s(\mu_R^2)) = \sum_{i=0}^\infty a_s^i(\mu_R^2) g^{\cal O}_{0i}\quad .
\end{eqnarray}
Inclusion of more and more terms in \eqref{resumgen} predicts the leading logarithms (LL), 
next to leading (NLL) etc logarithms of ${\cal O}$ to all orders in $a_s$.  
The functions $g^{\cal O}_i(\omega)$ are functions of process independent universal  
IR anomalous dimensions while  $g^{\cal O}_0$ depend on the hard process. 
For inclusive reactions such as DIS, invariant mass distribution of lepton pairs in DY, Higgs boson productions in various 
channels, all the ingredients to perform the  resummation of threshold logarithms in $N$ space up to third order
(next to next to next to leading logarithmic (N$^3$LL) accuracy) are available. 

While the resummed results provide reliable predictions that can be compared against the experimental data,
it is important to find out the role of sub leading terms namely $\ln^i(1-z),i=0,1,\cdot \cdot \cdot$,  
We call them by next to SV (NSV) contributions.  In addition to understand the role of NSV terms,
the question on weather these terms can also be resummed systematically to all orders exactly
like the way the leading SV terms are resummed remains unanswered.  \textcolor{black}{These questions have already
been addressed in great detail and remarkable progress has been made in recent times leading to a better understanding
of NSV terms.  For example, applying diagrammatic techniques and using factorisation properties  or through physical evolution equations, several
interesting results on both fixed order as well as resummed predictions for NSV terms are available for the production of a colorless state in hadron colliders.  See,  \cite{Laenen:2008ux,Laenen:2010uz,Bonocore:2014wua,Bonocore:2015esa,Bonocore:2016awd,DelDuca:2017twk,Bahjat-Abbas:2019fqa,Soar:2009yh,Moch:2009hr,deFlorian:2014vta,Beneke:2018gvs,Bahjat-Abbas:2019fqa,Beneke:2019mua,Beneke:2019oqx} for more details.}  In this paper, exploiting mass factorisation,
renormalisation group invariance and using Sudakov K plus G equation
we make an attempt to provide an all order result both in $z$ space and in $N$ space,  which can predict
NSV terms of \textcolor{black}{diagonal channels in} DY and Higgs boson production to all orders in perturbation theory.    

\section{Next to SV in $z$ space}
In the following, we study the inclusive \textcolor{black}{cross sections} for the production of a
pair of leptons in DY and the production of a single scalar Higgs boson in gluon fusion and in bottom quark annihilation.  
Let us
denote the corresponding inclusive cross sections generically by $\sigma(q^2,\tau)$.
In the QCD improved parton model, $\sigma$ is written in terms of 
parton level coefficient functions (CF) denoted by $\Delta_{ab}(q^2,\mu_R^2,\mu_F^2,z)$ convoluted with appropriate
parton distribution functions (PDFs), $f_c(x_i,\mu_F^2)$, of incoming partons:  
\begin{eqnarray}
\label{QCDpm}
\sigma(q^2,\tau) = \sigma_0(\mu_R^2) \sum_{ab} \int dx_1 \int dx_2 f_a(x_1,\mu_F^2) f_b(x_2,\mu_F^2) 
\Delta_{ab}(q^2,\mu_R^2,\mu_F^2,z) \,,
\end{eqnarray}
where $\sigma_0$ is the born level cross section.
The scaling variable $\tau$ is defined by $\tau=q^2/S$, \textcolor{black}{$S$ is hadronic center of mass energy}.  For DY, 
$q^2=M^2_{l^+l^-}$, the invariant mass of 
the final state leptons and  $q^2=m_H^2$ for the Higgs boson productions, with $m_H$ being
the mass of the Higgs boson.  
The subscripts $a,b$ in $\Delta_{ab}$
and $c$ in $f_c$ collectively denote the type of parton (quark, antiquark and gluon), 
their flavour etc.  The scaling variable $x_i$ is 
the  momentum fraction of the incoming partons.
In the CF, $z=q^2/\hat s$ is the partonic scaling variable and $\hat s$ is the partonic center of mass energy and is
related to hadronic $S$ by \textcolor{black}{$\hat{s}=x_1 x_2 S$} which implies  $z=\tau/(x_1 x_2)$.  The scale $\mu_F$ is factorisation scale 
which results from mass factorisation and the scale \textcolor{black}{$\mu_R$ is the renormalisation scale which results from
UV renormalisation of the theory}.  Both $\sigma_0$ and $\Delta_{ab}$ depend on the renormalisation scale,
however their product is independent of the scale if we include $\Delta_{ab}$ to all orders in perturbation
theory.

The partonic cross section is computable order by order in QCD perturbation theory.  Beyond leading order,
one encounters, UV, soft and collinear divergences at the intermediate stages of the computation.  If we use
dimensional regularisation to regulate all these divergences, the partonic cross sections depend on 
the  space time dimension $n=4+\epsilon$ and the divergences show up as poles in $\epsilon$.  The UV divergences
are removed by QCD renormalisation constants in modified minimal subtraction ($\overline {MS}$) scheme.  
The soft divergences from the gluons and the collinear divergence 
resulting from final state partons cancel independently when we perform the sum over all the degenerate states.  
Since the hadronic observables under study  are infrared safe, these partonic
cross sections are factorisable in terms of collinear singular Altarelli-Parisi \textcolor{black}{(AP) \cite{Altarelli:1977zs}} kernels $\Gamma_{ab}$ and finite
CFs at an arbitrary factorisation scale $\mu_F$.  
The factorised formula that relates the collinear finite CFs $\Delta_{ab}$ and  the parton level subprocesses 
is given by
\textcolor{black}{
\begin{eqnarray}
\label{MassFact}
{1 \over z} {\hat \sigma_{ab} (q^2,z,\epsilon) } = \sigma_0(\mu_R^2)\sum_{a' b'}  \Gamma^T_{a a'}(z,\mu_F^2,\epsilon) \otimes \left(
{\Delta_{a'b'}(q^2,\mu_R^2,\mu_F^2,z,\epsilon)}\right) \otimes \Gamma_{b'b}(z,\mu_F^2,\epsilon) \,. 
\end{eqnarray}}
These kernels are \textcolor{black}{then} absorbed into the bare PDFs 
to define collinear finite PDFs.  Note that the singular AP kernels  do not depend on the type of partonic reaction but 
depend only \textcolor{black}{on} the type of partons in addition to the scaling variable $z$ and scale \textcolor{black}{$\mu_F$}.  
The \textcolor{black}{symbol $\otimes$ refers to convolution, which is defined for functions, $f_i(x_i),i=1,2,\cdot \cdot \cdot,n$, as, 
\begin{eqnarray}
\label{conv}
	\left ( f_1 \otimes f_2 \otimes \cdot \cdot \cdot \otimes f_n\right)(z) 
= \prod_{i=1}^n \Bigg(\int dx_i f_i(x_i)\Bigg) \delta(z - x_1 x_2 \cdot\cdot \cdot x_n ) \,. 
\end{eqnarray} }
The partonic cross section in perturbation theory in QCD can be expressed in powers of \textcolor{black}{unrenormalised strong coupling constant $\hat{a}_s$:
\begin{eqnarray}
\label{Deltaab}
\hat \sigma_{ab}(q^2,z,\epsilon) = \sum_{i=0}^\infty \hat{a}_s^{i+\alpha}~\hat \sigma_{ab}^{(i)}(q^2,\mu_R^2,z,\epsilon) \,. 
\end{eqnarray}
where the value of $\alpha$ depends on the process under study.}
\textcolor{black}{Since the aim of this paper is to investigate the structure of NSV terms in diagonal channels,} we will restrict ourselves to $\Delta_{q \overline q}$ 
for DY, $\Delta_{b \overline b}$ for Higgs boson production in bottom quark annihilation and $\Delta_{gg}$ for
Higgs boson production in gluon fusion, \textcolor{black}{throughout the paper unless stated otherwise}.  
We call these CFs collectively by $\Delta_{c \overline c}$ with $c\overline c = q\overline q,
b \overline b, gg$.

\textcolor{black}{
Before we proceed further with the diagonal channels, let us study the structure of mass factorised results \eqref{MassFact}  for both
diagonal and off-diagonal channels in the threshold limit.  In particular, we would like to find out which are the terms
that survive if we want to retain only SV and/or NSV terms when we perform threshold expansion.}  
We begin with the mass factorisation formula for a diagonal channel.  We will show that to retain only SV and NSV terms in $\Delta_{c \overline c}$ using the mass factorised result,
it will be sufficient to keep only 
those components of AP kernels $\Gamma_{ab}$s and of $\hat \sigma_{ab}$s or $ \Delta_{ab}$s that 
upon convolution give SV and/or NSV terms. For definiteness, let us look at the mass factorised Drell-Yan result:
\begin{align}\label{massfactDY}
   \frac{\hat\sigma_{q \bar q}}{z\sigma_0}= & ~
    \Gamma_{qq}^T \otimes \Delta_{qq}
 \otimes \Gamma_{q \bar q}
 +
  \Gamma_{qq}^T \otimes \Delta_{qg}
 \otimes \Gamma_{g \bar q} 
+ \Gamma_{qq}^T \otimes  \Delta_{q\bar q}
 \otimes \Gamma_{\bar q \bar q}
\nonumber \\
 + &~ \Gamma_{qg}^T \otimes  \Delta_{g q}
 \otimes \Gamma_{ q \bar q}
+\Gamma_{qg}^T \otimes  \Delta_{gg}
 \otimes \Gamma_{g \bar q }
 + \Gamma_{qg}^T \otimes  \Delta_{g\bar q}
 \otimes \Gamma_{ \bar q \bar q }
 \nonumber \\
+& ~ \Gamma_{ q\bar q }^T \otimes  \Delta_{\bar q q}
 \otimes \Gamma_{  q \bar q }
  + \Gamma_{ q\bar q }^T \otimes  \Delta_{\bar q g}
 \otimes \Gamma_{  g \bar q }
+ \Gamma_{ q\bar q }^T \otimes  \Delta_{\bar q \bar q}
 \otimes \Gamma_{  \bar q \bar q }\,.
\end{align}
Here, we either have convolutions  with terms  involving only diagonal terms, such as $ \Gamma_{qq}^T \otimes \Delta_{q \overline q}
 \otimes \Gamma_{\bar q \bar q}$, or with 
terms involving one diagonal and a pair of
non-diagonal terms, for example $ \Gamma_{q q}^T\otimes  \Delta_{q g}  \otimes \Gamma_{g\bar q}$.
The former gives SV plus NSV terms  upon convolutions while the latter will 
give only beyond the NSV terms. 
And the diagonal $\Gamma_{cc}$s also contain convolutions with only diagonal AP splitting functions, $P_{cc}$,
or one diagonal and a pair of non-diagonal AP splitting functions $P_{ab},a\not = b$.
 We drop those terms in diagonal $\Gamma_{c c}$s that contain pair of non-diagonal $P_{ab}$s, \textcolor{black}{as they contribute to beyond NSV accuracy}.
This results in 
\begin{eqnarray}
{\hat \sigma_{q \bar q}^{\mathrm{sv+nsv}}  \over z\sigma_0} =\Gamma_{q q}^T\otimes \Delta_{q \bar q}^{\mathrm{sv+nsv}} \otimes \Gamma_{\bar q \bar q}\,.  
\end{eqnarray}
Similar argument will go through for $\hat \sigma_{b \overline b}$ and $\hat \sigma_{gg}$ as well. 
This allows us to write the mass factorised result given in \eqref{MassFact} in terms of only diagonal 
terms $\hat \sigma_{c \overline c}$, $\Delta_{c \overline c}$ and AP kernels $\Gamma_{c c}$ and 
the sum over $ab$ is dropped. 
\textcolor{black}{Hence, dropping beyond NSV terms and restricting to only diagonal terms results \eqref{MassFact} to take the simple form:}
\begin{align}
\label{MassFactRe}
{\Delta^{\rm sv+nsv}_{c\overline c}(q^2,\mu_R^2,\mu_F^2,z,\epsilon)} =  \sigma^{-1}_0(\mu_R^2)\Bigg(  \Big(\Gamma^T\Big)^{-1}_{cc}(z,\mu_F^2,\epsilon) &\otimes 
\frac{1}{z}\hat \sigma^{\rm sv+nsv}_{c\overline c} (q^2,z,\epsilon)  
\nonumber\\
&\otimes \Big(\Gamma\Big)^{-1}_{\bar c\bar c}(z,\mu_F^2,\epsilon) \Bigg)\,. 
\end{align}

In summary, since our main focus here is on SV and NSV terms  
resulting from quark initiated processes for DY and gluon or bottom quark initiated  processes for  Higgs boson production,
we can safely drop contributions from non-diagonal partonic channels in the mass 
factorised result of $\Delta_{c \overline c}$.  In addition,
gluon-gluon initiated channels which start contributing at NNLO onwards for DY and quark anti-quark initiated channels for Higgs
boson production are also dropped as they do not contribute \textcolor{black}{to} NSV of $\Delta_{c \overline c}$.  

\textcolor{black}{Turning our attention to off-diagonal terms,  for instance $\hat \sigma_{qg}$, we find
\begin{align}\label{massfactqg}
   \frac{\hat\sigma_{q g}}{z\sigma_0} = &~
    \Gamma_{qq}^T \otimes \Delta_{qq}
 \otimes \Gamma_{q g}
 +
  \Gamma_{qq}^T \otimes \Delta_{qg}
 \otimes \Gamma_{g g} 
 + \Gamma_{qq}^T \otimes  \Delta_{q\bar q}
 \otimes \Gamma_{\bar q g}
  \nonumber \\
 +&~ \Gamma_{qg}^T \otimes  \Delta_{g q}
 \otimes \Gamma_{ q g}
+\Gamma_{qg}^T \otimes  \Delta_{gg}
 \otimes \Gamma_{g g }
 + \Gamma_{qg}^T \otimes  \Delta_{g\bar q}
 \otimes \Gamma_{ \bar q g }
\nonumber \\
 +&~ \Gamma_{q\bar q }^T \otimes  \Delta_{\bar q q}
 \otimes \Gamma_{  q g }
  + \Gamma_{q\bar q }^T \otimes  \Delta_{\bar q g}
 \otimes \Gamma_{  gg }
 + \Gamma_{q\bar q }^T \otimes  \Delta_{\bar q \bar q}
 \otimes \Gamma_{  \bar q g }\,.
\end{align}
Like in the case of diagonal channels, the mass factorisation for the off-diagonal ones also contains both diagonal and off-diagonal
terms from $\Delta_{ab}$ and AP kernels, in different combinations.   As expected, in above result,  we find no single term
that can give pure SV contribution.  This is because every term contains at least one off-diagonal term.  Recall, this is not
the case for $\hat \sigma_{q \overline q}$.   Hence, mass factorised result for the off-diagonal channel
starts with NSV and beyond, where the former comes from terms  containing
at least two diagonal terms either from $\Delta_{ab}$ or $\Gamma_{ab}$.  
Since we are interested only in NSV terms, we drop terms that contain more than two off-diagonal
terms in the mass factorisation formula to obtain:
\begin{eqnarray}
\frac{\hat \sigma_{qg}^{\mathrm{sv+nsv}}}{z\sigma_0} =\Gamma_{q q}^T \otimes \Delta_{q \overline q}^{\mathrm{sv+nsv}}\otimes \Gamma_{\overline q g}  + \Gamma_{q q}^T\otimes\Delta_{qg}^{\mathrm{sv+nsv}} \otimes\Gamma_{gg} \,.
\end{eqnarray}
Note that the off-diagonal $\Delta_{qg}$ receives contribution from $\hat \sigma_{qg}$ as well as from $\Delta_{q \overline q}$
unlike the diagonal $\Delta_{q \overline q}$ which receives only from single $\hat \sigma_{q \overline q}$.  }

This analysis using mass factorisation formula and threshold expansion, which is valid to all orders in perturbation theory, demonstrates a simple structure for the diagonal $\Delta_{c \overline c}$ namely {it contains only one kind of term which comprises diagonal kernels and  $\hat \sigma_{c \overline c}$}.  
On the other hand, in the off-diagonal channel,  we have two kinds of terms containing  diagonal and off-diagonal $\Delta_{ab}$s 
which mix under factorisation.  
As we will see in the following,  due to the simple structure in the diagonal channels,  we can study  their all order structure of NSV logarithms using certain homogeneous differential equation.
However,  the off-diagonal ones pose challenge to such a study due to in-homogeneous terms present in the corresponding differential equation.  Hence, in the following, we will focus only on diagonal partonic channels.  

 Beyond leading order, the partonic channels that contribute to $\hat \sigma_{c \overline c}^{(i)}$ can be broadly
classified into two classes namely those containing no partonic final state/no emission and the ones with at least
one partonic final state.  The former ones are called form factor (FF) contributions while the latter ones are
called real emission contributions.  In FFs, the entire partonic center of mass energy goes into producing
a pair of leptons in DY or Higgs boson in Higgs boson production while in real emission processes, the initial state
energy is shared among all the final state particles.  Let us denote FF of DY by $\hat {F}_{q}$ and FF of
Higgs boson productions by $\hat {F}_b,\hat {F}_{g}$ respectively.  

{\color{black} Our next step is to  factor out the square of the \textcolor{black}{UV renormalised FF}  ($Z_{UV,c}\hat{F}_{c}$) with $c = q, \overline q, b,g$ from the partonic channels $\hat \sigma_{c \overline c }$. 
Here the $Z_{UV,c}$ is overall renormalisation constant that is required for  Higgs boson production from gluon fusion  and bottom quark annihilation. 
We call the resulting one by Soft-Collinear function, given by}
\begin{eqnarray}
\label{normS}
{\cal S}_{c}(\hat a_s,\mu^2,q^2,z,\epsilon) &=& \left(\sigma_0(\mu_R^2) \right)^{-1}
	\left(Z_{UV,c}(\hat{a}_s,\mu_R^2,\mu^2,\epsilon) \right)^{-2}
|\hat F_{c}(\hat a_s,\mu^2,Q^2,\epsilon)|^{-2} 
\nonumber \\
&& \times \delta(1-z) \otimes \hat \sigma^{\rm sv+nsv}_{c \overline c} (q^2,z,\epsilon) 
\end{eqnarray}
where $\hat a_s$ is the bare strong coupling constant, $Q^2 = -q^2$ 
Note that ${\cal S}_{c}$ does not depend on  $\mu_R^2$ and hence, ${\cal S}_{c}$ is RG invariant.  \textcolor{black}{The function ${\cal S}_{c}$
is computable in perturbation theory in powers of $\hat a_s$ and later in sec. \ref{sec:Sc} we discuss its perturbative structure and also how several of its coefficients can be determined from the fixed order results.}  
Since we have restricted ourselves to SV+NSV contributions to $\Delta_{c \overline c}$, 
that is those resulting from the phase space region in the limit $z\rightarrow 1$,
we keep only those terms that are proportional to distributions $\delta(1-z)$,
${\cal D}_i(z)$ and NSV terms of the kind of $\ln^i(1-z)$ with $i=0,1,\cdots$ and drop the rest of the terms resulting from
the convolutions.
Substituting  \textcolor{black}{for $\hat{\sigma}_{c\bar{c}}$ from \eqref{normS} in terms of ${\cal S}^c$, in \eqref{MassFactRe}}  and keeping only the diagonal terms in AP kernels, we find  
\textcolor{black}{
\begin{eqnarray}
\label{MasterF}
\Delta_c(q^2,\mu_R^2,\mu_F^2,z) &=& \Delta^{\rm{sv+nsv}}_{c \overline c}(q^2,\mu_R^2,\mu_F^2,z) \,, \nonumber \\
&=&\left(Z_{UV,c}(\hat{a}_s,\mu_R^2,\mu^2,\epsilon) \right)^2
|\hat F_{c}(\hat a_s,\mu^2,Q^2,\epsilon)|^2
\delta(1-z) 
\otimes 
\nonumber\\  &&
    \big(\Gamma^T\big)_{cc}^{-1}(z,\mu_F^2,\epsilon)\otimes\mathcal{S}_{c} (\hat a_s,\mu^2,q^2,z,\epsilon)\otimes
    \Gamma^{-1}_{\bar{c}\bar{c}}(z,\mu_F^2,\epsilon)\,.
\end{eqnarray}
The decomposition formula for $\Delta^{\rm{sv+nsv}}_{c \overline c}$ given in \eqref{MasterF}, is the first step towards obtaining the all order perturbative structure, which we are going to unravel in the subsequent section. It is to be noted that owing to the simplification in the mass factorised formula, given in \eqref{MassFactRe}, we obtain the above all order decomposition formula. It provides the pathway to study the partonic CFs in terms of certain building blocks namely the form factor $\hat F_{c}$, overall renormalisation constants $Z_{UV,c}$, the Soft-Collinear function $\mathcal{S}_{c}$ and the AP splitting kernels $\Gamma_{c c}$, which conspire  among themselves in such a way leading to a structure for $\Delta_c$ in terms of certain anomalous dimensions, universal and process dependent coefficients.   In the next sub-section, using differential equations that each   of these building blocks  satisfies, we obtain an all-order structure for $\Delta^{\rm{sv+nsv}}_{c \overline c}$.}

\subsection{Next to SV Formalism}
\textcolor{black}{
In this section we discuss the formalism which accounts for both SV and NSV corrections to $\Delta_c$ owing to the decomposition formula given in \eqref{MasterF}. We study the underlying evolution equations corresponding to each of the building blocks, namely $\{ \hat{F}_c, Z_{UV,c}, \Gamma_{cc}, \mathcal S_c\}$, with respect to the renormalisation and  factorization scales and also the energy scale of the process under study. Following this, we derive the perturbative structure of each of the components and thereby present the analytic structure of the partonic CF.}

In the master formula, \eqref{MasterF}, the form factor for the DY process is the
matrix element of vector current $\overline \psi_q \gamma_\mu \psi_q$ between on-shell quark states
and for the Higgs boson production in gluon fusion (bottom quark annihilation), 
it is the matrix element of $G_{\mu\nu}^a G^{\mu\nu a}$ ($\overline \psi_b \psi_b$) between on-shell gluon (bottom quark) 
states. 
Here $\psi_c$ is the $c$ type quark field operator and $G_{\mu \nu a}$ is the gluon field strength operator with $a$ being the 
$SU(N_c)$ gauge group index in the adjoint representation.  These FFs are known in QCD up to third
order in perturbation theory, 
\cite{vanNeerven:1985xr,Harlander:2000mg,Ravindran:2004mb,Moch:2005tm,Gehrmann:2005pd,Baikov:2009bg,Gehrmann:2010ue,Gehrmann:2014vha,vonManteuffel:2016xki,Henn:2016men,Henn:2019rmi,vonManteuffel:2020vjv,Gehrmann:2010tu}. 
\textcolor{black}{The evolution equation for the overall renormalisation constant with respect to the renormalisation scale reads as,
\begin{equation}\label{ZUVEq}
 \mu_R^2 \frac{d}{d\mu_R^2} \ln Z_{UV,c} (\hat{a}_s, \mu_R^2, \mu^2, \epsilon) =  
 \sum_{i=1}^{\infty}  a_s^i (\mu_R^2)~ \gamma^c_{i-1} \,,
\end{equation}
where $\gamma^c_{i}$ is the UV anomalous dimension. For the vector current, the UV anomalous dimension is zero to all orders in QCD while
for the Higgs boson productions, $\gamma^c_{i}$s are non-zero.  For $c=b$, see \cite{Vermaseren:1997fq} and
for $c=g$, it is expressed in terms of QCD beta function coefficients to all orders \cite{Chetyrkin:2005ia}.} 

Perturbative results of FF in renormalisable quantum field theory demonstrate rich structure, in particular,
one finds that they satisfy certain differential equations.  The simplest one is the RG equation that FFs satisfy, namely 
$\mu_R^2\frac{ d \hat F_c}{d\mu_R^2} = 0$, using which  
we can  predict the logarithms resulting from the UV sector, i.e., the logarithms of the form $\ln^k(\mu_R^2), k=1,\cdots$ at
every order in perturbation theory.  In addition, these FFs satisfy     
Sudakov differential equation \cite{Sudakov:1954sw,Sen:1981sd,Collins:1989bt,Magnea:1990zb,Magnea:2000ss,Sterman:2002qn,Moch:2005id,Ravindran:2005vv} which is used to study their IR structure in terms of certain IR 
anomalous dimensions such as cusp $A^c$,collinear $B^c$ and soft $f^c$ anomalous dimensions.    
In dimensional regularisation, the equation takes the following form:
\begin{align}\label{KGFF}
Q^2\frac{d}{dQ^2}\ln \hat{F}_c \big( \hat{a}_s, Q^2,\mu^2,\epsilon\big)= \frac{1}{2} \Big[ K^c \Big(& \hat{a}_s, \frac{\mu_R^2}{\mu^2},\epsilon\Big) 
+ G^c \Big( \hat{a}_s,\frac{Q^2}{\mu_R^2},\frac{\mu_R^2}{\mu^2},\epsilon \Big) \Big] \,,
\end{align}
where $Q^2=-q^2$.  The above equation is called K+G equation.  The unrenormalised FFs contain both UV and IR divergences.  The latter result from
soft gluons and massless partons which give soft and collinear divergences respectively.  
UV divergences go away after UV renormalisation.  The IR divergences of the FFs can be shown to
factorise.  The divergence of FFs are such that the factorised IR divergent part is $q^2$ dependent.
The consequence of these facts is that the right hand side of the differential equation can expressed 
in terms of two functions $K^c$ and $G^c$
in such a way that $K^c$ accounts for all the poles in $\epsilon$ whereas $G^c$ is finite term
in the limit $\epsilon\rightarrow 0$.   The RG invariance of FFs implies, in the limit $\epsilon \rightarrow 0$,
\begin{align}
\label{RGKG}
	\mu_R^2\frac{d}{d\mu_R^2}K^c\Big(\hat{a}_s,\frac{\mu_R^2}{\mu^2},\epsilon \Big) =-
	\mu_R^2\frac{d}{d\mu_R^2}G^c\Big(\hat{a}_s,\frac{Q^2}{\mu_R^2},\frac{\mu_R^2}{\mu^2},\epsilon \Big) = -A^c(a_s(\mu_R^2)).
\end{align}
The solutions to \eqref{RGKG} are given in \cite{Ravindran:2005vv,Ravindran:2006cg}. 
Substituting these solutions in \eqref{KGFF} one can find the structure of FF in 
terms of IR anomalous dimensions $A^c$ (cusp), $B^c$ (collinear) and $f^c$ (soft) as well as the process dependent quantities ($g^{c,k}_j$). 
A more elaborate discussion on the structure of FF can be found in \cite{Ravindran:2005vv}.  
The IR anomalous dimensions are known to three loops in QCD, 
see \cite{Kodaira:1981nh,Kodaira:1982az,Vogt:2004mw,Ravindran:2004mb,Moch:2005tm,Moch:2004pa,vonManteuffel:2016xki,Das:2020adl} and for beyond three loops, see \cite{vonManteuffel:2020vjv}.

The fact that the initial state collinear divergences in parton level cross sections factorises in terms of AP
kernels $\Gamma_{ab}(z,\mu_F^2,\epsilon)$ implies RG evolution equation with respect to
the scale $\mu_F$:
\begin{equation}\label{RGGam}
   \mu_F^2\frac{d}{d\mu_F^2}\Gamma_{ab}\big(z,\mu_F^2,\epsilon\big) = 
	\frac{1}{2}\sum_{a'=q,\overline q,g} P_{a a'} \big(z,a_s(\mu_F^2)\big)\otimes \Gamma_{a'b}\big(z,\mu_F^2,\epsilon\big) \,,
\quad \quad a,b = q,\overline q,g \,.
\end{equation}
Since we are interested only \textcolor{black}{in}  
diagonal Altarelli-Parisi kernels for our analysis, the corresponding AP splitting functions 
$ P_{c c}\big(z,\mu_F^2\big)$ are expanded  
around $z=1$ and all those terms that do not contribute to SV+NSV are dropped.  The AP splitting functions
near $z=1$ take the following form:
\begin{eqnarray}
\label{eq:KernelNSV}
	P_{cc}\big(z,a_s(\mu_F^2)\big) &=& 2 B^c(a_s(\mu_F^2)) \delta(1-z) + P^{\prime}_{cc}\big(z,a_s(\mu_F^2)\big)\,,
\end{eqnarray}
where,
\begin{eqnarray}
	P^{\prime}_{cc}\big(z,a_s(\mu_F^2)\big) &=& 2  \Bigg[ A^c(a_s(\mu_F^2)) {\cal D}_0(z)
                      + C^c(a_s(\mu_F^2)) \ln(1-z) + D^c(a_s(\mu_F^2)) \Bigg]
\nonumber\\&&
		      + {\cal O}((1-z)) \,.
\end{eqnarray}
In the rest of the paper, we drop the terms in $P^{\prime}_{cc}$ proportional to ${\cal O}((1-z))$ for our study.
The constants $C^c$ and $D^c$ can be obtained from the 
the splitting functions $P^{\prime}_{c c}$ which are known to three loops in QCD \cite{Moch:2004pa,Vogt:2004mw}
(see \cite{GonzalezArroyo:1979df,Curci:1980uw,Furmanski:1980cm,Hamberg:1991qt,Ellis:1996nn,Moch:2004pa,Vogt:2004mw,Soar:2009yh,Ablinger:2017tan,Moch:2017uml} for the lower order ones).
Similar to the cusp and the collinear anomalous dimensions, the constants $C^c$ and $D^c$ are also expanded
in powers of $a_s(\mu_F^2)$ as:
\begin{eqnarray}
C^c(a_s(\mu_F^2)) = \sum_{i=1}^\infty a_s^i(\mu_F^2) C_i^c,
\quad \quad 
D^c(a_s(\mu_F^2)) = \sum_{i=1}^\infty a_s^i(\mu_F^2) D_i^c \,,
\end{eqnarray}
where $C_i^c$ and $D_i^c$ to third order are available in \cite{Moch:2004pa,Vogt:2004mw}.
\subsubsection{The Soft-Collinear Function
}
\label{sec:Sc}
{\color{black}Our next task is to study the Soft-Collinear function, ${\cal S}_{c}$, in detail. 
The Eq.\eqref{normS}
can be used to compute this function order by order in QCD perturbation theory.  The ${\cal S}_{c}$ should contain right IR divergences
to cancel those resulting from FF and AP kernels to give IR finite $\Delta_{c}$.  
The IR structure of ${\cal S}_{c}$ in the SV limit was studied
in \cite{Ravindran:2005vv,Ravindran:2006cg} using a differential equation analogous to \eqref{KGFF} supplemented
with RG invariance.  
It was found that this function demonstrates a rich
infrared structure in the SV approximation. 
Further, it provides a suitable framework to obtain SV contribution order by order 
in perturbation theory. Since, the function ${\cal S}_c$ obtained in \cite{Ravindran:2005vv,Ravindran:2006cg} 
is an all order result in $z$ space which allows one to write the integral representation
suitable for studying resummation in Mellin $N$ space.   In the following, we proceed along this direction
to study NSV contributions in $z$ space to all orders in perturbation theory and to provide an integral
representation that can be used for performing  Mellin $N$ space resummation.} 
\textcolor{black}{Using \eqref{MasterF} and the K+G equation of FFs, \eqref{KGFF}, 
one can set up an evolution equation for the functions ${\cal S}_c$.  In other words, 
we can easily show that ${\cal S}_c$ satisfies  
K+G type of differential equation of the form
\begin{align}\label{KGSc}
q^2\frac{d}{dq^2}{\ln \mathcal S_c} = \Big[ \overline{K}^c \Big(& \hat{a}_s, \frac{\mu_R^2}{\mu^2},\epsilon,z \Big) 
+ \overline{G}^c \Big( \hat{a}_s,\frac{q^2}{\mu_R^2},\frac{\mu_R^2}{\mu^2},\epsilon,z \Big) \Big] \,,
\end{align}
where the right hand side of the above equation is written as a sum of 
$\overline{K}^c$ which accounts for all the divergent terms and $\overline{G}^c$ which is a finite 
function of $(z,\epsilon)$. In consequence to the above differential equation, \eqref{KGSc}, the Soft-Collinear Function, ${\cal S}_c$, admits an exponential solution given by:
\begin{equation}\label{RGSc}
    S_{c}(\hat{a}_s, q^2,\mu^2, \epsilon, z) = 
    \mathcal{C}\exp\Bigg(2 \mathrm{\Phi}^c(\hat{a}_s,q^2,\mu^2, \epsilon,z)\Bigg)\,,
\end{equation}}
where the exponent, ${\mathrm \Phi}^c$ gets contribution from $c \overline c$ initiated processes
containing at least one real radiation. The symbol ``$\mathcal{C}$" refers to convolution. For instance $\mathcal{C}$ acting on any exponential of a function has the following expansion:
\begin{eqnarray}
	\mathcal{C}e^{f(z)} = \delta(1-z) + \frac{1}{1!}f(z) + \frac{1}{2!}\big(f\otimes f\big)(z) + \cdots
\end{eqnarray}
In addition, ${\cal S}^c$s satisfy renormalisation group invariance namely
$\mu_R^2 \frac{d {\cal S}^c}{ d\mu_R^2} = 0$ which implies
\begin{eqnarray}
\label{eq:KGInv}
\mu_R^2\frac{d}{d\mu_R^2}\overline{K}^c(a_s(\mu_R^2),z) &=& - 
\mu_R^2\frac{d}{d\mu_R^2}\overline{G}^c(a_s(\mu_R^2),z) 
= -\overline A^c(a_s(\mu_R^2))\delta(1-z) \,,
\end{eqnarray}
where $\overline A^c$ is analogous of cusp anomalous dimension that appears in K+G equation of FFs.  
\textcolor{black}{The perturbative solution to \eqref{KGSc} can be obtained by integrating the differential equation after
substituting the fixed order solutions of RGs for $\overline K^c$ and $\overline G^c$.   Hence, we propose
an all order ansatz for the solution $\mathrm{\Phi}^c$ which takes the
general form } 
\begin{align}\label{PhiSV1}
\mathrm{\Phi}^{c}(\hat{a}_s, q^2,\mu^2,z, &\epsilon) = 
\sum_{i=1}^\infty\hat{a}_s^i\Big(\frac{q^2(1-z)^2}{\mu^2 z}\Big)^{i\frac{\epsilon}{2}} S_\epsilon^{i}
\Big(\frac{i\epsilon}{1-z}\Big)\hat{\phi}_{c}^{(i)}(z,\epsilon).
\end{align}
where $ S_\epsilon =\exp (\frac{\epsilon}{2}[\gamma_E - \ln (4\pi )])$
with $ \gamma_E $ being the Euler Mascheroni constant.  
The form of the solution given in \eqref{PhiSV1} is inspired by the  
result for the production of a pair of leptons in quark antiquark channel or Higgs boson
in gluon fusion at next to leading order in $a_s$.
The term \textcolor{black}{$\Big(\frac{q^2(1-z)^2}{\mu^2 z}\Big)^{\frac{\epsilon}{2}}$} in the parenthesis results from two body phase space while $\hat \phi_c(z,\epsilon)/(1-z)$ comes 
from the square of the matrix elements for  corresponding amplitudes. 
In general, the term $q^2 (1-z)^2/z$ inside the parenthesis is the hard scale in the problem and it 
controls the evolution of
$\mathrm{\Phi}^c$ at every order. The function $\hat \phi^{(i)}_c(z,\epsilon)$ is regular as $z \rightarrow 0$ but
contains poles in $\epsilon$.   
We have factored out $1/(1-z)$ explicitly so that it generates all the distributions ${\cal D}_j$ and $\delta(1-z)$ and 
NSV terms $\ln^k(1-z), k=0,\cdots$ when combined with the factor $((1-z)^2)^{i \epsilon/2}$ and 
$\hat \phi^{(i)}_c(z,\epsilon)$ at each order in $\hat a_s$.  Note that the term $z^{-i \epsilon/2}$ inside 
the parenthesis does not give distributions ${\cal D}_j$ and $\delta(1-z)$, however they can contribute to NSV terms
$\ln^j(1-z) ,j=0,1,\cdots$ when we expand around $z=1$.  
In addition the terms proportional to $(1-z)$ in $\hat \phi_c$ near $z=1$ also give NSV terms for $\mathrm{\Phi}^c$. 
Although the form of  solution for $\mathrm{\Phi}^c$ is good enough to study NSV terms, we rewrite this in a convenient form 
which separates SV terms from the NSV in $\mathrm{\Phi}^c$. Hence, 
we decompose $\mathrm{\Phi}^c$ as $\mathrm{\Phi}^c$ = $\mathrm{\Phi}_{A}^{c}$ + $\mathrm{\Phi}_{B}^{c}$ 
in such a way that $\mathrm{\Phi}_{A}^{c}$ contains only SV terms 
and the remaining $\mathrm{\Phi}_{B}^{c}$ contains next 
to soft-virtual terms in the limit $z \rightarrow 1$. 
The distribution $\mathrm{\Phi}_{A}^{c}$ satisfies K+G equation  
given in Eq.(35) 
of \cite{Ravindran:2005vv} also see  \cite{Ravindran:2006cg} for details. 
\textcolor{black}{The solution} for $\mathrm{\Phi}_{A}^{c}$ in powers of  $\hat a_s$ in dimensional regularisation
is given in \cite{Ravindran:2005vv}.  It is given by 
\begin{align}\label{PhiSV}
\mathrm{\Phi}_{A}^{c}(\hat{a}_s,
q^2,\mu^2, &\epsilon,z) = \sum_{i=1}^\infty\hat{a}_s^i\Big(\frac{q^2(1-z)^2}{\mu^2}\Big)^{i\frac{\epsilon}{2}}
 S_{\epsilon}^{i}
\Big(\frac{i\epsilon}{1-z}\Big)\hat{\phi}_{SV}^{c (i)}(\epsilon) \,, 
\end{align} 
where,
\begin{eqnarray}
\hat{\phi}_{SV}^{c(i)}(\epsilon) = \frac{1}{i\epsilon}\Big[ \overline{K}^{c(i)}(\epsilon) + \overline{G}_{SV}^{c(i)}(\epsilon) \Big]\,.
\end{eqnarray} 
The constants $\overline{K}^{c(i)}(\epsilon)$ and $\overline{G}_{SV}^{c(i)}(\epsilon) $ are known to third order in perturbation theory
\cite{Ravindran:2005vv,Ravindran:2006cg,deFlorian:2012za,Ahmed:2014cla,Li:2014afw}. {\color{black} For reader's convenience, we enlist the results of $\overline{K}^{c(i)}(\epsilon)$ and $\overline{G}_{SV}^{c(i)}(\epsilon) $ in Appendix[\ref{app:PhiSV}]. After substituting these perturbative constants one can get the  perturbative structure of the SV coefficients as:
\begin{eqnarray}
\hat{\phi}_{SV}^{c(1)}(\epsilon)&=&{1\over \epsilon^2} \Bigg(2 A_1^c\Bigg) 
              +{1 \over \epsilon} \Bigg({\cal \overline{G}}_1^c(\epsilon)\Bigg)
\nonumber\\[2ex]
\hat{\phi}_{SV}^{c(2)}(\epsilon)&=&{1\over \epsilon^3} \Bigg(-\beta_0 A_1^c\Bigg) 
                  +{1\over \epsilon^2} \Bigg({1 \over 2} A_2^c 
                  - \beta_0  {\cal \overline{G}}_1^c(\epsilon)\Bigg)
                  +{1 \over 2 \epsilon} {\cal \overline{G}}_2^c(\epsilon)
\nonumber\\[2ex]
\hat{\phi}_{SV}^{c(3)}(\epsilon)&=& {1\over \epsilon^4} \Bigg({8 \over 9}\beta_0^2 A_1^c\Bigg) 
                  + {1\over \epsilon^3} \Bigg(-{2 \over 9} \beta_1 A_1^c
                    -{8 \over 9} \beta_0 A_2^c -{4 \over 3} 
                     \beta_0^2 {\cal \overline{G}}_1^c(\epsilon)\Bigg) 
\nonumber\\[2ex]
&&                  +{1\over \epsilon^2} \Bigg({2 \over 9} A_3^c 
                   -{1 \over 3} \beta_1 {\cal \overline{G}}_1^c(\epsilon) 
                   -{4 \over 3}\beta_0 {\cal \overline{G}}_2^c(\epsilon)\Bigg)
                  +{1 \over \epsilon}\Bigg({1 \over 3} {\cal \overline{G}}_3^c(\epsilon)\Bigg)
\nonumber\\[2ex]
\hat{\phi}_{SV}^{c(4)}(\epsilon)&=& {1\over \epsilon^5} \Bigg(-\beta_0^3 A_1^c\Bigg) 
                  +{1 \over \epsilon^4} \Bigg({2 \over 3} \beta_0 \beta_1 A_1^c
                   +{3 \over 2}\beta_0^2 A_2^c -2 \beta_0^3 {\cal \overline{G}}_1^c(\epsilon)\Bigg)
\nonumber\\[2ex]
&&                  -{1 \over \epsilon^3} \Bigg({1 \over 12} \beta_2 A_1^c 
                     -{1 \over 4} \beta_1 A_2^c 
                    - {3 \over 4}\beta_0 A_3^c +{4 \over 3} 
                       \beta_0 \beta_1 {\cal \overline{G}}_1^c(\epsilon)
                    +3\beta_0^2 {\cal \overline{G}}_2^c(\epsilon)\Bigg)
\nonumber\\[2ex]
&&                  +{1 \over \epsilon^2} \Bigg({1 \over 8} A_4^c
                       -{1 \over 6} \beta_2 {\cal \overline{G}}_1^c(\epsilon)
                       -{1 \over 2} \beta_1 {\cal \overline{G}}_2^c(\epsilon) 
                    -{3\over 2} \beta_0 {\cal \overline{G}}_3^c(\epsilon)\Bigg)
                  +{1 \over \epsilon} \Bigg({1 \over 4} {\cal \overline{G}}_4^c(\epsilon)\Bigg)
\end{eqnarray}}
The integral representation for $\mathrm{\Phi}_{A}^{c}$ is given in  \cite{Ravindran:2006cg} and is
reproduced here for completeness: 
\begin{eqnarray}
\label{phiA}
    \mathrm{\Phi}_{A}^{c}(\hat{a}_s,\mu^2,q^2,z,\epsilon\big) &=
	& \bigg(\frac{1}{1-z}\bigg\{\int_{\mu_F^2}^{q^2(1-z)^2} \frac{d\lambda^2}{\lambda^2} A^c(a_s(\lambda^2)) + \overline{G}^c_{SV}\big(a_s(q^2(1-z)^2),\epsilon\big) \bigg\} \bigg)_+
   \nonumber \\
&&   + \delta(1-z) \sum_{i=1}^\infty \hat a_s^i \bigg(\frac{q^2}{\mu^2}\bigg)^{i\frac{\epsilon}{2}} 
S_\epsilon^i  \hat{\phi}^{c(i)}_{SV}(\epsilon) 
\nonumber\\ &&
+ {1 \over (1-z)_+}\sum_{i=1}^\infty \hat a_s^i  \bigg(\frac{\mu_F^2}{\mu^2}\bigg)^{i\frac{\epsilon}{2}}
S_\epsilon^i  \overline{K}^{c(i)}(\epsilon)     .
\end{eqnarray}

 Having all the information about the SV coefficients, let us now study in detail the structure of $\mathrm{\Phi}_{B}^{c}$ using the \eqref{KGSc}.
Subtracting out the K+G equation for the SV part $\mathrm{\Phi}_{A}^{c}$ from {\eqref{PhiSV}, 
we find  that $\mathrm{\Phi}_{B}^{c}$ satisfies
\begin{align}\label{KGphiNSV}
q^2\frac{d}{dq^2}\mathrm{\Phi}_{B}^{c}(q^2,z,\epsilon) = \frac{1}{2} \Big[ G_{L}^c \Big( \hat{a}_s,\frac{q^2}{\mu_R^2},\frac{\mu_R^2}{\mu^2},\epsilon,z \Big) \Big] \,,
\end{align}
where $G_{L}^{c} = \overline G^c - \overline G^c_{SV}$,
\begin{align}
G_{L}^{c} \Big( \hat{a}_s,\frac{q^2}{\mu_R^2},\frac{\mu_R^2}{\mu^2},z,\epsilon \Big) &= 
\sum_{i=1}^\infty a_s^i\big(q^2 (1-z)^2\big)  \mathcal{G}_{L,i}^{c}(z,\epsilon).
\end{align}
\textcolor{black}{The NSV part of the solution that satisfies \eqref{KGphiNSV} takes the following form}
\begin{equation}\label{phiB}
    \mathrm{\Phi}_{B}^{c}(\hat{a}_s,\mu^2,q^2,z,\epsilon) = \sum_{i=1}^{\infty} \hat a_s^i \bigg(\frac{q^2 (1-z)^2}{\mu^2}\bigg)^{i \frac{\epsilon}{2}} S_\epsilon^i {\color {black} \hat \varphi_c^{(i)}(z,\epsilon) }\,, 
\end{equation}
where the perturbative expansion of the NSV coefficient $\hat \varphi_{c}^{(i)}(z,\epsilon)$ reads as,
\begin{align}
\label{phifc}
    {\color{black} \hat \varphi_{c}^{(1)}(z,\epsilon)} &= \frac{1}{\epsilon} \mathcal{G}_{L,1}^c(z,\epsilon) \,, \nonumber \\
      {\color{black} \hat \varphi_{c}^{(2)}(z,\epsilon)} &= \frac{1}{\epsilon^2}(-\beta_0 \mathcal{G}_{L,1}^c(z,\epsilon)) + \frac{1}{2\epsilon}                      \mathcal{G}_{L,2}^c(z,\epsilon) \,, \nonumber \\
      {\color{black} \hat  \varphi_{c}^{(3)}(z,\epsilon) }&=  \frac{1}{\epsilon^3} \bigg(\frac{4}{3} \beta_0^2  \mathcal{G}_{L,1}^c(z,\epsilon)\bigg)
		    + \frac{1}{\epsilon^2} \bigg(-\frac{1}{3} \beta_1  \mathcal{G}_{L,1}^c(z,\epsilon) - \frac{4}{3} \beta_0 \mathcal{G}_{L,2}^c(z,\epsilon) \bigg)
                     + \frac{1}{3\epsilon}   \mathcal{G}_{L,3}^c(z,\epsilon) \,, \nonumber \\
      {\color{black} \hat \varphi_{c}^{(4)}(z,\epsilon)} &= \frac{1}{\epsilon^4} (-2\beta_0^3  \mathcal{G}_{L,1}^c(z,\epsilon))
                    + \frac{1}{\epsilon^3} \bigg( \frac{4}{3} \beta_0 \beta_1  \mathcal{G}_{L,1}^c(z,\epsilon) + 3 \beta_0^2  \mathcal{G}_{L,2}^c(z,\epsilon) \bigg)
                     \nonumber \\
                    & + \frac{1}{\epsilon^2} \bigg( -\frac{1}{6} \beta_2  \mathcal{G}_{L,1}^c(z,\epsilon) 
                    -\frac{1}{2} \beta_1  \mathcal{G}_{L,2}^c(z,\epsilon) -\frac{3}{2} \beta_0  \mathcal{G}_{L,3}^c(z,\epsilon) \bigg)
                    + \frac{1}{4\epsilon}  \mathcal{G}_{L,4}^c(z,\epsilon) \,.
\end{align}
The $\epsilon$-expansion of the renormalized NSV quantities $\mathcal{G}_{L,i}^c(z,\epsilon)$ can be further decomposed as,
\begin{equation}
    \mathcal{G}_{L,i}^c(z,\epsilon) =  L^c_i(z) + \overline \chi_{L,i}^c(z) + \sum_{j=1}^{\infty}\epsilon^j \mathcal{G}_{L,i}^{c,(j)}(z)\,,
\end{equation}
with
\begin{align}
\label{phisc}
  \overline \chi_{L,i}^c(z) =  \overline \chi_{i}^c\Big|_{\left(\overline {\cal G}^{c,(j)}_{i} \rightarrow \mathcal{G}_{L,i}^{c,(j)}(z)\right)} \,,
 \end{align}
where $\overline \chi^c_i$ is given in \eqref{eq:bBH-Cg}. \textcolor{black}{Unlike the SV renormalized coefficients $\overline{\mathcal{G}}_{i}^{c,(j)}$, the NSV coefficients} $\mathcal{G}_{L,i}^{c,(j)}(z)$ in the above equations are parametrised in terms of
$\ln^k(1-z), k=0,1,\cdots$ and all the terms that vanish as $z\rightarrow 1$ are dropped
\begin{eqnarray}
\label{GikLj}
\mathcal{G}_{L,i}^{c,(j)}(z) = \sum_{k=0}^{i+j-1} \mathcal{G}_{L,i}^{c,(j,k)} \ln^k(1-z) \,.
\end{eqnarray}
The highest power of the $\ln(1-z)$ at every order depends on the order of the perturbation, namely
the power of $a_s$ and also the power of $\epsilon$ at each order in $a_s$. \textcolor{black}{ We determine this highest power by studying results for the bare partonic cross sections $\hat \sigma_{c\overline c}$ at higher orders in $\hat a_s$, expanded in powers of $\epsilon$ to high accuracy.  Alternatively, we can use the known mass factorised results for $\Delta_{c \overline c}$ to obtain this power.  In the former approach, we used  the results for  $\hat \sigma_{c \overline c}$, computed up to second order in $\hat a_s$ , i.e, $i=1,2$ with $\epsilon$ expanded up to 3rd power for $i=1$ and first power for $i=2$.  In the case of
$\Delta_{c \overline c}$, we used the known results up to third order in $a_s$ to obtain the highest power of logarithms.  Extrapolating the findings from these two fixed order results to all orders in $\hat a_s$ and $\epsilon$, we obtain the highest power for $\ln(1-z)$ to be $i+j-1$.  We devote a separate sub-section, ( see \ref{ap:PhiLog}), to elaborate this peculiar structure of the logarithms}. 

\textcolor{black}{Similar to the SV case, the NSV function $\rm \Phi_B^c$ can be written in an integral form using \eqref{KGphiNSV} and  the perturbative structure given in \eqref{phifc} as,}

\begin{eqnarray}
\label{phiBint}
   \mathrm{\Phi}_{B}^{c}(\hat{a}_s,\mu^2,q^2,z,\epsilon\big) &= & \int_{\mu_F^2}^{q^2(1-z)^2} \frac{d\lambda^2}{\lambda^2} L^c(a_s(\lambda^2),z) + \varphi_{f,c}\big(a_s(q^2(1-z)^2),z,\epsilon\big) |_{\epsilon=0}  \nonumber\\ 
    &&+ \varphi_{s,c}\big(a_s(\mu_F^2),z,\epsilon\big) \,,
\end{eqnarray}
Here, the first line is completely finite as $\epsilon \rightarrow 0$
while the second line, $\varphi_{s,c}$, is divergent.  
The fact that the $\mathrm{\Phi}_{B}^{c}$ is RG invariant implies that $\varphi_{s,c}$ satisfies the renormalisation
group equation:
\begin{eqnarray}
\label{RGphis}
\mu_F^2 {d \over d\mu_F^2}  \varphi_{s,c}(a_s(\mu_F^2),z) = L^c (a_s(\mu_F^2),z).
\end{eqnarray}
Further the $\Delta_c$ in \eqref{MasterF} is finite at every order in $a_s$
in the limit $\epsilon \rightarrow 0$ allows us
to determine the coefficients $L^c$ in terms of the NSV coefficients $C^c$ and $D^c$ in splitting kernels, given in \eqref{eq:KernelNSV}.  We find, at each order in perturbative expansion
\begin{align}
L^c (a_s(\mu_F^2),z) &= \sum_{i=1}^{\infty}a_s^i(\mu_F^2)L_c^i(z)
\text{~~with},~~
L^c_i(z) = C_i^c \ln(1-z) + D_i^c \,,
\end{align}  {
where the coefficients $C^c_i$ and $D^c_i$ are related to those of cusp $A^c_i$ and collinear $B^c_i$ anomalous dimensions in the following way up to third order \cite{Moch:2004pa,Dokshitzer:2005bf}:
\begin{align}
     D^c_1 =&  -A_1^c\,, \qquad
 D_2^c = -A_2^c + A_1^c\left(B_1^c-\beta_0\right)\,,
 \nonumber\\
 D_3^c =&-A_3^c-A_1^c\left(-B_2^c+\beta_1\right)-A_2^c\left(-B_1^c+\beta_0\right)\,,
 \nonumber\\
 C_1^c =& 0\,, \qquad
 C_2^c = \Big(A_1^c\Big)^2\,, \qquad
 C_3^c = 2 A_1^c A_2^c\,.
\end{align}
{\color{black}
Having fixed the divergent part of $\mathrm{\Phi}_{B}^{c}$ completely, we turn to the structure of the finite piece $\varphi_{f,c}$. We first expand them in powers of renormalised coupling $a_s$:}
{\color{black}
\begin{align}
\label{varphiexp}
\varphi_{f,c}(a_s(q^2(1-z)^2),z) =&
\sum_{i=1}^\infty a_s^i(q^2(1-z)^2) \sum_{k=0}^i \varphi_{c,i}^{(k)} \ln^k(1-z) \,,
\end{align}}
{\color{black}
where the highest power of $\ln(1-z)$ are in accord with the same in Eq.\eqref{GikLj}. We will discuss more on this structure in sec. \ref{ap:PhiLog}. The coefficients $\varphi_{c,i}^{(k)}$ can be expressed in terms of their unrenormalised counter part ${\cal G}^{c,(j,k)}_{L,i}$'s in \eqref{GikLj} as:
\begin{eqnarray}\label{eq:varphi}
\varphi_{c,1}^{(k)} &=&  \mathcal{G}_{L,1}^{c,(1,k)}, \quad \quad k=0,1\nonumber\\
\varphi_{c,2}^{(k)} &=&  \bigg(\frac{1}{2}\mathcal{G}_{L,2}^{c,(1,k)} + \beta_0\mathcal{G}_{L,1}^{c,(2,k)}\bigg), 
k = 0,1,2\nonumber\\
\varphi_{c,3}^{(k)} &=&  \bigg(\frac{1}{3}\mathcal{G}_{L,3}^{c,(1,k)} + \frac{2}{3}\beta_1\mathcal{G}_{L,1}^{c,(2,k)} + \frac{2}{3}\beta_0\mathcal{G}_{L,2}^{c,(2,k)} + \frac{4}{3}
         \beta_0^2\mathcal{G}_{L,1}^{c,(3,k)}\bigg), \quad \quad k=0,1,2,3\nonumber\\
\varphi_{c,4}^{(k)} &=& \bigg( \frac{1}{4}\mathcal{G}_{L,4}^{c,(1,k)} + \frac{1}{2}\beta_2\mathcal{G}_{L,1}^{c,(2,k)} + \frac{1}{2}\beta_1\mathcal{G}_{L,2}^{c,(2,k)} + \frac{1}{2}
         \beta_0\mathcal{G}_{L,3}^{c,(2,k)} 
         + 2\beta_0\beta_1\mathcal{G}_{L,1}^{c,(3,k)} 
         + \beta_0^2\mathcal{G}_{L,2}^{c,(3,k)}\nonumber\\
         && + 
         2\beta_0^3\mathcal{G}_{L,1}^{c,(4,k)}\bigg), \quad \quad k=0,1,2,3,4
\end{eqnarray}
with $\mathcal{G}_{L,1}^{c,(2,3)},\mathcal{G}_{L,1}^{c,(2,4)},\mathcal{G}_{L,2}^{c,(2,4)},\mathcal{G}_{L,1}^{c,(3,4)}$ are all zero. The structure of divergent and finite pieces of ${\rm \Phi}_{B}^c$ allows us to determine the coefficients  ${\cal G}^{c,(j,k)}_{L,i}$ and $\varphi_{c,i}^{(k)}$ and we postpone the discussion on this to next section. }

\textcolor{black}{So far, we have discussed the logarithmic structure of the building blocks of mass factorised
CFs within the framework of perturbation theory.  We used respective first order differential equations satisfied by each of them  as given in $(\ref{ZUVEq}, \ref{KGFF}, \ref{RGGam}, \ref{RGSc})$.  We found that each of them admits the solution which is of the exponential form whose exponents are controlled by process independent anomalous dimensions as well as process dependent coefficients.  Substituting these solutions for the building blocks, we obtain
\begin{eqnarray}
\label{MasterF1}
\Delta^{\rm{sv+nsv}}_{c \overline c}(q^2,\mu_R^2,\mu_F^2,z) \,, 
= \mathcal{C}\exp \bigg( \Psi^c\big(q^2,\mu_R^2,\mu_F^2,z,\epsilon\big)\bigg)\bigg |_{\epsilon=0} \,,
\end{eqnarray}
where $\Psi^c$ is a finite function in the limit $\epsilon \rightarrow 0$ 
and is  given by  
\begin{eqnarray}\label{Psi}
    \Psi^c\big(q^2,\mu_R^2,\mu_F^2,z,\epsilon\big) = &\Bigg( \ln \bigg( Z_{UV,c}\big(\hat{a}_s,\mu^2,\mu_R^2,\epsilon\big)\bigg)^2 +   \ln \big| \hat{F}_{c}\big(\hat{a}_s,\mu^2,Q^2,\epsilon\big)\big|^2\Bigg) \delta\big(1-z\big)\nonumber \\
    &+2 \mathrm{\Phi}^c\big(\hat{a}_s,\mu^2,q^2,z,\epsilon\big) - 2\mathcal{C} \ln \Gamma_{cc}\big(\hat{a}_s,\mu^2,\mu_F^2,z,\epsilon\big) \,. 
\end{eqnarray}
This all order result is the master formula 
which can be used for obtaining SV+NSV contributions to $\Delta_c$ order by order in perturbation theory provided various functions that appear
in the \eqref{Psi} are known to desired accuracy.  In particular,
it can predict certain SV and NSV terms to all orders in $a_s$ in terms of lower order terms. We elaborate this in more detail in sec. \ref{sec:alldelta}.
In the above formula, we keep the entire FF and overall renormalisation constant \textcolor{black}{as they are proportional to only $\delta(1-z)$}. However, in the functions $\mathrm{\Phi}^c$ and $\ln \Gamma_{cc}$, we keep only SV and NSV terms. In conclusion we have presented a formula, given in \ref{Psi}, which gives the analytical structure of the partonic CF in terms of the anomalous dimensions and SV and NSV coefficients.}
\subsubsection{Results for NSV Coefficients}
\textcolor{black}{In this sub-section, we evaluate explicit expressions for the NSV coefficients, introduced  earlier, by comparing against the state-of-the-art results of CFs and their building blocks such as FF, AP kernels etc. 
At every order $a_s^i$, the coefficients $ {\cal G}^{c,(j,k)}_{L,i}$ for various values of
$(j,k)$ can be determined using \eqref{MasterF} and \eqref{Psi}
known to order $a_s^i$ expanded in double series expansion 
of $\epsilon^j \ln^k(1-z)$.  In order to do this we use the available information up 
to two loop level to obtain ${\cal G}^{c,(j,k)}_{L,i}$ for $i=1,2$ for all the allowed values 
of $(j,k)$.
}
}

\textcolor{black}{We find that unlike the SV coefficients $\overline {\cal G}_i^{c,j}$ (see  \eqref{app:SVGij}),  
the quark and gluon
NSV coefficients  \textcolor{black}{${\cal G}_{L,i}^{c,(j,k)}$} do not satisfy maximal non-Abelian relation beyond 
one loop.} Recall  that $\overline {\cal G}_i^{c,(j)}$ satisfy $\overline {\cal G}_i^{q,(j)} =( C_F/C_A  ) \overline {\cal G}_i^{g,(j)}$, 
confirmed up to third order in $a_s$ \textcolor{black}{as shown in \cite{Ravindran:2005vv,Ravindran:2006cg}}.  

Third order contributions to $\Delta_c$ for DY became available very recently 
in \cite{Duhr:2020seh} and for the Higgs boson productions in gluon fusion as well as in bottom quark annihilation
the third order results \textcolor{black}{were presented in} \cite{Anastasiou:2015ema, Mistlberger:2018etf,Duhr:2019kwi}.
The analytical results for FFs, over all renormalisation constants, the functions $\mathrm{\Phi}_{A}^{c}$ and 
$\Gamma_{c\overline c}$ are all available up to third order in the literature. 
Using these results, we can in principle extract the relevant 
coefficients  ${\cal G}^{q,(j,k)}_{L,i}$ to third order.
In the absence of analytical results for second order corrections to $\Delta_q$ for 
positive powers of $\epsilon$, 
we can not determine the coefficients ${\cal G}^{q,(j,k)}_{L,i}$ \textcolor{black}{at the third order}.   

However, the combination of these coefficients namely $\varphi_{f,c}$, given in \eqref{eq:varphi} can be extracted 
for $c=q$ (DY) and $c =b $ ($b\overline bH$) and $c=g$ ($ggH$) \big) up to third order using the available results to third order.  
We find for the DY,
\begin{eqnarray}
   \varphi_{q,1}^{(0)} &=&
        4 C_F \,,
\quad
   \varphi_{q,1}^{(1)} = 0\,,
\quad
\nonumber\\
\varphi_{q,2}^{(0)} &=&
        C_F  C_A    \bigg( {1402 \over 27} - 28  \zeta_3 - {112 \over 3}   \zeta_2 \bigg)
       + C_F^2    (  - 32  \zeta_2 )
       + n_f  C_F    \bigg(  - {328 \over 27} + {16 \over 3}  \zeta_2 \bigg)\,,
\nonumber\\
   \varphi_{q,2}^{(1)} &=&
        10 C_F  C_A  
       - 10 C_F^2   \,,
\quad
   \varphi_{q,2}^{(2)} =
       - 4 C_F^2  \,,
\nonumber\\
   \varphi_{q, 3}^{(0)} &=&  C_F C_A^2   \bigg( {727211 \over 729} + 192 \zeta_5 - {29876 \over 27} 
         \zeta_3 - {82868 \over 81} \zeta_2 + {176 \over 3} \zeta_2 \zeta_3 + 120 
         \zeta_2^2 \bigg)
\nonumber\\&&
       + C_F^2 C_A   \bigg(  - {5143 \over 27} - {2180 \over 9} \zeta_3 - {11584 \over 27} 
         \zeta_2 + {2272 \over 15} \zeta_2^2 \bigg)
       + C_F^3   \bigg( 23 + 48 \zeta_3 
\nonumber\\&&
       - {32 \over 3} \zeta_2 - {448 \over 15} 
         \zeta_2^2 \bigg)
       + n_f C_F C_A   \bigg(  - {155902 \over 729} + {1292 \over 9} \zeta_3 + {26312 \over 81} \zeta_2 - {368 \over 15} \zeta_2^2 \bigg)
\nonumber\\&&
       + n_f C_F^2   \bigg(  - {1309 \over 9} + {496 \over 3} \zeta_3 + {2536 \over 27} 
         \zeta_2 + {32 \over 5} \zeta_2^2 \bigg)
       + n_f^2 C_F   \bigg( {12656 \over 729} 
\nonumber\\&&
       - {160 \over 27} \zeta_3 - {704 \over 27} 
         \zeta_2 \bigg) \,,
\nonumber\\
   \varphi_{q, 3}^{(1)} &=&  C_F C_A^2   \bigg( {244 \over 9} + 24 \zeta_3 - {8 \over 9} \zeta_2 \bigg)
       + C_F^2 C_A  \bigg (  - {18436 \over 81} + {544 \over 3} \zeta_3 + {964 \over 9} 
         \zeta_2 \bigg)
\nonumber\\&&
       + C_F^3   \bigg(  - {64 \over 3} - 64 \zeta_3 + {80 \over 3} \zeta_2 \bigg)
       + n_f C_F C_A   \bigg(  - {256 \over 9} - {28 \over 9} \zeta_2 \bigg)
\nonumber\\&&
       + n_f C_F^2   \bigg( {3952 \over 81} - {160 \over 9} \zeta_2 \bigg) \,,
\nonumber\\
   \varphi_{q, 3}^{(2)} &=&  C_F C_A^2   \bigg( 34 - {10 \over 3} \zeta_2 \bigg)
       + C_F^2 C_A   \bigg(  - 96 + {52 \over 3} \zeta_2 \bigg)
       + C_F^3   \bigg( {16 \over 3} \bigg)
\nonumber\\&&
       + n_f C_F C_A  \bigg (  - {10 \over 3} \bigg)
       + n_f C_F^2   \bigg( {40 \over 3} \bigg) \,,
\nonumber\\
   \varphi_{q, 3}^{(3)} &=&  C_F^2 C_A   \bigg(  - {176 \over 27} \bigg) + n_f C_F^2   \bigg( {32 \over 27} \bigg) \,,
\end{eqnarray}
and for the Higgs boson production 
\begin{eqnarray}
   \varphi_{g,1}^{(0)} &=& 4 C_A \,,
\nonumber\\
   \varphi_{g,1}^{(1)} &=& 0\,,
\nonumber\\
   \varphi_{g,2}^{(0)} &=& C_A^2    \bigg( {1306 \over 27} - 28  \zeta_3 - {208 \over 3}  \zeta_2 \bigg)
       + n_f  C_A    \bigg(  - {196 \over 27} + {16 \over 3}  \zeta_2 \bigg)\,,
\nonumber\\
   \varphi_{g,2}^{(1)} &=&
        C_A^2    \bigg( {2 \over 3} \bigg)
       + n_f  C_A    \bigg(  - {2 \over 3} \bigg)\,,
\nonumber\\
   \varphi_{g,2}^{(2)} &=&
        -4 C_A^2 \,,
\nonumber\\
   \varphi_{g,3}^{(0)} &=&
        C_A^3    \bigg( {563231 \over 729} + 192  \zeta_5 - {34292 \over 27}  \zeta_3
          - {113600 \over 81}  \zeta_2 + {176 \over 3}  \zeta_2  \zeta_3 + {3488 \over 15}  
         \zeta_2^2 \bigg)
\nonumber\\&&
       + n_f  C_A^2    \bigg(  - {117778 \over 729} + {1888 \over 9}  \zeta_3 + {26780 \over 
         81}  \zeta_2 - {232 \over 15}  \zeta_2^2 \bigg)
\nonumber\\&&
       + n_f  C_F  C_A    \bigg(  - {2653 \over 27} + {616 \over 9}  \zeta_3 + {40 \over 3}  \zeta_2
          + {32 \over 5}  \zeta_2^2 \bigg)
       + n_f^2  C_A    \bigg( {1568 \over 729} - {160 \over 27}  \zeta_3 - {152 \over 9}  \zeta_2
          \bigg)\,,
\nonumber\\
   \varphi_{g,3}^{(1)} &=&
        C_A^3    \bigg(  - {18988 \over 81} + {448 \over 3}  \zeta_3 + {1280 \over 9}  \zeta_2
          \bigg)
       + n_f  C_A^2    \bigg( {1528 \over 81} - 8  \zeta_3 - {248 \over 9}  \zeta_2 \bigg)
\nonumber\\&&
       + n_f  C_F  C_A    \bigg( 4 - {8 \over 3}  \zeta_2 \bigg)
       + n_f^2  C_A    \bigg( {56 \over 27} \bigg)\,,
\nonumber\\
   \varphi_{g,3}^{(2)} &=&
        C_A^3    \bigg(  - {1432 \over 27} + {40 \over 3}  \zeta_2 \bigg)
       + n_f  C_A^2    \bigg( {164 \over 27} + {2 \over 3}  \zeta_2 \bigg)
       + n_f^2  C_A    \bigg( {8 \over 27} \bigg)\,,
\nonumber\\
   \varphi_{g,3}^{(3)} &=&
        C_A^3    \bigg(  - {176 \over 27} \bigg)
       + n_f  C_A^2    \bigg( {32 \over 27} \bigg).
\end{eqnarray}

While the NSV function $\mathrm \Phi_B^c$ for quarks and gluon are not related,
they are found to be universal up to second order in the sense that they do not depend on the hard process. For example,
to second order in $a_s$, $\mathrm \Phi_B^q$ of DY is found to be identical to that of Higgs boson production in bottom quark annihilation 
\cite{Harlander:2003ai}.  In addition, we find that they agree with that of Graviton (G) production in quark annihilation 
processes 
\cite{deFlorian:2013sza,deFlorian:2013wpa,Ahmed:2015qia,Ahmed:2016qjf,Ahmed:2016qhu,Banerjee:2017ewt}.  In terms of $\varphi_{q,i}^{(k)}$ it translates to
\begin{eqnarray}
\varphi_{q,i}^{(k)}\Big|_{q+ \overline q \rightarrow l^+l^-+X}  = \varphi_{q,i}^{(k)}\Big|_{b+ \overline b \rightarrow H+X}
=\varphi_{q,i}^{(k)}\Big|_{q+ \overline q \rightarrow G+X}  \quad \quad i=1,2,k=0,i
\end{eqnarray}
Similarly, to second order in $a_s$,  $\mathrm{\Phi}_B^g$ from Higgs boson production in gluon fusion is found to be identical to 
that of graviton production in gluon fusion channel and pseudo scalar Higgs boson production 
\cite{Anastasiou:2002wq,Harlander:2002vv,Ravindran:2003um,Ahmed:2015qpa,Ahmed:2015qda,Ahmed:2016otz} in gluon fusion.
That is, 
\begin{eqnarray}
\varphi_{g,i}^{(k)}\Big|_{g + g \rightarrow H+X}  = \varphi_{g,i}^{(k)}\Big|_{g+g \rightarrow A+X}
=\varphi_{g,i}^{(k)}\Big|_{g + g \rightarrow G+X}  \quad \quad i=1,2,k=0,i
\end{eqnarray}
However, the universality breaks at third order,
namely, we find that the $\varphi_{b,3}^{(k)}$ for $k=0,1$ differs from that of DY production while
for $k=2,3$ they agree. 
\begin{eqnarray}
\varphi_{b,3}^{(0)} &=& \varphi_{q,3}^{(0)} - 16 C_A C_F \big(C_A-2 C_F\big)\,,
\nonumber\\
\varphi_{b,3}^{(1)} &=& \varphi_{q,3}^{(1)} + 8 C_A C_F \big(C_A- 2 C_F\big) \,,
\nonumber\\
\varphi_{b,3}^{(k)} &=& \varphi_{q,3}^{(k)}  \quad\quad \quad k=2,3.
\end{eqnarray}
The origin of this violation for $k=0,1$ at third order, \textcolor{black}{which has been evaluated using the state-of-art results \cite{Anastasiou:2015ema, Mistlberger:2018etf,Duhr:2019kwi,Duhr:2020seh}}, needs to be understood within the
framework of factorisation. 
\section{More on the Soft-Collinear Function, $\rm \Phi^c_B$}
\subsection{On the form of the solution}
In this section, we discuss in detail the peculiar structure of 
SV and NSV solutions 
given in \eqref{PhiSV} and \eqref{phiB} respectively,
that satisfy the K+G equation. 
Both of them contain divergent as well as finite terms at every order.
For example, the SV part of the solution, $\rm \Phi_A^c$, contains the right soft and collinear divergences
proportional to distributions $\delta(1-z)$ and ${\cal D}_0(z)$
to cancel those from the FF entirely and from the AP kernels partially
and the $z$ dependent finite terms
to correctly reproduce all the distributions in the SV part of CFs $\Delta_c$.
The NSV part,  $\rm \Phi_B^c$, removes the 
remaining collinear divergences of the AP kernels.  The finite part of it
when combined with SV counterpart of  $\rm \Phi_A^c$ contributes to next to SV terms to
CFs $\Delta_c$.  As we mentioned in the previous section, the $z$ dependence of the solution 
is inspired from the structure of
various contributions that constitute the next to leading order contributions to variety
of inclusive reactions, namely production of a pair of leptons in quark anti-quark annihilation,
a Higgs boson in gluon fusion or in bottom quark annihilation at hadron colliders.
In addition, the renormalisation group equation, Eq.(\ref{RGphis}), brings in 
additional $z$ dependent logarithmic structure through the anomalous dimensions $C^c(a_s)$ and $D^c(a_s)$. 

Note that the solution given in \eqref{PhiSV1} is organised 
in such a way that the term {\color{black}$\rm \Phi_A^c$} contains only leading contributions namely 
the distributions 
such as $\delta(1-z)$ and ${\cal D}_j(z)$, the so called SV terms 
and the term $\rm \Phi_B^c$, the sub-leading terms, i.e., the next to SV logarithms $\ln^k(1-z), k=0,1,\cdots$.
Even though $\color{black}\rm \Phi_A^c$ does not contain next to SV terms, they contribute 
to next to SV terms to $\Delta_c$, when the exponential is expanded in powers of $a_s$.
Not only do distributions result from the convolutions of two or more distributions, they
also give next to SV logarithms.  In addition, the convolution of distributions with next to SV terms
in turn give pure NSV logarithms.  Hence, the leading solution $\color{black} \rm \Phi_A^c$ 
plays an important role for generating next to SV terms for the CFs $\Delta_c$ at every order in perturbation theory.     

The solution $\rm \Phi_A^c$ (see \eqref{PhiSV}) at every order in $\hat a_s$ is found to factorise into  
$z$ dependent piece, $\big((1-z)^{m} \big)^{i \epsilon/2} {1\over 1-z}$ with $m=2$,
and the $z$ independent coefficients $\hat \phi^{c(i)}_{SV}(\epsilon)$.
The peculiarity of this solution is that we can retain the independence of $\hat \phi^{c(i)}_{SV}(\epsilon)$ with respect to
the variable $z$ at every order in $\hat a_s$, thanks to presence of the factor 
$\big((1-z)^{m}\big)^{i \epsilon/2}\frac{1}{(1-z)}$ which not only ensures the finiteness of SV part of CFs $\Delta_c$ but also
gives right distributions at every order.  The factor $m$ takes the value $m=2$ for DY and Higgs productions 
as observed in \eqref{PhiSV} and the origin of it can be traced to the number of external legs 
that require mass factorisation \cite{Ravindran:2006cg}.  It was observed in \cite{Ravindran:2006cg,Blumlein:2006pj} that the parameter $m$ takes the value $m=1$ for the SV part of the solutions to CFs of structure functions of Deep Inelastic Scattering (DIS) and of Semi-Inclusive Annihilation (SIA) of hadron production and the reason is that only one of the external legs requires mass factorisation. 
The uniqueness of  the structure of $\hat \phi^{c(i)}_{SV}$  
may be attributed to the fact that the entire $z$ dependence of the
solution factorises at every order \textcolor{black}{as $\big((1-z)^{m} \big)^{i \epsilon/2} {1\over 1-z}$ leaving $\hat \phi^{c(i)}_{SV}(\epsilon)$ $z$-independent}.

Like SV part, the NSV part of the solution is also determined by demanding that 
it should contain the right divergences to cancel those present in AP kernels.  
{The structure of the finite part of the solution is determined by \eqref{phiBint}, which when combined with
SV part of the solution, reproduces the correct NSV terms for $\Delta_c$.}
The perturbative structure of higher order results allows only certain powers of logarithms at every order
in perturbation theory thanks to inherent transcendentality structure of Feynman integrals that appear at 
every order in $a_s$ and in $\epsilon$ in the  dimensionally  regularised theory.  We find that the coefficients $\varphi_{c}^{(i)}(z,\epsilon)$  are consistent with this
expectation.
In addition, the solution demonstrates an interesting structure that deserves a mention.
\subsubsection*{Class Of Solutions-Class I}
We find that the K+G equation allows
us to construct not just one solution  but a class of solutions,  a minimal class, satisfying the right divergent structure as well as the dependence on $\ln^k(1-z), k=0,1,\cdots$:
\begin{eqnarray}
\label{alpha}
{\rm{\Phi}}^c_{B,\alpha} = \sum_{i=1}^\infty \hat a_s^i \left( {q^2 (1-z)^\alpha \over \mu^2}\right)^{{i \epsilon  \over 2}} S_{\epsilon}^i
\varphi_{c,\alpha}^{(i)}(z,\epsilon)  \,.
\end{eqnarray}       
The predictions from the solutions ${\rm{\Phi}}^c_{B,\alpha}$ are 
found to be independent of choice of $\alpha$ owing to the explicit $z$-dependence of 
the coefficients $\varphi_{c,\alpha}^{(i)}(z,\epsilon)$ at every order in $\hat a_s$ and in $\epsilon$. 
It is straightforward to show that any variation of $\alpha$ in the factor $(1-z)^{i \alpha \epsilon}$ can always 
be compensated by suitably adjusting the $z$ independent coefficients of
$\ln(1-z)$ terms in $\varphi_{c,\alpha}^{(i)}(z,\epsilon)$ 
at every order in $\hat a_s$. 
The reason for this is the invariance of the solution under certain  ``gauge like" transformations 
on both $(1-z)^{i \alpha \epsilon}$
and $\varphi_{c,f,\alpha}(z,\epsilon)$ at every order in $\hat a_s$.  
Note that the logarithmic structure of $\varphi_{c,\alpha}^{(i)}(z,\epsilon)$
plays an important role.
Because of this invariance, these transformations neither affect the divergent structure nor the finite parts of 
$ {\rm{\Phi}}^c_{B,\alpha}$.    We find that the invariance can be realised through the renormalisation group equation of strong coupling
constant.   To end, the solution given in Eq.(\ref{alpha}) takes the following integral form:
\begin{eqnarray}
\label{phiBintalpha}
   \mathrm{\Phi}_{B,\alpha}^{c} &= & \int_{\mu_F^2}^{q^2(1-z)^\alpha} \frac{d\lambda^2}{\lambda^2} L^c(a_s(\lambda^2),z) + \varphi_{f,c,\alpha}\big(a_s(q^2(1-z)^\alpha),z,\epsilon\big) |_{\epsilon=0}  \nonumber\\ 
    &&+ \varphi_{s,c}\big(a_s(\mu_F^2),z,\epsilon\big) \,,
\end{eqnarray}
The finite part $\varphi_{f,c,\alpha}$ can be expanded as
\begin{eqnarray}
\label{varphiexp1}
\varphi_{f,c,\alpha}(a_s(q^2(1-z)^\alpha),z) = \sum_{i=1}^\infty a_s^i(q^2(1-z)^\alpha) \sum_{k=0}^i \varphi_{c,\alpha,i}^{(k)} \ln^k(1-z) \,.
\end{eqnarray}
The fact that the predictions are insensitive to $\alpha$ relates the coefficient $\varphi_{c,\alpha,i}^{(k)}$  
to $\varphi_{c,i}^{(k)}$, the solution corresponding to $\alpha=2$, through
\begin{eqnarray}
\varphi_{c,\alpha,1}^{(0)}  &=& \varphi_{c,1}^{(0)},\qquad \quad
\varphi_{c,\alpha,1}^{(1)}  = -D^c_1 {\overline \alpha} + \varphi_{c,1}^{(1)}, \qquad \quad
\varphi_{c,\alpha,2}^{(0)}  = \varphi_{c,2}^{(0)}
\nonumber\\
\varphi_{c,\alpha,2}^{(1)}  &=&  -{\overline \alpha} \Big(D^c_2 - {\beta_0} \varphi_{c,1}^{(0)}\Big) + \varphi_{c,2}^{(1)} ,\quad \quad
\nonumber\\
\varphi_{c,\alpha,2}^{(2)}  &=& -  \frac{1}{2}{\overline \alpha}^2 \beta_0 D^c_1 -  {\overline \alpha} \Big(C^c_2   - {\beta_0} \varphi_{c,1}^{(1)} \Big) + \varphi_{c,2}^{(2)}
\nonumber\\
\varphi_{c,\alpha,3}^{(0)}  &=& \varphi_{c,3}^{(0)},
\qquad \quad
\varphi_{c,\alpha,3}^{(1)}  = -{\overline \alpha} \Big(D^c_3  - {\beta_1}  \varphi_{c,1}^{(0)} - 2 {\beta_0}  \varphi_{c,2}^{(0)}\Big) + \varphi_{c,3}^{(1)}
\nonumber\\
\varphi_{c,\alpha,3}^{(2)}  &=&- {\overline \alpha}^2 \Big( \frac{1}{2}{\beta_1} D^c_1 + {\beta_0} D^c_2  - {\beta_0}^2  \varphi_{c,1}^{(0)} \Big)-{\overline \alpha} \Big(C^c_3 {\overline \alpha}  - {\beta_1}  \varphi_{c,1}^{(1)} - 2 {\beta_0}  \varphi_{c,2}^{(1)}\Big) + \varphi_{c,3}^{(2)}
\nonumber\\
\varphi_{c,\alpha,3}^{(3)}  &=& {\beta_0}^2 \bigg(-\frac{1}{3}D^c_1 {\overline \alpha}^3 + {\overline \alpha}^2 \varphi_{c,1}^{(1)}\bigg) +
     {\beta_0} {\overline \alpha} \bigg(-C^c_2 {\overline \alpha} + 2 \varphi_{c,2}^{(2)}\bigg) + \varphi_{c,3}^{(3)}
\nonumber\\
\varphi_{c,\alpha,4}^{(0)}  &=& \varphi_{c,4}^{(0)}, \qquad \quad
\varphi_{c,\alpha,4}^{(1)}  = - D^c_4 {\overline \alpha} + {\beta_2} {\overline \alpha} \varphi_{c,1}^{(0)} + 2 {\beta_1} {\overline \alpha} \varphi_{c,2}^{(0)} +
     3 {\beta_0} {\overline \alpha} \varphi_{c,3}^{(0)} + \varphi_{c,4}^{(1)}
\nonumber\\
\varphi_{c,\alpha,4}^{(2)}  &=& -C^c_4 {\overline \alpha} - \frac{1}{2}{\beta_2} D^c_1 {\overline \alpha}^2 - {\beta_1} D^c_2 {\overline \alpha}^2 -
     \frac{3}{2} {\beta_0} D^c_3 {\overline \alpha}^2 + \frac{5}{2} {\beta_0} {\beta_1} {\overline \alpha}^2 \varphi_{c,1}^{(0)}
     \nonumber\\&&
     + {\beta_2} {\overline \alpha} \varphi_{c,1}^{(1)} +
     3 {\beta_0}^2 {\overline \alpha}^2 \varphi_{c,2}^{(0)} + 2 {\beta_1} {\overline \alpha} \varphi_{c,2}^{(1)} + 3 {\beta_0} {\overline \alpha} \varphi_{c,3}^{(1)} + \varphi_{c,4}^{(2)}
\nonumber\\
\varphi_{c,\alpha,4}^{(3)}  &=& {\beta_0}^3 {\overline \alpha}^3 \varphi_{c,1}^{(0)} + {\beta_0}^2 {\overline \alpha}^2 \bigg(-D^c_2 {\overline \alpha} + 3 \varphi_{c,2}^{(1)}\bigg) -
     \frac{1}{6}{\beta_1} {\overline \alpha} \bigg(6 C^c_2 {\overline \alpha} + 5 {\beta_0} {\overline \alpha} \bigg(D^c_1 {\overline \alpha} - 3 \varphi_{c,1}^{(1)}\bigg) 
     \nonumber\\&&
     - 12 \varphi_{c,2}^{(2)}\bigg)
      - \frac{3}{2} {\beta_0} {\overline \alpha} \bigg(C^c_3 {\overline \alpha} - 2 \varphi_{c,3}^{(2)}\bigg) + \varphi_{c,4}^{(3)}
\nonumber\\
\varphi_{c,\alpha,4}^{(4)}  &=& {\beta_0}^3\bigg( - \frac{1}{4}D^c_1 {\overline \alpha}^4 + {\overline \alpha}^3 \varphi_{c,1}^{(1)}\bigg) +
     {\beta_0}^2 {\overline \alpha}^2 \bigg(-C^c_2 {\overline \alpha} + 3 \varphi_{c,2}^{(2)}\bigg) + 3 {\beta_0} {\overline \alpha} \varphi_{c,3}^{(3)} + \varphi_{c,4}^{(4)}
\end{eqnarray}
where $\overline \alpha = \alpha-2$.  The above relations are the transformations for 
$\varphi_{c,\alpha,i}^{(k)}$ that are required to compensate the contributions resulting from the 
change in the exponent of $(1-z)$ from $i \epsilon$ to ${i \alpha \epsilon}$.  This 
invariance property with respect to the parameter $\alpha$  makes the solution a peculiar one compared to
SV counter part.  
\subsubsection*{Class Of Solutions-Class II}
We would like to point out that the class of solutions parametrised by $\alpha$ is not the only one that satisfies K+G equation.  
For example, if we do not restrict $z$-dependence in $\varphi_c^{(i)}$ , 
we can obtain different kinds of solution.  
Then for such solution, we need to add more terms on the right hand side of \eqref{alpha} 
in such a way that all the requirements  are fulfilled.  
In other words if we assume the following form for the
solution,
\begin{eqnarray}
 \tilde \Phi^c_B = \sum_{i=1}^\infty \hat a_s^i   \sum_{\alpha=2}^{2i} \left( {q^2 (1-z)^\alpha \over \mu^2}\right)^{{i \epsilon  \over 2}}S_\epsilon^i
 \tilde \varphi_{c,\alpha}^{(i)}(\epsilon) 
 \end{eqnarray}
 \textcolor{black}{with various $\tilde \varphi_{c,\alpha}^{(i)}(\epsilon)$s to contain right divergent as well as  finite 
terms which when we sum them up over $\alpha$s, we can obtain $\Delta_c$ that agrees with the known result.}
\textcolor{black}{In the following we explain this using an example that can provide the justification for
the proposed solution.  We use  \cite{Anastasiou:2014lda} for this purpose.   In 
\cite{Anastasiou:2014lda} inclusive production of Higgs boson was computed using the method of threshold expansion up to third order in $a_s$ in dimensional regularisation.  For the diagonal channel,   $\hat \sigma_{gg}$, the results to third order show remarkable structure in terms of $z$ and $\epsilon$ namely the factorisation of terms of the form $(1-z)^{\epsilon}$ and functions that depend only on $\epsilon$.  Generalising this structure to $i$th order in $a_s$,  one obtains the factorisation of the form, $\sum_{\alpha=2}^{2 i} (1-z)^{\alpha i \epsilon/2} \chi_i^\alpha(\epsilon)$.  The factor $(1-z)^{ \alpha i \epsilon/2}$ originates from soft and collinear 
configurations of partons. The corresponding soft and collinear scales are given by $(q^2 (1-z))^{i \alpha \epsilon/2}$ and hence one can conclude that the threshold expansion beyond SV approximation contains multiple scales parametrised by $\alpha$.     From the explicit computations one finds that every collinear parton gives $(1-z)^{\epsilon/2}$ and soft parton gives $(1-z)^\epsilon$ \textsuperscript{1}\footnotetext{\textsuperscript{1} We thank Claude Duhr for explaining this point to us}.  Pure virtual contributions to born amplitude give $\delta(1-z)$ and the hard part from the real emissions gives terms proportional to $(1-z)^\eta, \eta\ge 0 $.  For a given process, we can determine the values of $\alpha$ by studying the number of soft and collinear configurations.  This way we can find out the allowed values of $\alpha$ for every process at every order in $a_s$.   The highest power of $\alpha$ at a given order is determined by the number of allowed soft and collinear configurations in that order. The values of $\alpha$ extracted from results known to third order can be used to extrapolate 
to obtain the upper limit on $\alpha$ at $i$th order in $a_s$ and it turns out  to be $2 i$.  The coefficients of the scales $\chi_i^\alpha(\epsilon)$ can be expanded in powers of $\epsilon$.  The singularity structure in $\epsilon$ is completely determined by the finiteness of mass factorised result.  Note that the remarkable multi-scale structure of the
fixed order results \cite{Anastasiou:2014lda}  for the cross sections confirms the structure of $\tilde \Phi^c_B$ given above.
}

\textcolor{black}{While these two classes of solutions may look different in the structure, both of them
give identical predictions to all orders for CFs and in addition, it is easy to
relate the coefficients of these solutions by finite transformations.  Hence, they
are equivalent. } In the present paper, we use class-I solution with the choice $\alpha=2$ in (\ref{alpha}) so that the solution 
resembles more like the SV part.  
Thanks to the invariance property of the solution, the 
different choices for $\alpha$ neither alter the qualitative behaviour 
nor the quantitative predictions for $\Delta_c$ to all orders.
For example, an alternate choice, say $\alpha=1$ can only
change the coefficients of $\ln^k(1-z)$ in the $\varphi_{f,c}$ without affecting the all order
structure and the predictions for $\Delta_c$.  
With our choice of $\alpha=2$, the all order solution, equivalently integral representation   resembles that of SV part.  We will see later
that this choice will allow us to study $N$ space resummation for both SV and NSV terms 
with single order one term namely $\omega = 2 a_s \beta_0 \ln N$.  

\subsection{On the Logarithmic Structure} \label{ap:PhiLog}
In the last section, we derived $z$ space result
that can correctly predict certain SV and NSV 
terms to all orders from the knowledge of 
previous orders.  This was possible due to a peculiar
logarithm structure of the solution to K+G equation
at every order in $\hat a_s$ and $\epsilon$, see \eqref{GikLj}. 
In this sub section, we present an explicit result for $\mathrm{\Phi}^c, ~c=b$ to second order in
perturbation theory in order to explain the structure of SV and NSV logarithms at a given order in $\hat a_s$ with 
 an accuracy of $\epsilon^{n}$.  We have used inclusive cross section for the production of Higgs boson in
bottom quark annihilation for this purpose.  The conclusions remain unchanged as long as
color neutral production in diagonal channels are considered.
To order $\hat a_s^2$, the inclusive cross section for the production of Higgs boson in
bottom quark annihilation receives contributions from  a) pure real emissions
\begin{eqnarray}
b+\overline b \rightarrow H+g,
\quad
b+\overline b \rightarrow H+g+g,
\quad
b+\overline b \rightarrow H+b+\bar b ,\quad  b+\overline b \rightarrow H+q+\bar q,
\nonumber
\end{eqnarray}
b) pure virtual corrections through one and two loop corrections to leading order
$b+\overline b \rightarrow H$ 
and  c) interference of pure real emission process
$b+\overline b \rightarrow H +g$ with the loop corrected process $b+\overline b \rightarrow H+g$. Here, $q$ refers light quarks leaving $t$- and $b$-quarks. We compute these parton level sub processes using the standard Feynman diagram approach.
Beyond the leading order in strong coupling, all these sub processes develop UV and IR divergences and they are regulated
in dimensional regularisation.
As we encounter large number of Feynman diagrams, we use
QGRAF to generate them and an in-house FORM
routine to perform all the symbolic manipulations, e.g. for Dirac,
$SU(N_c)$ color and Lorentz algebra.  We use the integration-by-parts (IBP) identities through
a \textit{Mathematica} based package, LiteRed, to
reduce Feynman integrals to a minimum set of master integrals. In addition,
for real emission and real-virtual processes the method of reverse
unitarity is used along with IBP identities to reduce the resulting phase-space integrals to a set of few
master integrals.  The master integrals for the virtual processes can be found in \cite{Gehrmann:2010tu,Anastasiou:2012kq}
and for the real emission in \cite{Anastasiou:2012kq} up to desired accuracy in $\epsilon$.
While individual sub processes contain UV, soft and collinear divergences, after renormalising
the strong coupling constant $\hat a_s$ and the Yukawa coupling $\lambda$, the sum becomes UV finite.  In addition, the soft
and final state collinear divergences cancel in real and virtual sub processes leaving only initial
state collinear divergences in 
$\hat \sigma_{b \bar b}$.  

Since we are interested only in 
those terms that are proportional to distributions and NSV logarithms $\ln^k(1-z)$, we expand \textsuperscript{1}\footnotetext{\textsuperscript{1}We thank Claude Duhr for helping us with the expansion of Harmonic Polylogs \cite{Duhr:2019tlz}.}
$\hat \sigma_{b \bar b }$ around $z=1$ and drop those terms that vanish when $z \rightarrow 1$.
In order to extract $\mathrm{\Phi}^c$ from the latter, we follow \eqref{normS}, where  the virtual contributions are factored out from  $\hat \sigma_{c\bar c}$ giving rise to the function $\mathcal{S}_c$.
Owing to \eqref{KGSc}, ${\cal S}_b$ has an exponential structure
\begin{align}
{\cal S}_{b }(z,q^2,\epsilon) = {\cal C} \exp\left(2\mathrm{\Phi}^b ( z,q^2,\epsilon) \right)
\end{align}
where $\mathrm{\Phi}^b = \mathrm{\Phi}^b_A+\mathrm{\Phi}^b_B$.  Expanding $\mathrm{\Phi}^b_B$ in powers of $\hat a_s$ as
\begin{align}\label{ap:phiB}
    \mathrm{\Phi}_{B}^{b}(\hat{a}_s,\mu^2,q^2,z,\epsilon) &= \sum_{i=1}^{\infty} \hat a_s^i \bigg(\frac{q^2 (1-z)^2}{\mu^2}\bigg)^{i \frac{\epsilon}{2}} S_\epsilon^i \varphi_b^{(i)}(z,\epsilon)\nonumber\\
    &= \sum_{i=1}^{\infty} \hat a_s^i \bigg(\frac{q^2}{\mu^2}\bigg)^{i \frac{\epsilon}{2}} S_\epsilon^i ~ \mathrm{\hat{ \Phi}}_{NSV,b}^{(i)}(z,\epsilon) \,,
\end{align}
and using explicit results for $\hat \sigma_{b \bar b}^{SV+NSV}$, $Z_{UV,b}$ and $\hat F$,
we obtain $\mathrm{\hat{ \Phi}}_{NSV,b}^{(i)}$ for $i=1,2$ in powers of $\epsilon$.  They are given by
\begin{small}
\begin{align}
    \mathrm{\hat{ \Phi}}_{NSV,b}^{(1)} = C_F\bigg\{&\frac{1}{\epsilon} \bigg(-8\bigg)
                                        + \bigg(-8 L_z + 4\bigg)
                                        + \epsilon\bigg( -4L_z^2 + +4 L_z + 3\zeta_2\bigg)
                                        +\epsilon^2\bigg( -\frac{4}{3}L_z^3 +2L_z^2
                                        \nonumber\\&
                                        +3\zeta_2 L_z
                                        -\bigg(\frac{7}{3}\zeta_3 +\frac{3}{2}\zeta_2\bigg) \bigg)

                                        +\epsilon^3\bigg( -\frac{1}{3}L_z^4 +\frac{2}{3}L_z^3 +\frac{3}{2}\zeta_2L_z^2 - \bigg(\frac{7}{3}\zeta_3 + \frac{3}{2}\zeta_2\bigg)L_z
                                        \nonumber\\&
                                        +\bigg(\frac{7}{6}\zeta_3 +\frac{3}{16}\zeta_2^2\bigg) \bigg) \bigg\}\nonumber\\
    \mathrm{\hat{ \Phi}}_{NSV,b}^{(2)} = C_FC_A&\bigg\{\frac{1}{\epsilon^2} \bigg(\frac{88}{3}\bigg)                                              + \frac{1}{\epsilon} \bigg(\frac{176}{3} L_z+8\zeta_2-\frac{664}{9} \bigg)
                                        + \bigg(\frac{176}{3} L_z^2 +\bigg(16\zeta_2-\frac{1238}{9}\bigg)L_z
                                        \nonumber\\&~
                                        +\frac{1402}{27}-28\zeta_3-\frac{178}{3}\zeta_2\bigg)
                                        + \epsilon\bigg( \frac{352}{9}L_z^3 +\bigg(16\zeta_2-\frac{2341}{18}\bigg)L_z^2
 +\bigg(\frac{2750}{27}
 \nonumber\\&
 -56\zeta_3-\frac{356}{3}\zeta_2\bigg) L_z + \frac{934}{9}\zeta_3 -\frac{4021}{81} + \frac{982}{9}\zeta_2-4\zeta_2^2\bigg)\bigg\}
                                     + C_F^2\bigg\{\frac{1}{\epsilon} \bigg(16 L_z
                                     \nonumber\\&
                                     +12 \bigg)
                                        + \bigg(28 L_z^2 +14L_z

                                        -32\zeta_2\bigg)
                                        + \epsilon\bigg( \frac{74}{3}L_z^3 + \frac{13}{2}L_z^2                         +\bigg(6
                                        -76\zeta_2\bigg) L_z
 \nonumber\\&
-8 + 48\zeta_3  -\zeta_2\bigg)\bigg\}
 +C_Fn_f\bigg\{\frac{1}{\epsilon^2} \bigg(\frac{-16}{3}\bigg)                                               +\frac{1}{\epsilon} \bigg(\frac{-32}{3} L_z+\frac{112}{9} \bigg)
                                        + \bigg(\frac{-32}{3} L_z^2
                                        \nonumber\\&
                                        +\bigg(\frac{224}{9}\bigg)L_z
                                        +\frac{28}{3}\zeta_2-\frac{328}{27}\bigg)

                                        + \epsilon\bigg( \frac{-64}{9}L_z^3 +\frac{224}{9}L_z^2
 +\bigg(\frac{56}{3}\zeta_2-\frac{656}{27}\bigg) L_z
 \nonumber\\&
 +\frac{1030}{81}- \frac{124}{9}\zeta_3  - \frac{196}{9}\zeta_2\bigg)\bigg\}\,.
\end{align}
\end{small}
As can be seen from the above results,  at order $\hat {a}_s$, the leading pole in $\epsilon$ is of order one and it is two at $\hat {a}_s^2$ and the increment of one unit for the leading poles is expected to continue with the order of perturbation.  However, the pole structure for $\hat \sigma_{b\bar b}$  shows an increment of two units.  In addition, at every order in $\hat{a}_s$, for a given  color factor, the combination of $\epsilon$ and the leading logarithm shows uniform transcendentality weight.  In other words, if we assign $n_\epsilon$ weight  for $\epsilon^{-n_\epsilon}$ and $n_L$ for  $\ln^{n_L}(1-z)$, then the highest weight at every order in $\epsilon$ shows uniform transcendentality  $w=n_{\epsilon}+n_{L}$. For instance, at one loop, we find $w=1$ at  every order of $\epsilon$ and at two loops it is two ($w=2$).
This clearly explains that the highest power of $\ln(1-z)$ at every order in $\epsilon$ is constrained by the order of $\hat a_s$ and the accuracy in $\epsilon$ and is found to be $i+j$ for the term $\hat a_s^i \epsilon^j$.  This translates to $i+j-1$ for ${\cal G}_{L,i}^{(j)}$ in \eqref{GikLj} as the latter is the coefficient of $\epsilon^{j-1}$.  This exercise provides an explanation for the logarithmic structure given in \eqref{GikLj}, in particular the upper limit of the summation. This logarithmic structure determines the structure of $\varphi_{f,c}$ given in  \eqref{varphiexp}.  In Appendix~\ref{ap:phic}, we present
$\mathcal{G}_{L,i}^{c,(j,k)}$ up to
second order in $\hat a_s$ with $i=1,2$. 

Precisely because of the logarithmic structure
of the exponents, namely, increment by one unit, we get
logarithms in CFs with increment of two units. It is easy
to understand this structure if we observe that when we expand 
the exponents containing ${\cal D}_k$ and $\ln^k(1-z)$ to
obtain CFs,
the resulting convolutions between various orders in $a_s$ will be of the form
${\cal D}_k \otimes {\cal D}_l$ and/or ${\cal D}_k \otimes \ln^l(1-z)$ which will result in leading distributions 
${\cal D}_{k+l+1}$ and leading NSV logarithms $\ln^{k+l+1}(1-z)$. 


%
\section{All order predictions for $\Delta_c$}\label{sec:alldelta}
In this section, we discuss the predictive power of the master formula \eqref{MasterF}.  
In other words, given $Z_{UV,c}$,$\hat F_c$, $\mathrm \Phi^c$ and
the $\Gamma_{c c}$ up to a certain order in perturbation theory, we show that the master formula can
predict certain SV and NSV terms to all orders in perturbation theory.        
The partonic coefficient function $\Delta_c$ can be expanded order by order in
perturbation theory in powers of $a_s(\mu_R^2)$ as
\begin{align}
\label{Delexp}
\Delta_c(q^2,\mu_R^2,\mu_F^2,z) = \sum_{i=0}^\infty a_s^i(\mu_R^2) \Delta_c^{(i)}(q^2,\mu_R^2,\mu_F^2,z)\,,
\end{align}
where the coefficient $\Delta_c^{(i)}$ can be obtained by first expanding the exponential 
given in \eqref{Psi} in powers of $a_s(\mu_R^2)$ and then performing all the resulting convolutions  
in $z$ space.  Note that $\Delta_c^{(0)} = \delta(1-z)$.  We have dropped all those terms that are 
of order ${\cal O}((1-z)^\alpha), \alpha > 0$. {{Finally, we write the following decomposition , 
\begin{equation}
    \Delta_{c}^{(i)}(q^2,\mu_R^2,\mu_F^2,z) =  \Delta_{c}^{SV,(i)}(q^2,\mu_R^2,\mu_F^2,z) +  \Delta_{c}^{NSV,(i)}(q^2,\mu_R^2,\mu_F^2,z).
\end{equation}
Here $\Delta_c^{SV,(i)}$ 
contains only SV terms, such as the distributions ${\cal D}_i,~( i=0,1,\cdots$) and $\delta(1-z)$ and 
next to SV terms, i.e., the logarithms $\ln^i(1-z),~(i=0,1,\cdots) $ are embedded within $\Delta_c^{NSV,(i)}$. Now given the distribution function $\rm \Phi^c$, upto a certain order in $a_s$, there are several SV and NSV logarithms which can be predicted 
to all orders in $a_s$. 
For example, we observe that if $ \Psi^c$ is known at leading order in $a_s$, we can predict all the leading
distributions ${\cal D}_i$ and leading NSV terms $\ln^i(1-z)$ to all orders in $a_s$.
In the following, we elaborate on this by comparing our predictions with the available N$^3$LO results  and also predict N$^4$LO and some higher order results for few observables. }}

{Given $\Psi^c$ at order $a_s$, by expanding the master formula \eqref{MasterF} in powers of strong coupling constant, 
we obtain the leading SV terms $({\cal D}_3,{\cal D}_2)$,
$({\cal D}_5,{\cal D}_4),\cdots,({\cal D}_{2i-1},{\cal D}_{2i-2})$ and the leading NSV terms
$\ln^3(1-z),\ln^5(1-z),\cdots,\ln^{2i-1}(1-z)$  at $a_s^2,a_s^3,\cdots,a_s^i$ 
respectively for all $i$.
Since $C^c_1$ is identically zero, $\ln^{2i}(1-z)$ terms do not contribute for all $i$. 
Hence we predict,
\begin{eqnarray}
   \Delta_{c}^{NSV} &=& a_s \Delta_{c}^{NSV(1)}+a_s^2\Big[ - 128 C_i^2 L^3_z  + \mathcal{O}(L^2_ z) \Big]
       + a_s^3 \Big[  - 512 C_i^3 L^5_z  
       + \mathcal{O}(L^4_z) \Big] 
       \nonumber\\
       &&
+ a_s^4 \Big[  - {4096 \over 3} C_i^4 L^7_z + \mathcal{O}(L^6_z) \Big] + \mathcal{O}(a_s^5)
\end{eqnarray}
Here we write $\ln^i(1-z)\equiv L^i_z$ for brevity. Also $C_i=C_F$ for $c=\{q,b\}$ i.e. for DY and Higgs production through bottom quark annihilation. And for Higgs production through gluon fusion i.e. $c=g$, we have $C_i = C_A$.  
Thus with the knowledge of one loop anomalous dimensions $\{C_1^c,D_1^c,A_1^c,B_1^c,f_1^c\} $ and one-loop $\varphi_{c,1}^{(k)}$, we predicted the above NSV logarithms and the known NNLO, N$^3$LO results \cite{Anastasiou:2015ema, Mistlberger:2018etf,Duhr:2019kwi} for 
DY and Higgs boson productions confirm this. 

Similarly from $\Psi^c$ to order $a_s^2$, we can predict
the tower consisting of  $({\cal D}_3$,${\cal D}_2)$, $({\cal D}_5,{\cal D}_4)$, 
$\cdots$,$({\cal D}_{2i-3},{\cal D}_{2i-4})$ 
and of $L^4_z,L^6_z,\cdots,L^{2i-2}_z$ at $a_s^3,a_s^4,\cdots,a_s^i$ respectively for all $i$. 
For the DY and Higgs production in bottom quark annihilation, our prediction reads as:
\begin{align}
   \Delta_{q(b)}^{NSV}& = a_s \Delta_{q(b)}^{NSV (1)}+a_s^2 \Delta_{q(b)}^{NSV (2)} 
       + a_s^3 \bigg[  - 512 C_F^3 L^5_z
       + \bigg( {7040 \over 9} C_F^2 C_A  
       - {1280 \over 9} n_f C_F^2 
       \nonumber\\&
       + 1728 C_F^3 \bigg) L^4_z 
       +\mathcal{O}(L^3_z) \bigg]
       + a_s^4 \bigg[  - {4096 \over 3} C_F^4  L^7_z
                +\bigg( {39424 \over 9} C_F^3 C_A 
                + {19712 \over 3} C_F^4
                 \nonumber\\&
                 - {7168 \over 9} n_f C_F^3 \bigg) L^6_z +\mathcal{O}(L^5_z) \bigg] +\mathcal{O}(a_s^5)
\end{align}
and for the Higgs production in gluon fusion, 
\begin{align}
   \Delta_g^{NSV}& = a_s\Delta_g^{NSV (1)}+a_s^2 \Delta_g^{NSV (2)} 
       + a_s^3 \bigg[  - 512 C_A^3 L^5_z
       + \bigg( {22592 \over 9} C_A^3 
       - {1280 \over 9} n_f C_A^2 \bigg)
       L^4_z 
       \nonumber\\&
       + \mathcal{O}(L^3_z) \bigg]
       + a_s^4 \bigg[  - {4096 \over 3} C_A^4  L^7_z
                +\bigg( {98560 \over 9} C_A^4  
                 - {7168 \over 9} n_f C_A^3 \bigg) L^6_z
                 + \mathcal{O}(L^5_z) \bigg] 
                 + \mathcal{O}(a_s^5)
\end{align}
Our predictions for $L^i_z,i=5,4$ agree with the those obtained by explicit computation 
\cite{Anastasiou:2014lda,Duhr:2019kwi}.  For the comparison purpose, we have presented the logarithms only
upto order $a_s^4$, however, the master formula can predict such logarithms to all orders in $a_s$.   Thanks to \cite{Anastasiou:2014lda,Duhr:2019kwi,Duhr:2020seh}, the third order results are now available for all these processes allowing us to   determine $\varphi_{f,c}$ for  $c=q,b,g$  till third order.  Using this, we can predict a tower of  $({\cal D}_3,{\cal D}_2),({\cal D}_5,{\cal D}_4) \cdots,({\cal D}_{2i-5},{\cal D}_{2i-6})$
and of $L^5_z,\cdots,L^{2i-3}_z$ at $a_s^4,a_s^5,\cdots,a_s^i$ respectively for all $i$.   In the following for the illustrative purpose, we have presented the NSV terms $L_z$ till seventh order in $a_s$.  For DY, we find
\begin{small}
\begin{align}\label{DY567}
\begin{autobreak}
   \Delta_q^{NSV} = 
   a_s\Delta_q^{NSV (1)}
   +a_s^2 \Delta_q^{NSV (2)} 
   + a_s^3 \Delta_q^{NSV (3)}
        + \textcolor{blue}{\pmb{a_s^4}}\bigg[ \bigg\{  - {4096 \over 3} C_F^4 \bigg\} L^7_z
       +  \bigg\{ {39424 \over 9} C_F^3 C_A 
       + {19712 \over 3} C_F^4 
       - {7168 \over 9} n_f C_F^3 \bigg\} 
       L^6_z
       +  \bigg\{  - {123904 \over 27} C_F^2 C_A^2 
       - \bigg({805376 \over 27} 
       - 3072 \zeta_2\bigg) C_F^3 C_A
          + \bigg(9088 +
           20480 \zeta_2\bigg) C_F^4
          + {45056 \over 27} n_f C_F^2 C_A 
        + {139520 \over 27} n_f C_F^3 - {4096 \over 27} n_f^2 C_F^2  \bigg\} L^5_z 
        + \mathcal{O}\big(L^{4}_z\big) \bigg]
        + \textcolor{blue}{\pmb {a_s^5}} \bigg[ \bigg\{-\frac{8192}{3} C_F^5\bigg\}L^9_z + \bigg\{   \frac{51200}{3} C_F^5 -\frac{8192}{3} C_F^4n_f 
        +  \frac{45056}{3}C_F^4C_A 
        \bigg\} L^8_z
        + \bigg\{ \bigg(\frac{72704}{3} + \frac{229376}{3}\zeta_2\bigg) C_F^5 
-  \bigg(\frac{1120256}{9} - \frac{32768}{3}\zeta_2\bigg)
C_F^4C_A 
-  \frac{81920}{81} C_F^3n_f^2 
+ \frac{194560}{9}C_F^4n_f
+ \frac{901120}{81} C_F^3C_An_f
-\frac{2478080}{81} C_F^3C_A^2\bigg\}L^7_z
+\mathcal{O}\big(L^{6}_z\big)\bigg]
+ \textcolor{blue}{\pmb {a_s^6}} \bigg[ \bigg\{-\frac{65536}{15} C_F^6\bigg\}  L^{11}_z
+ \bigg\{ \frac{167936}{5} C_F^6 -\frac{180224}{27} C_F^5n_f
+ \frac{991232}{27} C_F^5C_A \bigg\}
L^{10}_z
+ \bigg\{ \bigg(\frac{145408}{3} + 196608\zeta_2\bigg) C_F^6 
+  \frac{5054464}{81} C_F^5n_f
-\frac{327680}{81} C_F^4n_f^2 
- \bigg(  \frac{28997632}{81} 
-  \frac{81920}{3}\zeta_2\bigg) C_F^5C_A   
+ \frac{3604480}{81} C_F^4C_An_f
-\frac{9912320}{81} C_F^4C_A^2 \bigg\}
 L^9_z+ \mathcal{O}\big(L^{8}_z\big)\bigg] 
+ \textcolor{blue}{\pmb {a_s^7}}\bigg[\bigg\{ -\frac{262144}{45} C_F^7 \bigg\}L^{13}_z 
+ \bigg\{ \frac{2392064}{45}C_F^7 
-\frac{1703936}{135} C_F^6n_f 
+ \frac{9371648}{135}C_F^6C_A \bigg\}
L^{12}_z
+ \bigg\{ \bigg( \frac{1163264}{15} + \frac{5767168}{15}\zeta_2\bigg)C_F^7 
+ \frac{55115776}{405}C_F^6n_f 
- \bigg( \frac{ 315080704}{405} 
- \frac{262144}{5}\zeta_2\bigg) C_F^6C_A 
-\frac{ 917504}{81} C_F^5n_f^2 
+ \frac{10092544}{81} C_F^5C_An_f 
-\frac{ 27754496}{81} C_F^5C_A^2 \bigg\}L^{11}_z
+\mathcal{O}\big(L^{10}_z\big)\bigg]
+\mathcal{O}\big(a_s^8)\,,
\end{autobreak}
\end{align}
\end{small}
for the Higgs production in bottom quark annihilation, 
\begin{small}
\begin{align} \label{bB567}
   \Delta_b^{NSV} =& ~a_s\Delta_b^{NSV (1)}+a_s^2 \Delta_b^{NSV (2)} + a_s^3 \Delta_b^{NSV (3)}
       + {\color{blue} \pmb{a_s^4}}
       \Big[ \Delta_q^{NSV (4)} - 6144 C_F^4 L^5_{z} + \mathcal{O}\big(L^4_{z}\big) \Big] \qquad
      \nonumber\\&
        +{\color{blue} \pmb{a_s^5}}
       \Big[  \Delta_q^{NSV (5)} - 16384 C_F^5L^7_{z} + \mathcal{O}\big(L^6_{z}\big) \Big] 
       +{\color{blue} \pmb{a_s^6}}
       \Big[  \Delta_q^{NSV (6)} - 32768 C_F^6 L^9_{z}+ \mathcal{O}\big(L^8_{z}\big) \Big] 
       \nonumber\\&
       +{\color{blue} \pmb{a_s^7}}
       \Big[  \Delta_q^{NSV (7)} - \frac{262144}{5} C_F^7L^{11}_{z}
       + \mathcal{O}\big(L^{10}_{z}\big) \Big]
        + \mathcal{O}\big( a_s^8\big) \
       \,,
\end{align}
\end{small}
and for the Higgs production in gluon fusion,
\begin{small}
\begin{align} \label{gg567}
   \Delta_g^{NSV}& =
  ~ a_s\Delta_g^{NSV (1)}
   +a_s^2 \Delta_g^{NSV (2)} 
   + a_s^3 \Delta_g^{NSV (3)}
\nonumber\\&
       +{\color{blue} \pmb{a_s^4}} \bigg[\bigg\{  - {4096 \over 3} C_A^4 \bigg\}L_z^7
   +  \bigg\{ {98560 \over 9} C_A^4 - {7168 \over 9} n_f C_A^3 \bigg\}L_z^6
       +  \bigg\{  \bigg(- \frac{298240 }{ 9} + 23552 \zeta_2 \bigg) C_A^4 
  \nonumber\\ &    
       + {174208 \over 27} n_f C_A^3 - {4096 \over
         27} n_f^2 C_A^2  \bigg\}L_z^5 
         +\mathcal{O}\big(L_z^4\big) \bigg]
     +   {\color{blue} \pmb{a_s^5}} \bigg[
     \bigg\{  -\frac{8192}{3} C_A^5 \bigg\}L_z^9
       +  \bigg\{ \frac{96256}{3} C_A^5 
\nonumber \\ &       
       -\frac{8192}{3}   C_A^4 n_f  \bigg\}L_z^8
       +  \bigg\{  \bigg(-\frac{12283904}{81} + \frac{262144}{3}\zeta_2 \bigg) C_A^5 
       + \frac{2569216}{81}   C_A^4 n_f  -\frac{81920}{81} n_f^2 C_A^3 
          \bigg\}L_z^7
          \nonumber\\&
          +\mathcal{O}\big(L_z^6\big) \bigg]
     +   {\color{blue} \pmb{a_s^6}} \bigg[
     \bigg\{  -\frac{65536}{15} C_A^6 \bigg\}L_z^{11}
       +  \bigg\{ \frac{9490432}{135} C_A^6  -\frac{180224}{27 }  C_A^5 n_f  \bigg\}L_z^{10}
       +  \bigg\{  \bigg( \frac{671744}{3}\zeta_2
\nonumber\\&
       -\frac{4261888}{9} \bigg) C_A^6 + \frac{8493056}{81}   C_A^5 n_f  -\frac{327680}{81} n_f^2 C_A^4 
          \bigg\}L_z^9
          +\mathcal{O}\big(L^{8}_z\big) \bigg]
         \nonumber\\&
     +   {\color{blue} \pmb{a_s^7}} \bigg[
     \bigg\{  -\frac{262144}{45} C_A^7 \bigg\}L_z^{13}
       +  \bigg\{ \frac{3309568}{27}  C_A^7  -\frac{1703936}{135}   C_A^6 n_f  \bigg\}L_z^{12}
       +  \bigg\{  \bigg(-\frac{449429504}{405} 
       \nonumber\\&
       + \frac{1310720}{3}\zeta_2 \bigg) C_A^7 + \frac{11583488}{45}  C_A^6 n_f  - \frac{917504}{81} n_f^2 C_A^5 
          \bigg\}L_z^{11}
          +\mathcal{O}\big(L_z^{10}\big) \bigg]+\mathcal{O}\big(a_s^8)\,.
\end{align}
\end{small}
Our predictions for  $L^7_z, L^6_z$ and $L^5_z$ terms at fourth order for $\Delta_c$ agree 
with those of \cite{Moch:2009hr,Soar:2009yh,deFlorian:2014vta,Das:2020adl} predicted using physical evolution
equations. As can be seen from (\ref{DY567}-\ref{gg567}), given the third order results,
our master formula can predict three highest logarithms for fifth order onwards in $a_s$. For instance at $a_s^5$, we can predict $L^9_z,L^8, L^7_z$.
Generalising this, if we know $\Psi^c$ up to $n$th order,
we can predict $({\cal D}_{2i-2n+1},{\cal D}_{2i-2n})$ and $L^{2i-n}_z$ at every order in $a_s^i$ for all $i$.
\begin{table}[htp!]
\begin{center}
\begin{small}
\begin{tabular}{|p{1.2cm}|p{1.5cm}|p{1.5cm}|p{1.5cm}||p{1.5cm}|p{1.5cm}|p{3.15cm}|}
 \hline
 \multicolumn{4}{|c||}{GIVEN} & \multicolumn{3}{c|}{PREDICTIONS}\\
 \hline
 \hline
 \rowcolor{lightgray}
 $\Psi_c^{(1)}$ & $\Psi_c^{(2)}$ &$\Psi_c^{(3)}$&$\Psi_c^{(n)}$&$\Delta_c^{(2)}$&$\Delta_c^{(3)}$& \quad \quad $\Delta_c^{(i)}$\\
 \hline
 ${\cal D}_0,{\cal D}_1,\delta$   &      &&   & ${\cal D}_3,{\cal D}_2$ &${\cal D}_5,{\cal D}_4$ &${\cal D}_{(2i-1)},{\cal D}_{(2i-2)}$\\
	$L_{z}^{1},L_{z}^{0}$ &  & &  &$L_{z}^{3}$ & $L_{z}^{5}$&  $L_{z}^{(2i-1)}$\\
 \hline
  &  ${\cal D}_0,{\cal D}_1,\delta$ &   &&&${\cal D}_3,{\cal D}_2$&${\cal D}_{(2i-3)},{\cal D}_{(2i-4)}$\\
	& $L_{z}^{2},L_{z}^{1},L_{z}^{0}$ & & & &$L_{z}^{4}$ &$L_{z}^{(2i-2)}$ \\
  \hline
  & & ${\cal D}_0,{\cal D}_1,\delta$& &&&${\cal D}_{(2i-5)},{\cal D}_{(2i-6)}$ \\
	& &$L_{z}^{3},\cdots,L_{z}^{0}$ & & & & $L_{z}^{(2i-3)}$ \\
  \hline
 & & &  ${\cal D}_0,{\cal D}_1,\delta$ &&&${\cal D}_{(2i-(2n-1))},{\cal D}_{(2i-2n)}$  \\
	& & & $L_{z}^{n},\cdots,L_{z}^{0}$ & &  &$L_{z}^{(2i-n)}$ \\
  \hline
\end{tabular}
\end{small}
\end{center}
	\caption{\label{tab:Table1} Towers of Distributions ($\mathcal{D}_i$) and 
	 NSV logarithms ($\ln^i(1-z)$) that can be predicted for $\Delta_c$ using \eqref{MasterF}. Here $\Psi_c^{(i)}$ and $\Delta_c^{(i)}$ denotes $\Psi_c$ and $\Delta_c$ at order $a_s^i$ respectively. Also the symbol $L_{z}^{i}$ denotes $\ln^i(1-z)$.}
\end{table}}

Table[\ref{tab:Table1}] is devoted to summarise the predictions from the master formula for any given order of $a_s$. We also present the explicit structure of $\Delta_c$ till four loop in Appendix[\ref{ap:Delta}] as well as in the ancillary files with the \arXiv  \ submission.

The predictive power of the master formula to all orders in $a_s$ in terms of distributions
and $\ln(1-z)$ terms in $\Delta_c$ is due to the all order structure of the exponent $\Psi^c$ and 
this can be further 
exploited to resum them.  We devote a separate section for this.}\\
So far, we have compared our higher predictions for SV and NSV logarithms obtained using the lower order results against those available in the literature and found that our all order master formula correctly predicts these  logarithms.  
{\color{black}For example, from the knowledge of the 
second order result for $\Psi^c$, we can correctly predict $\ln^5(1-z)$ and $\ln^4(1-z)$ terms} at third order. 
Even though this second order information is not sufficient 
to predict the lower order  NSV logarithms, namely $\ln^k(1-z)$ for $k=3,2,1,0$ at $a_s^3$ level,  we observe
that our predictions 
for these logarithms agree with the known results for several color factors.  
\begin{table}[htp!] 
\begin{center}
\begin{small}
{\renewcommand{\arraystretch}{1.7}
\begin{tabular}{|p{0.7cm}||P{1.5cm}|P{1.5cm}||p{1.2cm}||P{1.5cm}|P{1.5cm}||P{1.5cm}|P{1.5cm}|}
 \rowcolor{lightgray}
    & \multicolumn{2}{c||}{$gg\rightarrow H$}   
    &&\multicolumn{2}{c||}{Drell-Yan (DY)}  
   & \multicolumn{2}{c|}{$b\overline b \rightarrow H$} 
     \\ 
  \hline
$C_A^3$ & $\frac{-111008}{27}+3584\zeta_2$ & $\frac{-110656}{27} + 3584\zeta_2 + \chi_1$&$C_F^3$ &$2272 + 3072\zeta_2$ & $2272+3072\zeta_2$   & $736 + 3072 \zeta_2$   &   $736 + 3072 \zeta_2$ \\
$C_A^2n_f$ &$\frac{6560}{9}$&$\frac{19616}{27} + \chi_2$&$C_F^2 n_f$ & $\frac{19456}{27}$  & $\frac{6464}{9}+\chi_3$ &  $\frac{19456}{27}$  & $\frac{6464}{9}+\chi_3$ \\
$C_An_f^2$ &$\frac{-256}{27}$&$\frac{-256}{27}$&$C_A C_F^2$   &$\frac{-111904}{27} + 512\zeta_2$   &   $\frac{-37184}{9} + 512\zeta_2+\chi_4$   &$\frac{-111904}{27} + 512\zeta_2$& $\frac{-37184}{9} + 512\zeta_2+\chi_4$  \\ 
&&&$C_F n_f^2$   &$\frac{-256}{27}$  &   $\frac{-256}{27}$   &   $\frac{-256}{27}$   &   $\frac{-256}{27}$   \\
&&&$C_A C_F n_f$   &$\frac{2816}{27}$   &   $\frac{2816}{27}$   &  $\frac{2816}{27}$   &   
   $\frac{2816}{27}$   \\ 
 &&&  $C_A^2 C_F $   &$\frac{-7744}{27}$   &   $\frac{-7744}{27}$   &   $\frac{-7744}{27}$   &   $\frac{-7744}{27}$  \\
 \hline
\end{tabular}}
\end{small}
\end{center}
	\caption{\label{tab:Table6} Comparison of $\ln^3(1-z)$ coefficients at the third order against exact results.  The left column stands for the exact results and the right column gives the respective contributions when $\Psi^c$ is taken till two loop.}
\end{table}

In Table [\ref{tab:Table6}] we compare our predictions for 
$\ln^3(1-z)$ terms at the third order, which are obtained using 
$\Psi^c$ considered till $a_s^2$, against the 
known results for the DY production, Higgs productions in 
bottom quark annihilation and gluon fusion.
As can be seen from the table, the master formula correctly predicts the results for many color factors. For instance, for DY, the predictions for color factors $C_F^3, C_F n_f^2, C_AC_Fn_f$ and $C_A^2C_F$ are matching with the exact results. However for the other color factors, certain third order information are required, which is represented as $\chi_i$ which when taken into account will reproduce the exact  $\ln^3(1-z)$ terms at third order.

\section{Resummation of next to SV in $N$ space}
To study all order behavior of $\Delta_c$ in Mellin space,  it is convenient to use the 
integral representations of both $\mathrm{\Phi}_{A}^{c}$ and $\mathrm{\Phi}_{B}^{c}$ given in 
\eqref{phiA} and  \eqref{phiBint} respectively.  Substituting the solutions for $\hat F_c$ and renormalisation constant
$Z_{UV,c}$ and the $\ln \Gamma_{cc}$ along with the integral representations for $\mathrm{\Phi}_{A}^{c}$ and $\mathrm{\Phi}_{B}^{c}$ in 
\eqref{MasterF},
we find   
\begin{eqnarray}
\label{resumz}
\Delta_c(q^2,\mu_R^2,\mu_F^2,z)= C^c_0(q^2,\mu_R^2,\mu_F^2) 
~~{\cal C} \exp \Bigg(2 \Psi^c_{\cal D} (q^2,\mu_F^2,z) \Bigg)\,,
\end{eqnarray}
where
\begin{eqnarray}
\label{phicint}
\Psi^c_{\cal D} (q^2,\mu_F^2,z) &=& {1 \over 2}
\int_{\mu_F^2}^{q^2 (1-z)^2} {d \lambda^2 \over \lambda^2} 
	P^{\prime}_{cc} (a_s(\lambda^2),z)  + {\cal Q}^c(a_s(q^2 (1-z)^2),z)\,,
\end{eqnarray}
with
\begin{eqnarray}
\label{calQc}
{\cal Q}^c (a_s(q^2(1-z)^2),z) &=&  \left({1 \over 1-z} \overline G^c_{SV}(a_s(q^2 (1-z)^2))\right)_+ + \varphi_{f,c}(a_s(q^2(1-z)^2),z).
\end{eqnarray}
The coefficient $C_0^c$ is $z$ independent coefficient and is expanded in powers of $a_s(\mu_R^2)$ as
\begin{eqnarray}
\label{C0expand}
C_0^c(q^2,\mu_R^2,\mu_F^2) = \sum_{i=0}^\infty a_s^i(\mu_R^2) C_{0i}^c(q^2,\mu_R^2,\mu_F^2)\,,
\end{eqnarray}
where the coefficients $C^c_{0i}$ are presented in the ancillary files along with the \arXiv \ submission. Also one can find $C^c_{0}$ for DY and Higgs production in \cite{Catani:2014uta}. 
\eqref{resumz} is our $z$ space resummed result for $\Delta_c$ in integral representation which
can be used to predict SV and NSV terms to all orders in perturbation theory in terms of universal anomalous dimensions,
$A^c,B^c,C^c,D^c,f^c$, SV coefficients $\overline {\cal G}_{i}^{c,(j)}$, NSV coefficients
${\cal G}_{L,i}^{c,(j,k)}$  and process dependent $C_{0i}^c$.  We have few comments in order.  
The next to SV corrections to various inclusive processes were studied in a series of papers \cite{Laenen:2008ux,Laenen:2010kp,Laenen:2010uz,Bonocore:2014wua,Bonocore:2015esa,Beneke:2019oqx,Beneke:2019mua} and 
lot of progress have been made which lead to better understanding of the underlying physics.
Our result has close resemblance with 
the one  which  was conjectured in \cite{Laenen:2008ux}
and indeed there are few terms which are common in both the results.  Our result, \eqref{phicint} 
differs from Eq.(31) in \cite{Laenen:2008ux}, in the upper limit of the integral, 
the presence of extra term $\varphi_{f,c}$ and the dependence on the variable $z$.
These differences do not alter the SV predictions but 
will give NSV terms different from those obtained using Eq.(31) of \cite{Laenen:2008ux}. 

The Mellin moment of $\Delta_c$ is now straight forward to compute using the integral representation
given in \eqref{phicint}.  Note that the \eqref{phicint} is suitable for obtaining
only SV and NSV terms while the predictions using this formula beyond NSV terms 
such as those proportional to ${\cal O}((1-z)^n \ln^j(1-z));n,j\ge 0$ in $z$ space and terms of 
${\cal O}(1/N^2)$ in $N$ space will not be correct! 
Hence, we compute the Mellin moment of \eqref{resumz} in the appropriate limit of N such that the resulting expression
in $N$ space correctly predicts all the SV and NSV terms.  The limit $z\rightarrow 1$ translates
to $N\rightarrow \infty$ and if one is interested to include NSV terms, we need to keep ${\cal O}(1/N)$
corrections in the large $N$ limit.  
The Mellin moment of $\Delta_c$ is given by
\begin{eqnarray}
\label{DeltaN}
\Delta_{c,N}(q^2,\mu_R^2,\mu_F^2) = C_0(q^2,\mu_R^2,\mu_F^2) \exp\left(
\Psi_N^c (q^2,\mu_F^2) 
\right)\,,
\end{eqnarray}
where
\begin{eqnarray}\label{eq:Mellin}
\Psi_N^c(q^2,\mu_F^2) = 2 \int_0^1 dz z^{N-1}\Psi_{\cal D}^c (q^2,\mu_F^2,z) .
\end{eqnarray}
The computation of Mellin moment in the large $N$ limit which retains SV and NSV terms 
involves two major steps: 1. following \cite{Laenen:2008ux} and we replace $\int dz ({z^{N-1}-1})/(1-z)$ 
and $\int dz z^{N-1}$ by a theta function
$\theta(1-z-1/N)$ and apply the operators $\Gamma_A(N \frac{d}{dN})$ and $\Gamma_B(N \frac{d}{dN})$ on the integrals
respectively; 2. we perform the integrals over $\lambda^2$ after expressing $a_s(\lambda^2)$ in terms of
$a_s(\mu_R^2)$ obtained using resummed solution to RG equation of $a_s$ in \eqref{resumas}.  Step 1 makes sure that
we retain only  $\ln^j(N)$ and $(1/N) \ln^j(N)$ terms and step 2 guarantees the resummation
of $2 \beta_0 a_s(\mu_R^2) \ln N$ terms to all orders and also the organisation of the result in
powers of $a_s(\mu_R^2)$ .  The details of the computation are 
described in the Appendix[\ref{ap:PsiN}] .  The Mellin moment of the exponent
takes the following form: 
\begin{eqnarray}
\label{eq:Psi}
\Psi_N^c = \Psi_{\rm{sv},N}^c + \Psi_{\rm{nsv},N}^c 
\end{eqnarray}
where we have split $\Psi_N^c$ in such a way that all those terms that are functions of $\ln^j(N), j=0,1,\cdots$ are
kept in $\Psi_{{\rm sv},N}^c$ and the remaining terms that are proportional to $(1/N) \ln^j(N), j=0,1,\cdots$ are contained
in $\Psi_{\rm{nsv},N}^c$.  Hence,
\begin{eqnarray}
\label{PsiSVN}
	\Psi_{\rm{sv},N}^c = \ln(g_0^c(a_s(\mu_R^2))) + g_1^c(\omega)\ln N + \sum_{i=0}^\infty a_s^i(\mu_R^2) g_{i+2}^c(\omega) \,,
\end{eqnarray}
where $g^c_i(\omega)$ are identical to those in \cite{Catani:1989ne,Moch:2005ba,H:2019dcl} obtained from the resummed formula for SV terms. {\color{black} It is to be noted that $g^c_i(\omega)$ vanishes in the limit $\omega \rightarrow 0$}. 
The coefficients $g^c_0(a_s)$ is expanded in powers of $a_s$ as (see \cite{Moch:2005ba})
\begin{eqnarray}
	\ln(g_0^c(a_s(\mu_R^2))) = \sum_{i=1}^\infty a_s^i(\mu_R^2) g^c_{0,i}\quad \,.
\end{eqnarray}
We also provide $g_0^c(a_s(\mu_R^2))$ in the ancillary files along with the \arXiv \ submission. {\color{black} The $N$-independent coefficients $C^c_0$ and $g_0^c$ are related to the coefficients $\tilde g_0^c$ given in the paper \cite{H:2019dcl,H.:2020ecd} using the following relation,
\begin{equation}
\tilde g_0^c(q^2, \mu_R^2, \mu_F^2) = C_0^c(q^2,\mu_R^2,\mu_F^2) \  g_0^c(a_s(\mu_R^2))
\end{equation}
which can be expanded in terms of $a_s(\mu_R^2)$ as,
\begin{equation}
  \tilde g_0^c(a_s(\mu_R^2)) = \sum_{i=0}^\infty a_s^i(\mu_R^2) \tilde g^c_{0,i}\quad \,.
\end{equation}
The function $\Psi_{\rm{nsv},N}^c$ is given by}
\textcolor{black}{
\begin{align}
\label{PsiNSVN}
 \Psi_{\rm{nsv},N}^c = {1 \over N} 
\sum_{i=0}^\infty a_s^i(\mu_R^2) \bigg ( \bar g_{i+1}^c(\omega)    
+ h^c_{i}(\omega,N) \bigg)\,,
\end{align}
with 
\begin{align}\label{hNSV}
h^c_i(\omega,N) = \sum_{k=0}^{i} h^c_{ik}(\omega)~ \ln^k(N).
\end{align}}
where $\bar g^c_i(\omega)$ and $h^c_{ik}(\omega)$ are presented in the Appendix[\ref{ap:gbar}] and Appendix[\ref{ap:hij}] respectively. We also provide these coefficients till four loop in the ancillary files with the \arXiv  \ submission.
{\textcolor{black} {We can see that in each coefficient, say $g_i^c(\omega),\overline g_i^c(\omega), h_{ik}^c(\omega)$ from the SV as well as the NSV, we are resumming in Mellin space ``order one" term $\omega$ to all orders in perturbation theory.  This is the consequence of the argument  in the coupling constant $a_s (q^2(1-z)^2)$ resulting from the integral over $\lambda$ and the function ${\cal Q}^c$.  
The peculiarity of the series is that the SV $g_1^c(\omega)$ comes with $\ln N$ and hence it starts with a double logarithm.   This extra $\ln N$ arises from  the Mellin moment of the factor $1/(1-z)_+$ appearing in the exponent.
Similarly for $\Psi_{\rm{nsv},N}^c$ we note that it is proportional to $1/N$ at every order as expected.
Explicit $\ln N $ that appear with $h_{ik}^c(\omega)$ results from the explicit $\ln(1-x)$ appearing in the exponent.  
The sum containing $\bar g^c_i, i= 1,2,\cdots$ results entirely from $A^c$ coefficients of $P^{\prime}_{cc}$ and 
from the function $\bar G_{SV}^c$ of \eqref{calQc}.  We find that none of the coefficients $\bar g^c_i(\omega)$ contains
explicit $\ln N$. The second sum comes from $C^c,D^c$ coefficients of $P^{\prime}_{cc}$ and 
from $\varphi_{f,c}$ and each term in this expansion contains explicit $\ln^k(N), k=0,\cdots,i$.
 We find that coefficient of $h^c_{01}$ is proportional to $C^c_1$ which is identically zero.
 Hence, at order $a_s^0$, there is no $(1/N) \ln N$ term.
}}   

Summarising, we find that in Mellin $N$ space one obtains compact expression for the exponent in terms of quantities that
are functions of $\omega = 2 a_s(\mu_R^2) \beta_0 \ln N$ as we use resummed $a_s$ to perform the integral.  
In addition, the resummed $a_s$ allows us to organise the $N$ space perturbative expansion in such 
a way that  $\omega$ is treated as order one at every order in $a_s(\mu_R^2)$.  
Both integral representation in $z$-space
and Mellin moment of the integral in $N$ space contain exactly same information and hence predict 
SV and NSV logarithms to all orders in perturbation theory.  The 
all order structure is more transparent in $N$ space compared to $z$ space result and 
it is technically easy to use resummed result in $N$ space for any phenomenological studies.

Let us first consider $\Psi_{\rm{sv},N}^c$ given in \eqref{PsiSVN}.  If we keep only $\tilde g_{0,0}$ and $g_1$ terms  in 
\eqref{PsiSVN} and expand the exponent in powers of $a_s=a_s(\mu_R^2)$, we can 
can predict leading $a_s^i \ln^{2i} (N)$ terms 
for all $i>1$.  This happens because of the all order structure of $\mathrm{\Phi}_A^c$ in $z$ space.  For example if 
we know $\mathrm \Phi_A^c$ to order $a_s$, we can predict  
rest of the other terms of the form  $a_s^i {\cal D}_{2i-1}(z)$ in $\mathrm{\Phi}_A^c$ for all $i>1$.
If we further include $\tilde g_{0,1}$ and $g_2$ terms, then we can predict
next to leading $a_s^i \ln^{2i-1} (N)$ terms for all $i>2$.  
Again this is due to the fact that in $z$ space, knowing $\mathrm{\Phi}_A^c$ to second order one
can predict $a_s^i {\cal D}_{2i-2}(z)$ terms for all $i>3$.
In general, resummed 
result with terms {\color{black} $\tilde g^c_{0,0},\cdot\cdot \cdot \tilde g^c_{0,n-1}$ and $g^n_1,\cdot \cdot \cdot, g^c_{n}$ can predict
$a_s^i \ln^{2i-n+1}(N)$ or $a_s^i {\cal D}_{2i-n}(z)$ terms for $i>n$}.  

The inclusion of sub leading terms through $\exp(\Psi_{\rm{nsv},N}^c)$ gives  
additional $(1/N) \ln^j(N)$ terms in $N$ space or $\ln^j(1-z)$ terms in $z$ space.  
In perturbative QCD, $C^c_1=0$, where $c=q,g$ and we use this in the rest of our analysis.  
Like the $\Psi_{\rm{sv},N}^c$ exponent, $\Psi_{\rm{nsv},N}^c$ also organises the perturbation theory by keeping  
$2 a_s(\mu_R^2) \beta_0 \ln N$ terms as order one at every order in $a_s$. However these terms are suppressed by
$1/N$ factor at every order in $a_s$. 

\textcolor{black}{We find that if we keep $\{\tilde g^c_{0,0},g^c_1\}$ in $\Psi_{\rm{sv},N}^c$ and  $\{\overline g^c_1,  h^c_{0}\}$
in $\Psi_{\rm{nsv},N}^c$ and drop the rest, one
can predict $(a_s^i/N) \ln^{2i-1}(N)$ terms for CFs for all $i>1$.
We call this tower of logarithms by NSV-Leading Logarithm (NSV-LL).
Similarly, knowing, along with the previous ones, $\{\tilde g^c_{0,1}$,$g^c_2\}$  in $\Psi_{\rm{sv},N}^c$ and $\{\overline g^c_2,
 h^c_1\}$ in $\Psi_{\rm{nsv},N}^c$, one
can predict $(a_s^i /N) \ln^{2i-2} (N)$ for CFs for all $i>2$.
This belongs to NSV-Next-to-Leading Logarithm (NSV-NLL).
In general, resummed result
with $\overline g^c_1 , \cdots,\overline g^c_{n+1}$ and $ h^c_0 , \cdot \cdot \cdot , h^c_{n}$ in $\Psi_{\rm{nsv},N}^c$ along with
$\tilde g^c_{0,0},\cdot\cdot \cdot, \tilde g^c_{0,n}$ and $g^c_1,\cdot \cdot \cdot, g^c_{n+1}$ in $\Psi_{\rm{sv},N}^c$ can
predict $(a_s^i/N) \ln^{2i - (n+1)}(N)$ for all $i>n$ in Mellin space $N$ and it is NSV-N$^n$LL. We summarise our findings in the Table [\ref{tab:Table2}] below.}
\begin{table}[h!]
\centering
\begin{center}
\begin{small}
{\renewcommand{\arraystretch}{1.6}
\begin{tabular}{|P{2.9cm}|P{3.2cm}||P{1.7cm}|P{1.7cm}|P{2.0cm}|}
 \hline
 \multicolumn{2}{|c||}{GIVEN} & \multicolumn{3}{c|}{PREDICTIONS}\\
 \hline
 \hline
 \rowcolor{lightgray}
 Logarithmic & Resummed     &    $\Delta^{(2)}_{c,N}$&$\Delta^{(3)}_{c,N}$&  $\Delta^{(i)}_{c,N}$
  \\
  \rowcolor{lightgray}
                  Accuracy &   Exponents  & & &\\
 \hline
NSV-LL&	$\tilde g^c_{0,0},g^c_1,\overline g^c_1,h^c_0$              & $ L_N^3$ &$ L_N^5$&$  L_N^{2i-1}$ \\
 \hline
NSV-NLL &$\tilde g^c_{0,1},g^c_2,\overline g^c_2, h^c_1$   &&$L_N^4$&$ L_N^{2i-2}$\\
  \hline
NSV-N$^2$LL&$\tilde g^c_{0,2},g^c_3,\overline g^c_3, h^c_2$    &&&$L_N^{2i-3}$ \\
  \hline
NSV-N$^n$LL&$\tilde g^c_{0,n},g^c_{n+1},\overline g^c_{n+1}, h^c_n$   &&&$ L_N^{2i-(n+1)}$  \\
  \hline
\end{tabular}}
\end{small}
\end{center}
	\caption{\label{tab:Table2} The all order predictions for $1/N$ coefficients of $\Delta_{c,N}$ for a given set of resummation coefficients \Big\{$\tilde g^c_{0,i},g^c_i(\omega),\overline{g}^c_i(\omega), h^c_{i}(\omega)$\Big\} at a given order. Here $L_N^i = {1\over N}\ln^i(N)$ }
\end{table}

\textcolor{black}{We find that unlike SV resummed terms, which result from only $\mathcal{D}_0$ and $a_s(q^2(1-z)^2)$, the resummation of NSV terms is controlled in addition by $\ln(1-z)$ at each order in $a_s$  as can be seen from \eqref{phicint}. This logarithmic dependence in $\rm{\Phi}_B^c$ at each order along with resummed $a_s(q^2(1-z)^2)$, allows one to reorganize order one terms differently from SV case. 
Hence, the resulting NSV resumed result has different logarithmic structure  in terms of order one $\omega$ compared to that of SV. }

\textcolor{black}{
Few remarks on the resummed result are in order in the light of previous section.  
Note that we considered a particular solution $\rm \Phi^c_B$ that
corresponds to the case $\alpha=2$ and summed up order
one terms $\omega$ in Mellin $N$ space using the resummed solution to
RGE of $a_s$.  While the SV part is insensitive to $\alpha$, the NSV terms, namely
the resummation exponents   $h^c(\omega)$   
depend on $\alpha$ ($\alpha=2$) 
through $\omega$ resulting from $a_s(q^2/N^\alpha)$ 
and the coefficients $\varphi_{c,\alpha,i}^{(k)}$.  We had already seen how $\varphi^{(k)}_{c,\alpha,i}$ transforms with respect to $\alpha$.  The resummed result in the $N$ space for arbitrary $\alpha$ will be function of $a_s(q^2/N^\alpha)$.  This will lead to resummation of 
order one $\omega_\alpha = \alpha \beta_0 a_s(\mu_R^2) \ln N$ to all orders in $a_s$.  Hence, the summation of order one $\omega_\alpha$ terms with $\alpha$ dependent coefficients $\varphi_{c,\alpha,i}$
leads variety of resummed predictions each depending on the choice of $\alpha$.  However, the fixed order predictions for the CFs $\Delta_c$ will be unaffected, thanks to the invariance in NSV solution.  This invariance has allowed us to choose $\alpha=2$ to resum order one $\omega$ terms analogous to SV counterpart.} 

 {There have been several attempts \cite{DelDuca:2017twk,Bahjat-Abbas:2019fqa,Beneke:2018gvs,Beneke:2019mua,Beneke:2019oqx}
 in the past
to understand the structure of NSV logarithms of
inclusive cross sections and its all order structure and in this context, we compare our prediction at LL level for CF
of DY, $\Delta_{q}^{LL}$ against that of \cite{Beneke:2018gvs}. Note that
the \cite{Beneke:2018gvs} contains NSV terms only to LL accuracy.
In \cite{Beneke:2018gvs} , within the framework of soft-collinear effective theory (SCET), the authors have obtained leading logarithmic terms at NSV for the quark-antiquark production channel of the DY process to all orders in $a_s$. This was achieved by extending the factorisation properties of the
cross section to NSV level and using renormalisation group equations of NSV operators and soft functions.  Using our $N$ space result, in the LL approximation, that is for DY
\begin{align}
    \Delta_{c,N}^{LL} &= \tilde g^c_{0,0} \exp\bigg[\ln N ~ g^c_1(\omega)+ \frac{1}{N}\bigg(\overline g^c_1(\omega) + h^c_0(\omega,N)\bigg)\bigg]\bigg|_{LL}
\end{align}
we obtain,
\begin{align}
   \Delta_{c,N}^{LL} & =\exp\bigg[8C_Fa_s\bigg(\ln^2 N + \frac{\ln N}{N}\bigg)\bigg]
\end{align}
where we have expanded the exponents in powers of $a_s$
and kept only terms of ${\cal O}(1/N)$.   The above $N$ space result can be Mellin transformed to $z$ space and it reads as
\begin{align}
   \Delta_{c,N}^{LL}= \Delta_{c,SV}^{LL}  -16C_Fa_s\exp\bigg[8C_Fa_s\ln^2(1-z)\bigg]\ln(1-z)
\end{align}
The above result agrees exactly with Eq.(4.2) of \cite{Beneke:2018gvs} for  $\mu = Q$.   Our result given in \eqref{eq:Psi} contains terms that can in principle resum N$^n$LL, $n\ge 0$ provided
the universal anomalous dimensions and process dependent
coefficients are available to desired accuracy in $a_s$.  Hence given three loop results, which are available
for several observables, we can perform N$^2$LL resummation taking into
account NSV logarithms.
}

\section{Physical Evolution Kernel}
 {In the past, in \cite{Grunberg:2009yi},  the scheme invariant approach through physical evolution equation 
was explored to understand the structure of NSV terms for the coefficient functions of DIS cross section.  The physical evolution kernel that controls the evolution of the physical obervables with respect to external scale $q^2$ is invariant under scheme transformations with respect to renormalisation and factorisation.  This property can be exploited to
understand certain universal structure of perturbative predictions.   By suitably modifying
physical evolution kernel (PEK) \cite{Grunberg:2009yi} with the help of scales in the strong coupling constant
and using the renormalisation group invariance, predictions at second and third orders for
the CFs of DIS structure functions were made, given the known lower order results for CFs.  Even though, the predictions
did not agree for some of the color factors, it was found that they were
very close to the known results.}  
Using the second order results for DIS, semi-inclusive $e^+ e^-$ annihilation and DY, 
a striking observation was made by Moch and Vogt 
in \cite{Moch:2009hr} (and \cite{deFlorian:2014vta,Das:2020adl})
on the  PEK namely the enhancement of a single-logarithms at large $z$ to all order in $1-z$.  
It was found that if one conjectures that it will hold true at every order in $a_s$, the structure of corresponding  
leading $\ln(1-z)$ terms in the kernel can be constrained.  This allowed them to predict 
certain next to SV logarithms at  higher orders in $a_s$ which are in agreement with the known results up to third order.   

Motivated by this approach, we use our formulation that describes 
next to SV  logarithms both in $z$ and $N$ spaces to study the
structure of physical evolution equation and present
 our findings on the structure of leading logarithms in the PEK.
For convenience we work in Mellin space.
The Mellin moment of hadronic cross section  $\sigma(q^2,\tau)$ is given by 
\begin{eqnarray}
\sigma_N(q^2) = \int_0^1 d\tau \tau^{N-1} \sigma(q^2,\tau)
\end{eqnarray}
The  hadronic observable $\sigma(q^2,\tau)$ is renormalisation scheme (RS) independent 
namely it does not depend on the scheme in which CFs $\Delta_{ab}$ and the structure functions $f_c$ are defined. 
The fact that $f_c$ is independent of $q^2$, the first derivative of $\sigma$ with respect to $q^2$ 
will not depend on  $f_c$.  Restricting ourselves to SV and NSV terms, we can define physical evolution kernel
${\cal K}^c$ by  
\begin{eqnarray}
{\cal K}^c(a_s(\mu_R^2),N) &=&q^2 {d \over d q^2} \ln\left({ \sigma_N(q^2)\over \sigma_0(q^2)}\right)\Bigg|_{\rm{sv+nsv}} \,,
\nonumber\\
&=&  q^2 {d \over d q^2} \ln \Delta_{c,N}(q^2) .
\end{eqnarray}
which is independent of any renormalisation scheme.   
The kernel ${\cal K}^c(a_s(\mu_R^2),N)$ can be computed  order by order in perturbation theory using $\ln \Delta_{c,N}$.
\begin{eqnarray}
	{\cal K}^c(a_s(\mu_R^2),N) = \sum_{i=1}^\infty a_s^i(\mu_R^2) {\cal K}^c_{i-1}(N)
\end{eqnarray}
As in \cite{Moch:2009hr}, the leading $(1/N) \ln^i(N)$ terms at every order
defined by ${\cal K}^c$:
\begin{eqnarray}
\overline {\cal K}_i^c = {\cal K}_i^c\bigg|_{(1/N) \ln^i(N) }
\end{eqnarray}
can be obtained.   
Using \eqref{DeltaN}, we find that these terms can be obtained directly from $\Psi_{\rm{nsv},N}^c$ alone 
and are given by 
\begin{align}
\label{PEK}
\overline {\cal K}^c_0 &= A_1^c +2 D^c_1\,,
\nonumber\\
\overline {\cal K}^c_1 &= 2 A_1^c \beta_0 -2 C^c_2 + 4 \beta_0~ D^c_1 + 2 \beta_0~ \varphi_{c,1}^{(1)}\,,
\nonumber\\
\overline {\cal K}^c_2&=4 A_1^c \beta_0^2 -8 \beta_0 ~C^c_2 + 8 \beta_0^2~ D^c_1 + 8 \beta_0^2~ \varphi_{c,1}^{(1)} - 4 \beta_0~ \varphi_{c,2}^{(2)}\,,
\nonumber\\
\overline {\cal K}_3&=8 A_1^c \beta_0^3 -24 \beta_0^2~ C^c_2 + 16 \beta_0^3~ D^c_1 + 24 \beta_0^3~ \varphi_{c,1}^{(1)} 
- 24 \beta_0^2~ \varphi_{c,2}^{(2)} + 6 \beta_0~ \varphi_{c,3}^{(3)}
\,,\nonumber\\
\overline {\cal K}^c_4&=16 A_1^c \beta_0^4 -64 \beta_0^3~ C^c_2 + 32 \beta_0^4~ D^c_1 + 64 \beta_0^4~ \varphi_{c,1}^{(1)} 
- 96 \beta_0^3~ \varphi_{c,2}^{(2)} 
+ 48 \beta_0^2~ \varphi_{c,3}^{(3)} - 8 \beta_0~ \varphi_{c,4}^{(4)}.
\end{align}
We find that the structure of $\overline {\cal K}_i^c$ resembles very much like that of \cite{Moch:2009hr}.
{Interestingly, the leading logarithms at every order depends only on the universal
anomalous dimensions $A_1^c,D_1^c$ and $C_2^c$, and the diagonal coefficients $\varphi_{c,k}^{k}$ with $k<i$,
where $i$ is the order of the perturbation.  In addition, if we substitute the known values for these
quantities in the \eqref{PEK}, we obtain
\begin{eqnarray}
\overline {\cal K}^c_1  &=& -8 \beta_0 C_i - 32 C_i^2
\nonumber\\
\overline {\cal K}^c_2  &=& -16\beta_0^2 C_i - 112\beta_0 C_i^2
\nonumber\\
\overline {\cal K}^c_3  &=& -32\beta_0^3 C_i - {896 \over 3} \beta_0^2 C_i^2
\nonumber\\
\overline {\cal K}^c_4  &=& -64\beta_0^4 C_i - {2176 \over 3} \beta_0^3 C_i^2 - 8\beta_0 \varphi_{c,4}^{(4)}
\end{eqnarray}
where $C_i=C_F$ for $c=q,b$ and $C_i=C_A$ for $c=g$.}

{
The reason for the agreement of our predictions for PEK to third order with
those of \cite{Moch:2009hr} is simply because of the K+G equation that $\rm \Phi^c$ satisfies.  In fact, K+G equation is partonic version of
the  physical evolution equation and the partonic PEK given by
$\overline K^c + \overline G^c$.  The logarithm structure of PEK
is controlled by  the upper limit $i$ in the summation over the index $k$ 
in \eqref{varphiexp}.  
In $N$-space, the highest power of corresponding   $\ln N$ in the $1/N$ coefficient of ${\cal K}^c$ is in turn controlled by  the upper limit on the summation in \eqref{GikLj}.  
Our predictions based on the inherent transcendentality structure of perturbative results are in complete agreement with
the logarithmic structure of CFs or  PEKs obtained from explicit results.}  
Note that the structure of PEK \eqref{PEK} expressed in terms of $A_1^c, C_2^c$, $D_1^c$ and $\varphi_{c,i}^{(i)}$ is straightforward to understand from K+G equations and renormalisation group invariance.  {\color{black}However,
as was already noted in \cite{Moch:2009hr}, the coefficient of the leading logarithms contains peculiar structure containing only $\beta_0^i$ and $\beta_0^{i-1}$
at every order in $a_s^i$.    In addition, if the structure continues to be true at every order, the coefficients $\varphi_{c,i}^{(i)}$ has to be proportional to $\beta_0^{i-2}$  for every $i$ which can be tested when results beyond third order become available. }

\section{Conclusions}
Understanding the structure of threshold logarithms in inclusive reactions such as 
production of a pair of leptons in Drell-Yan process and of Higgs boson in gluon annihilation
as well as bottom quark annihilation is important because they not only dominate
but also become large in certain kinematical regions spoiling the reliability of the
perturbative predictions.
The soft plus virtual contributions that dominate in the threshold region
are well understood in terms of certain IR anomalous dimensions and process independent 
soft distributions.  A systematic way of resumming SV logarithms to all orders
exists in Mellin $N$ space.  While SV contributions dominate, the next to SV contributions are as important as
SV for any precision studies and hence can not be ignored. 
Next to SV terms also can give large contributions at every order, thereby
spoiling the reliability of the perturbation series.  The canonical resolution through resummation for the next to SV terms 
is unfortunately hard to achieve.  
In this article, we have studied the structure of next to SV logarithms both in
$z$ and $N$ spaces  for the diagonal partonic channels.  Using IR factorisation and UV renormalisation group invariance,
we show that both SV and next to SV contributions satisfy  Sudakov differential
equation whose solution provides an all order perturbative result in strong coupling constant.   
We show that like SV contributions, next to SV contributions also demonstrate
IR structure in terms of certain infrared anomalous dimensions.  However, NSV terms depend, in addition, on certain process  dependent functions.   
The underlying universal IR structure of NSV terms can
be further unravelled when results for variety of inclusive reactions become
available.  In $z$ space, we show that the next to SV contributions
do  exponentiate allowing us to predict the corresponding next to SV logarithms to all orders.   We find that the NSV part of the solution is invariant under gauge like transformations allowing us to construct class of solutions, all giving identical fixed order
predictions for NSV terms of CFs $\Delta_c$.   We show that the exponent in the $z$ space has an integral representation
which can be used to study these threshold logarithms in Mellin $N$ space.
We also show that the NSV logarithms in $N$ space organise themselves exactly
like the SV ones in such a way so as to keep $2 a_s(\mu_R^2)\beta_0 \ln N$ as a order one term to all orders in $a_s(\mu_R^2)$.  Unlike
the SV part of the resummed result, the resummation coefficients for NSV terms are found to be controlled not only
by process independent anomalous dimensions but also by process dependent
$\varphi_{c,i}^{(k)}$.

The all order master formula that we obtain in $z$ space demonstrates a
perturbative structure which can predict certain SV and NSV logarithms 
to all orders in strong coupling constant $a_s$, given the 
lower order results .  From the available results
at $a_s$ and at $a_s^2$ for the CFs, our predictions
for third order NSV logarithms are in complete agreement with the known results available for variety of inclusive reactions, namely DY production and Higgs productions in bottom quark annihilation and gluon fusion.  
Using the corresponding CFs that are known to third order, our formalism allows us to predict  three leading NSV logarithms to all orders starting from fourth order, of which, we reported here the results 
to order $a_s^7$.   We have studied the logarithmic
structure of physical evolution kernel, in particular the leading logarithms, and found that they are controlled only by process independent anomalous dimensions $\beta_0, A_1^c,C_2^c, D_1^c$ and diagonal coefficients $\varphi_{c,i}^{(i)}$ at every order 
$a_s^i$.   We conclude by noting  that the structure of NSV logarithms demonstrates a rich perturbative structure that need to be explored further.

\section{Acknowledgements}
We thank Claude Duhr for useful discussion and his constant help throughout this project.  
We thank Claude Duhr and Bernhard Mistlberger for providing third order results for the inclusive reactions.  VR thanks G. Grunberg for useful discussions.
We would like to thank L. Magnea and E. Laenen for their encouragement to work on this area.   Finally we thank the referee for valuable questions and suggestions.  

\appendix
\section{Details of the Mellin moment of $\Psi^c_{\mathcal D}$} \label{ap:PsiN}
In this section, we evaluate the Mellin moment of $\Psi^c_{\mathcal D}$ in the following way. At first,  following Eq. (\ref{eq:Mellin}) we decompose $\Psi^c_{N}$ into $\Sigma^c_{\rm{sv},N}$ and $\Sigma^c_{\rm{nsv},N}$. So,  
we begin with
\begin{eqnarray}
    \Sigma^c_{\rm{sv},N} &=&  \int_0^1 dz \bigg(\frac{z^{N-1}-1}{1-z}\bigg)\bigg(
\int_{\mu_F^2}^{q^2 (1-z)^2} {d \lambda^2 \over \lambda^2}
	 2 A^c (a_s(\lambda^2)  +  2 \overline G^c_{SV}(a_s(q^2 (1-z)^2))\bigg)
\end{eqnarray}
We follow the method described in \cite{Laenen:2008ux} to perform Mellin moment.
In the large $N$,  keeping ${1 \over N}$ corrections, we replace
\begin{eqnarray}
\label{replaceA}
\int_0^1 dz \left({z^{N-1}-1}\right) &&\rightarrow \Gamma_A\left(N{d \over dN}\right) \int_0^1 dz~ \theta\left(1-z-{1\over N}\right) 
\end{eqnarray} 
where $\Gamma_A(N{d \over dN})$ is given in Appendix[\ref{ap:Gamma}]. We expand $\Gamma_A$ in powers of  $Nd/dN$ and  apply on the integral.
We then  make appropriate change of variables and interchange of integrals to obtain
\begin{eqnarray}\label{eq:SigmaSVN}
   \Sigma_{\rm{sv},N}^c &=& 
	- \int^{q^2}_{q^2/N^2} \frac{d\lambda^2}{\lambda^2} \bigg\{
	     \bigg(\ln\frac{q^2}{\lambda^2 N^2} - 2 \gamma_{1}^A \bigg) A^c(a_s(\lambda^2)) + \overline{G}^c_{SV}\big(a_s(\lambda^2)\big) 
\nonumber\\&&
	+ \lambda^2 {d \over d \lambda^2} \mathcal{F}_A^c (a_s(\lambda^2)) \bigg\} 
+ \mathcal{F}_A^c(a_s(q^2))
\nonumber\\&&
     - 2 \bigg(\gamma_{1}^A + \ln N\bigg) \int_{\mu_F^2}^{q^2} \frac{d\lambda^2}{\lambda^2} A^c(a_s(\lambda^2))\,,
\end{eqnarray}
where 
\begin{eqnarray}
	\mathcal{F}_A^c(a_s(\lambda^2)) & =& -2 \gamma_{1}^A  \overline{G}^c_{SV}\big(a_s(\lambda^2)\big) 
	+ 4 \sum_{i=0}^\infty \gamma_{{i+2}}^A \Big(-2 \beta (a_s(\lambda^2))\frac{\partial}{\partial a_s(\lambda^2)}\Big)^i 
	\bigg\{   A^c(a_s(\lambda^2)) 
\nonumber\\&&
	+ \beta(a_s(\lambda^2)) {\partial \over \partial a_s(\lambda^2)}  
	\overline{G}^c_{SV}\big(a_s(\lambda^2)\big) \bigg \}.
\end{eqnarray}
Here $\beta(a_s(\lambda^2))$ is defined as, $\beta(a_s(\lambda^2)) = - \sum \limits_{i=0}^{\infty}\beta_i~a_s^{i+2}(\lambda^2)$ (also see  \cite{Chetyrkin:2017bjc,Luthe:2017ttg,Herzog:2017ohr} for QCD) .
\\
Replacing $a_s(\lambda^2)$ by
\begin{eqnarray}
\label{resumas}
a_s(\lambda^2)&=&\bigg({a_s(\mu_R^2) \over l}\bigg) \bigg[1-{a_s(\mu_R^2)\over l} {\beta_1 \over \beta_0} \ln l
   + \bigg({a_s(\mu_R^2) \over l}\bigg)^2 \bigg({\beta_1^2\over \beta_0^2} (\ln^2 l-\ln l
\nonumber\\
&&+l-1)-{\beta_2\over \beta_0} (l-1) \bigg)
   + \bigg({a_s(\mu_R^2)\over l}\bigg)^3 \bigg({\beta_1^3\over \beta_0^3} \Big(2 (1-l) \ln l + {5\over 2} \ln^2 l
\nonumber \\
&&- \ln^3 l -{1\over 2} + l - {1\over 2} l^2\Big)
             +{\beta_3 \over 2 \beta_0} (1-l^2) + {\beta_1 \beta_2 \over \beta_0^2} \Big(2 l \ln l
\nonumber\\
&&- 3 \ln l - l (1-l)\Big)\bigg)\bigg ]\,,
\end{eqnarray}
where $l = 1 - \beta_0 a_s(\mu_R^2) \ln(\mu_R^2/\lambda^2)$ and performing the integrals over $\lambda^2$
we obtain the result. The entire result is decomposed into two parts. The one proportional to $\frac{1}{N}$, are expressed in terms of $\bar{g}_i^c(\omega)$  given in Eq. (\ref{PsiNSVN}). And the remaning part is embedded in Eq. (\ref{PsiSVN}).  

Similarly we define,
\begin{eqnarray}
     \Sigma_{\rm{nsv},N}^c &=&  2 \int_0^1 dz\hspace{3mm} z^{N-1} \bigg\{ \int_{\mu_F^2}^{q^2(1-z)^2} \frac{d\lambda^2}{\lambda^2} L^c(a_s(\lambda^2),z) + \varphi_{f,c}\big(a_s(q^2(1-z)^2),z\big)  \bigg\}.
\end{eqnarray}
Following \cite{Laenen:2008ux}, in the large $N$ and keeping ${1 \over N}$ corrections, we replace
\begin{eqnarray}
\label{replaceB}
\int_0^1 dz z^{N-1} &&\rightarrow \Gamma_B\left(N{d \over dN}\right) \int_0^1 {dz \over 1-z}~ \theta\left(1-z-{1\over N}\right) \,,
\end{eqnarray} 
where $\Gamma_B(N{d \over dN})$ is given in Appendix[\ref{ap:Gamma}] and we replace $Nd/dN$ by  
\begin{eqnarray}
	N{d \over dN} = N {\partial \over \partial N} - 2 \beta(a_s(\lambda^2)) {\partial \over \partial a_s(\lambda^2)} \,,
\end{eqnarray} 
to deal with $N$ appearing in the argument of $a_s(q^2/N^2)$ and also the explicit ones present in $\varphi_{f,c}$.
After a little algebra, we obtain
\begin{eqnarray}
     \Sigma_{\rm{nsv},N}^c&=& 
     -\frac{1}{N}\int_{\frac{q^2}{N^2}}^{q^2} \frac{d\lambda^2}{\lambda^2}\bigg\{\xi^c(a_s(\lambda^2),N) + 
     \lambda^2 {d \over d\lambda^2} \mathcal{F}^c_B\big(a_s(\lambda^2),N\big)\bigg\}
     +\frac{1}{N}\mathcal{F}^c_B\big(a_s(q^2),N\big)
\nonumber\\ &&
     +\frac{1}{N}\int_{\mu_F^2}^{q^2} \frac{d\lambda^2}{\lambda^2}
     {\xi}^c(a_s(\lambda^2),N)\,,
\end{eqnarray}
where the functions $\xi^c$ is defined as
\begin{eqnarray}
	\xi^c(a_s,N) &=& -2 \bigg( -\gamma_{1}^B\big( D^c(a_s) - C^c(a_s) \ln N \big) 
+ \gamma_{2}^BC^c(a_s)\bigg)\,,
\end{eqnarray}
and 
\begin{eqnarray}
  \mathcal{F}^c_B\big(a_s(\lambda^2),N\big) &=& 
2\gamma_{1}^B\varphi_{f,c}(a_s(\lambda^2),N) - 4\gamma_{2}^B\bigg(\lambda^2 \frac{d}{d\lambda^2}\varphi_{f,c}(a_s(\lambda^2),N) 
+ \tilde\xi^c(a_s(\lambda^2),N)\bigg) \nonumber\\
   &&  + 8\bigg( \gamma_{3}^B + \tilde{\gamma}^{B}\bigg) \bigg(\lambda^2\frac{d}{d\lambda^2}
\bigg\{
\lambda^2\frac{d}{d\lambda^2}\varphi_{f,c}(a_s(\lambda^2),N) 
+ \tilde \xi^c(a_s(\lambda^2),N) \bigg\}
\nonumber\\
&&+ {1 \over 2} C^c(a_s(\lambda^2))\bigg).
\end{eqnarray}
where
\begin{eqnarray}
\tilde\xi^c(a_s,N) &=& 
 \big(D^c(a_s) -C^c(a_s) \ln N\big),
\nonumber\\
	\varphi_{f,c}(a_s(\lambda^2),N) &=& \sum \limits_{i=1}^{\infty} \sum \limits_{k=0}^{i} a_s^i(\lambda^2) \varphi_{c,i}^{(k)} (-\ln N) ^k\,, \quad
\tilde \gamma^B = \sum_{i=4}^\infty \gamma_{i}^B \left(N {d\over d N} \right)^{i-3}.
\end{eqnarray}
Using Eq.(\ref{resumas}), we perform $\lambda^2$ integrations to obtain the result in terms of $h_{ij}^c(\omega)$ given in Eq. (\ref{PsiNSVN}). 
\section{Perturbative constant of $\mathrm{\Phi^c_A}$}
\label{app:PhiSV}
In this section, we present the SV coefficients $\overline{K}^{c(i)}(\epsilon)$ to fourth order :
\begin{align}
\label{eq:App-SolnK}
\overline{K}^{c(1)}(\epsilon) &= \frac{1}{\epsilon} \Bigg\{ 2 A^c_{
                       1}\Bigg\}\,,
\nonumber\\
\overline{K}^{c(2)}(\epsilon) &= \frac{1}{\epsilon^2} \Bigg\{ -2 \beta_0
                       A^c_{1} \Bigg\} + \frac{1}{\epsilon} \Bigg\{
                       A^c_{ 2}\Bigg\}\,,
\nonumber\\
\overline{K}^{c(3)}(\epsilon) &= \frac{1}{\epsilon^3} \Bigg\{ \frac{8
                       }{3} \beta_0^2 A^c_{1} \Bigg\} +
                       \frac{1}{\epsilon^2} \Bigg\{ -\frac{2}{3}  \beta_1
                       A^c_{1} - \frac{8}{3}  \beta_0 A^c_{2}
                       \Bigg\} + \frac{1}{\epsilon} \Bigg\{ \frac{2
                       }{3} A^c_{ 3} \Bigg\}\,,
\nonumber\\
\overline{K}^{c(4)}(\epsilon) &= \frac{1}{\epsilon^4} \Bigg\{ -4 \beta_0^3
                       A^c_{1} \Bigg\} + \frac{1}{\epsilon^3}
                       \Bigg\{ \frac{8}{3} \beta_0 \beta_1 A^c_{1}
                       +6 \beta_0^2 A^c_{2} \Bigg\} +
                       \frac{1}{\epsilon^{2}} \Bigg\{ -\frac{1}{3} \beta_2
                       A^c_{1} - \beta_1 A^c_{2} - 3 \beta_0
                       A^c_{3} \Bigg\}
\nonumber\\
&+ \frac{1}{\epsilon} \Bigg\{
                       \frac{1}{2} A^c_{4} \Bigg\}
\end{align}
where the $A^c_i$ are the $i^{\rm th}$-order cusp anomalous dimensions:
\begin{equation}
    A^c(a_s(\mu_R^2)) = \sum_{i} ~a_s^i(\mu_R^2)~ A^c_i \,.
\end{equation}
The finite quantity $\overline{G}_{SV}^{c(i)}(\epsilon) $ are related to its renormalised counter parts $\overline {\cal G}_i^c(\epsilon)$ in the following way:
\begin{eqnarray}
\sum_{i=1}^\infty \hat a_s^i 
\left( {q^2 (1-z)^{2 } \over \mu^2}\right)^{i{\epsilon\over 2}} 
S^i_{\epsilon}
\overline G^{c(i)}_{{SV}}(\epsilon)
&=&
\sum_{i=1}^\infty a_s^i\left(q^2 (1-z)^{2}\right) 
\overline {\cal G}^{c}_{i}(\epsilon)
\label{Gbar1}
\end{eqnarray}
we find
\begin{eqnarray}
\overline { G}^{c(i)}_{{SV}}(\epsilon)&=&\overline {\cal G}_{1}^{c}(\epsilon)
\nonumber\\
\overline { G}^{c(2)}_{{SV}}(\epsilon)&=&{1\over \epsilon} \Bigg(
                  - 2 \beta_0  \overline {\cal G}_{1}^{c}(\epsilon)\Bigg)
                  +\overline {\cal G}_{2}^{c}(\epsilon)
\nonumber\\
\overline { G}^{c(3)}_{{SV}}(\epsilon)&=&  {1\over \epsilon^2} \Bigg(
                    4 \beta_0^2 \overline {\cal G}_{1}^{c}(\epsilon)\Bigg)
                  +{1\over \epsilon} \Bigg(
                  - \beta_1 \overline {\cal G}_{1}^{c}(\epsilon)
                  -4\beta_0 \overline {\cal G}_{2}^{c}(\epsilon)\Bigg)
                  +\overline {\cal G}_{3}^{c}(\epsilon)
\nonumber\\
\overline {G}^{c(4)}_{{SV}}(\epsilon)&=& {1 \over \epsilon^3} \Bigg(
                    -8 \beta_0^3 \overline {\cal G}_{1}^{c}(\epsilon)\Bigg)
                  +{1 \over \epsilon^2} \Bigg(
                    {16 \over 3} \beta_0 \beta_1 \overline {\cal G}_{1}^{c}(\epsilon)
                    +12\beta_0^2 \overline {\cal G}_{2}^{c}(\epsilon)\Bigg)
\nonumber\\
&&                  +{1 \over \epsilon} \Bigg(
                      -{2 \over 3} \beta_2 \overline {\cal G}_{1}^{c}(\epsilon)
                      -2 \beta_1 \overline {\cal G}_{2}^{c}(\epsilon)
                    -6 \beta_0 \overline {\cal G}_{3}^{c}(\epsilon)\Bigg)
                  +\overline {\cal G}_{4}^{c}(\epsilon)\,.
\end{eqnarray}
Through explicit determination of the quantity ${\cal \overline{G}}_i^I(\epsilon)$, it was found that it is dependent only on the initial partons and can be further decomposed as:
\begin{align}\label{eq:chap2-gbar}
\overline {\cal G}^{c}_{i}(\epsilon)=- f_i^c+ \overline \chi_i^c +
\sum_{j=1}^\infty \epsilon^j \overline {\cal G}^{c,(j)}_{i}\,,
\end{align}
where ,
\begin{align}
  \label{eq:bBH-Cg}
  \overline \chi^{c}_{1} &= 0\, ,
                \nonumber\\
  \overline\chi^{c}_{2} &= - 2 \beta_{0} \overline {\cal G}^{c,(1)}_{1}\, ,
                \nonumber\\
 \overline\chi^{c}_{3} &= - 2 \beta_{1}\overline {\cal G}^{c,(1)}_{1} - 2
                \beta_{0} \left(\overline {\cal G}^{c,(1)}_{2}  + 2 \beta_{0}\overline {\cal G}^{c,(2)}_{1}\right)
                \nonumber\\
\overline\chi^{c}_{4} &= - 2 \beta_{2}\overline {\cal G}^{c,(1)}_{1} - 2
                \beta_{1} \left(\overline {\cal G}^{c,(1)}_{2}  + 4 \beta_{0}\overline {\cal G}^{c,(2)}_{1}\right)
                - 2\beta_{0} 
                \left(\overline {\cal G}^{c,(1)}_{3}  + 2 \beta_{0}\overline {\cal G}^{c,(2)}_{2} + 4 \beta_{0}^2\overline {\cal G}^{c,(3)}_{1}\right)
                \,.
\end{align}
The SV coefficients ${\overline{\cal G}}^{c,k}_{i}$ in Eq.\eqref{eq:chap2-gbar} are found to exhibit Casimir scaling principle upto three loop. Hence, these coefficients for the Drell-Yan and Higgs production from gluon and bottom quark annihilation channels can be together expressed in the following way:
\begin{align}\label{app:SVGij}
\overline{\cal G}^{c,(1)}_{1}
&=C_R~ \Big(-3 \zeta_2\Big)\,,
\qquad \qquad
\overline{\cal G}^{c,(2)}_{1}
= C_R~ \Bigg({7 \over 3}  \zeta_3\Bigg)\,,
\nonumber\\
\overline{\cal G}^{c,(3)}_{1}
&= C_R~ \Bigg(-{3 \over 16}  \zeta_2^2\Bigg)\,,
\qquad \qquad
\overline{\cal G}^{c,(4)}_{1} =  C_R\left(-{7 \over 8}\zeta_2\zeta_3 + {31\over 20}\zeta_5\right)\,,
\nonumber\\
\overline{\cal G}^{c,(1)}_{2}
&= C_R C_A~ \Bigg({2428 \over 81} -{469 \over 9} \zeta_2
              +4 \zeta_2^2 -{176 \over 3} \zeta_3\Bigg)
+C_R n_f~ \Bigg(-{328 \over 81} + {70 \over 9} \zeta_2
                +{32 \over 3} \zeta_3 \Bigg)\,,
            \nonumber    \\
\overline{\cal G}^{c,(2)}_{2} &=
C_R n_f\left({976\over 243} - {196 \over 27}\zeta_2 - {1 \over 20}\zeta_2^2 - {310 \over 27}\zeta_3 \right)
\nonumber\\&
 ~~~ + C_RC_A\left(-{7288\over 243} + {1414 \over 27}\zeta_2 + {11 \over 40}\zeta_2^2 + {2077\over 27}\zeta_3 -
   {203\over 3}\zeta_2\zeta_3 + 43\zeta_5\right)\,,
   \nonumber\\
 \overline{\cal G}^{c,(1)}_{3} &= 
C_R  {C_A}^2 \Bigg(\frac{152}{63} \;{\zeta_2}^3 + \frac{1964}{9} \;{\zeta_2}^2
+ \frac{11000}{9} \;{\zeta_2} {\zeta_3} - \frac{765127}{486} \;{\zeta_2}
+\frac{536}{3} \;{\zeta_3}^2 - \frac{59648}{27} \;{\zeta_3} 
\nonumber\\
&
- \frac{1430}{3} \;{\zeta_5}
+\frac{7135981}{8748}\Bigg)
+ C_R {C_A} {n_f} \
\Bigg(-\frac{532}{9} \;{\zeta_2}^2 - \frac{1208}{9} \;{\zeta_2} {\zeta_3}
+\frac{105059}{243} \;{\zeta_2} 
\nonumber\\
&+ \frac{45956}{81} \;{\zeta_3} 
+\frac{148}{3} \;{\zeta_5} - \frac{716509}{4374} \Bigg)
+ {C_RC_F} {n_f} \
\Bigg(\frac{152}{15} \;{\zeta_2}^2 
- 88 \;{\zeta_2} {\zeta_3} 
+\frac{605}{6} \;{\zeta_2} + \frac{2536}{27} \;{\zeta_3}
\nonumber\\
&+\frac{112}{3} \;{\zeta_5} 
- \frac{42727}{324}\Bigg)
+ C_R {n_f}^2 
\Bigg( \frac{32}{9} \;{\zeta_2}^2 - \frac{1996}{81} \;{\zeta_2}
-\frac{2720}{81} \;{\zeta_3} + \frac{11584}{2187}\Bigg) \,.
\end{align}
Here, $C_R=C_A$ for $c=g$ and $C_R=C_f$ for $c=q,b$, with $C_A\equiv N_c$ and $C_F \equiv \frac{N_c^2-1}{2N_c}$ the Casimirs of adjoint and fundamental representations. Also, $\overline{G}^c_{SV}\big(a_s(q^2(1-z)^2),\epsilon\big)$ are related to the threshold exponent $\textbf{D}^c\big(a_s(q^2(1-z)^2)\big)$ via Eq.(46) of \cite{Ravindran:2006cg}.
\section{Perturbative constant of $\mathrm{\Phi^c_B}$} \label{ap:phic}
In this appendix, we present the relations between the expansion coefficients $\varphi_{c,i}^{(k)}$ appearing
in Eq.(\ref{varphiexp}) and the coefficients $\mathcal{G}_{L,i}^{c,(j,k)}$s:
\begin{eqnarray}
\varphi_{c,1}^{(k)} &=&  \mathcal{G}_{L,1}^{c,(1,k)}, \quad \quad k=0,1\nonumber\\
\varphi_{c,2}^{(k)} &=&  \bigg(\frac{1}{2}\mathcal{G}_{L,2}^{c,(1,k)} + \beta_0\mathcal{G}_{L,1}^{c,(2,k)}\bigg), 
k = 0,1,2\nonumber\\
\varphi_{c,3}^{(k)} &=&  \bigg(\frac{1}{3}\mathcal{G}_{L,3}^{c,(1,k)} + \frac{2}{3}\beta_1\mathcal{G}_{L,1}^{c,(2,k)} + \frac{2}{3}\beta_0\mathcal{G}_{L,2}^{c,(2,k)} + \frac{4}{3}
         \beta_0^2\mathcal{G}_{L,1}^{c,(3,k)}\bigg), \quad \quad k=0,1,2,3\nonumber\\
\varphi_{c,4}^{(k)} &=& \bigg( \frac{1}{4}\mathcal{G}_{L,4}^{c,(1,k)} + \frac{1}{2}\beta_2\mathcal{G}_{L,1}^{c,(2,k)} + \frac{1}{2}\beta_1\mathcal{G}_{L,2}^{c,(2,k)} + \frac{1}{2}
         \beta_0\mathcal{G}_{L,3}^{c,(2,k)} 
         + 2\beta_0\beta_1\mathcal{G}_{L,1}^{c,(3,k)} 
         + \beta_0^2\mathcal{G}_{L,2}^{c,(3,k)}\nonumber\\
         && + 
         2\beta_0^3\mathcal{G}_{L,1}^{c,(4,k)}\bigg), \quad \quad k=0,1,2,3,4
\end{eqnarray}
with $\mathcal{G}_{L,1}^{c,(2,3)},\mathcal{G}_{L,1}^{c,(2,4)},\mathcal{G}_{L,2}^{c,(2,4)},\mathcal{G}_{L,1}^{c,(3,4)}$ are all zero. 
We also present the explicit results for  $\mathcal{G}_{L,i}^{c,(j,k)}$ for bottom quark annihilation which is found to be same as Drell-Yan till 
second order in $\hat a_s$: 
\begin{eqnarray}
\mathcal{G}_{L,1}^{b,(1,0)} &=& 4 C_F,   \quad
\mathcal{G}_{L,1}^{b,(2,0)}  = 3 C_F\zeta_2,  \quad
\mathcal{G}_{L,1}^{b,(3,0)}  = -C_F\bigg(\frac{3}{2}\zeta_2 + \frac{7}{3}\zeta_3 \bigg), \quad
\nonumber\\
\mathcal{G}_{L,2}^{b,(1,0)} &=& C_AC_F\bigg(\frac{2804}{27} - \frac{290}{3}\zeta_2  - 56\zeta_3   \bigg)
   + C_F n_f \bigg( -\frac{656}{27} +  \frac{44}{3}\zeta_2 \bigg)
             - 64 C_F^2\zeta_2, \nonumber \\
\mathcal{G}_{L,2}^{b,(1,1)} &=& 20 C_F (C_A - C_F), \quad
\mathcal{G}_{L,2}^{b,(1,2)} = -8C_F^2,\nonumber\\
\end{eqnarray}
and for Higgs boson production in gluon fusion:
\begin{eqnarray}
\mathcal{G}_{L,1}^{g,(1,0)} &=& 4 C_A \,, \quad
\mathcal{G}_{L,1}^{g,(2,0)} = 3 C_A \zeta_2, \quad
\mathcal{G}_{L,1}^{g,(3,0)} = -C_A\bigg(\frac{3}{2}\zeta_2 + \frac{7}{3}\zeta_3 \bigg), \nonumber\\
\mathcal{G}_{L,2}^{g,(1,0)} &=& C_A^2\bigg(\frac{2612}{27} - \frac{482}{3}\zeta_2  - 56\zeta_3   \bigg)
+ C_A n_f \bigg( -\frac{392}{27} +  \frac{44}{3}\zeta_2 \bigg),\nonumber\\
\mathcal{G}_{L,2}^{g,(1,1)} &=& \frac{4}{3} C_A (C_A - n_f),\quad 
\mathcal{G}_{L,2}^{g,(1,2)}  = -8C_A^2.
\end{eqnarray}
and the remaining coefficients up to second order are identically zero.

\section{Expansion coefficients of $\Gamma_A(x)$ and $\Gamma_B(x)$}\label{ap:Gamma}
In the section, we present the expansion coefficients of $\Gamma_A(x)$ and $\Gamma_B(x)$ used in the Eqs.(\ref{replaceA},\ref{replaceB}) 
of the Appendix[\ref{ap:PsiN}] .  As in \cite{Laenen:2008ux}, the operators $\Gamma_A(x)$ and $\Gamma_B(x)$ are expanded in powers of $x$ as
\begin{eqnarray}
   \Gamma_A\big(x\big) =  \sum\limits_{k=0} - \gamma_{k}^Ax^k, 
\end{eqnarray}
where coefficients $\gamma_{k}^A$ are given by \cite{Laenen:2008ux}
\begin{eqnarray}
\gamma_{k}^A = {\Gamma_k(N) \over k!} (-1)^{k-1} \,,
\end{eqnarray}
See Eq.(25) of \cite{Laenen:2008ux} for the definition of $\Gamma_k(N)$. We find,
\begin{align}
\gamma_{0}^A&= 1\,,
\nonumber\\
\gamma_{1}^A&= \EuGa - \frac{1}{2N}\,,
\nonumber \\
\gamma_{2}^A&= \frac{1}{2}\bigg(\gamma_E^2 + \zeta_2\bigg) -\frac{1}{2N}\bigg(1+\EuGa\bigg) \,,
\nonumber \\
\gamma_{3}^A& = \frac{1}{6}\EuGa^3 + \frac{1}{2}\bigg(\EuGa\zeta_2\bigg) + \frac{1}{3}\zeta_3- \frac{1}{4N} \bigg(\EuGa^2 + 2 \EuGa + \zeta_2 \bigg) \,,
\nonumber \\
\gamma_{4}^A& = \frac{1}{24}\gamma_E^4 + \frac{1}{4}\bigg(\gamma_E^2\zeta_2\bigg) + \frac{9}{40}\zeta_2^2 + \frac{1}{3}\bigg(\gamma_E\zeta_3\bigg) - \frac{1}{12N}\bigg( \EuGa^3 + 3 \EuGa^2 + 
      3 \zeta_2 + 3\EuGa\zeta_2  + 2\zeta_3 \bigg)\,,
      \nonumber \\
\gamma_{5}^A&= \frac{1}{120}\gamma_E^5 + \frac{1}{12}\bigg(\gamma_E^3\zeta_2\bigg) + \frac{1}{40}\bigg(9\gamma_E\zeta_2^2\bigg) + 
     \frac{1}{6}\bigg(\gamma_E^2\zeta_3\bigg) + \frac{1}{6}\bigg(\zeta_2
    \zeta_3\bigg) + \frac{1}{5}\zeta_5
    \nonumber \\
&  -\frac{1}{240N} \bigg(  20 \EuGa^3 +5\EuGa^4 + 
      30  \EuGa^2\zeta_2 + 
      27 \zeta_2^2  +40  \zeta3 
    +   20 \EuGa \Big(3 \zeta_2 + 2 \zeta_3 \Big) \bigg) \,,
\nonumber \\
\gamma_{6}^A&= \frac{1}{720}\gamma_E^6 + \frac{1}{48}\bigg(\gamma_E^4\zeta_2\bigg) + \frac{9}{80}\bigg(\gamma_E^2\zeta_2^2\bigg) + 
     \frac{61}{560}\zeta_2^3 + \frac{1}{18}\bigg(\gamma_E^3\zeta_3\bigg) + \frac{1}{6}\bigg(\gamma_E\zeta_2\zeta_3\bigg) 
     \nonumber\\&
     + 
     \frac{1}{18}\zeta_3^2 + \frac{1}{5}\gamma_E\zeta_5
-\frac{1}{240N}\bigg(
5 \EuGa^4 + \EuGa^5 +10 \EuGa^3\zeta_2            + 27 \zeta_2^2 
 + 20 \zeta_2\zeta_3 + 10 \EuGa^2 \Big(3 \zeta_2 + 2\zeta_3\Big)
 \nonumber\\&
 +\EuGa \Big( 27 \zeta_2^2 + 40 \zeta_3 \Big)
 +24 \zeta_5\bigg)\,,
\nonumber \\
\gamma_{7}^A&= \frac{1}{5040}\gamma_E^7 + \frac{1}{240}\bigg(\gamma_E^5\zeta_2\bigg) + \frac{3}{80}\bigg(\gamma_E^3\zeta_2^2\bigg) + 
     \frac{61}{560}\bigg(\gamma_E\zeta_2^3\bigg) + \frac{1}{72}\bigg(\gamma_E^4\zeta_3\bigg) \nonumber\\&+ \frac{1}{12}\bigg(\gamma_E^2\zeta_2\zeta_3\bigg) + 
     \frac{3}{40}\bigg(\zeta_2^2\zeta_3\bigg) + \frac{1}{18}\bigg(\gamma_E\zeta_3^2\bigg) + \frac{1}{10}\bigg(\gamma_E^2\zeta_5\bigg) + 
     \frac{1}{10}\bigg(\zeta_2\zeta_5\bigg) + \frac{1}{7}\zeta_7
   \nonumber \\  
&-\frac{1}{10080N}\bigg(
42 \EuGa^5 +7  \EuGa^6  + 105\EuGa^4\zeta_2 +
      549\zeta_2^3  
     + 840\zeta_2\zeta_3 
     +140 \EuGa^3 \Big(3 \zeta_2 + 2\zeta_3\Big)
     \nonumber \\
    & + 21 \EuGa^2 \Big(27 \zeta_2^2 + 40\zeta_3 \Big) + 56 \Big(5 \zeta_3^2 + 18\zeta_5 \Big)
     + 42 \EuGa \Big(27 \zeta_2^2  + 20\zeta_2\zeta_3 + 24\zeta_5\Big) \bigg)\,,
\end{align}
and similarly $\Gamma_B\big(x\big)$ is given by \cite{Laenen:2008ux}
\begin{eqnarray}
	\Gamma_B\big(x\big) = \sum\limits_{k=1}\gamma_{k}^Bx^{k}, 
\end{eqnarray}
where $\gamma_{k+1}^B$ are given by \cite{Laenen:2008ux}
\begin{eqnarray}
	\gamma_{k+1}^B = {\Gamma^{k}(1) \over k!} (-1)^{k} \,,
\end{eqnarray}
explicitly we find,
\begin{align}
\gamma_{1}^B &= 1, \nonumber\\
 \gamma_{2}^B &= \gamma_E, \nonumber\\
 \gamma_{3}^B &= \frac{1}{2}\bigg(\gamma_E^2 + \zeta_2\bigg), \nonumber\\
 \gamma_{4}^B &= \frac{1}{6}\gamma_E^3 + \frac{1}{2}\bigg(\gamma_E\zeta_2\bigg) + \frac{1}{3}\zeta_3, \nonumber\\
 \gamma_{5}^B &= \frac{1}{24}\gamma_E^4 + \frac{1}{4}\bigg(\gamma_E^2\zeta_2\bigg) + \frac{9}{40}\zeta_2^2 + \frac{1}{3}\bigg(\gamma_E\zeta_3\bigg), \nonumber\\
 \gamma_{6}^B &= \frac{1}{120}\gamma_E^5 + \frac{1}{12}\bigg(\gamma_E^3\zeta_2\bigg) + \frac{1}{40}\bigg(9\gamma_E\zeta_2^2\bigg) + 
     \frac{1}{6}\bigg(\gamma_E^2\zeta_3\bigg) + \frac{1}{6}\bigg(\zeta_2
    \zeta_3\bigg) + \frac{1}{5}\zeta_5, \nonumber\\
 \gamma_{7}^B &= \frac{1}{720}\gamma_E^6 + \frac{1}{48}\bigg(\gamma_E^4\zeta_2\bigg) + \frac{9}{80}\bigg(\gamma_E^2\zeta_2^2\bigg) + 
     \frac{61}{560}\zeta_2^3 + \frac{1}{18}\bigg(\gamma_E^3\zeta_3\bigg) + \frac{1}{6}\bigg(\gamma_E\zeta_2\zeta_3\bigg) \nonumber\\&
     + 
     \frac{1}{18}\zeta_3^2 + \frac{1}{5}\gamma_E\zeta_5,\nonumber\\
\gamma_{8}^B &= \frac{1}{5040}\gamma_E^7 + \frac{1}{240}\bigg(\gamma_E^5\zeta_2\bigg) + \frac{3}{80}\bigg(\gamma_E^3\zeta_2^2\bigg) + 
     \frac{61}{560}\bigg(\gamma_E\zeta_2^3\bigg) + \frac{1}{72}\bigg(\gamma_E^4\zeta_3\bigg) \nonumber\\&+ \frac{1}{12}\bigg(\gamma_E^2\zeta_2\zeta_3\bigg) + 
     \frac{3}{40}\bigg(\zeta_2^2\zeta_3\bigg) + \frac{1}{18}\bigg(\gamma_E\zeta_3^2\bigg) + \frac{1}{10}\bigg(\gamma_E^2\zeta_5\bigg) + 
     \frac{1}{10}\bigg(\zeta_2\zeta_5\bigg) + \frac{1}{7}\zeta_7.
\end{align}
\section{Analytical structure of NSV Coefficients of $\Delta_{c\overline{c}}$ till Four Loop} \label{ap:Delta}
The partonic coefficient function given in Eq.(\ref{Delexp}) can be written as,
\begin{equation}
    \Delta_{c}^{(i)}(q^2,\mu_R^2,\mu_F^2,z) =  \Delta_{c}^{SV,(i)}(q^2,\mu_R^2,\mu_F^2,z) +  \Delta_{c}^{NSV,(i)}(q^2,\mu_R^2,\mu_F^2,z)
\end{equation}
where $\Delta_{c}^{SV,(i)}(q^2,\mu_R^2,\mu_F^2,z)$ can be found in \cite{Ravindran:2005vv,Ravindran:2006cg,deFlorian:2012za,Ahmed:2014cla,Li:2014afw}. Here we present $\Delta_{c}^{NSV,(i)}$ to fourth order where we set $\mu_R^2 = \mu_F^2 = q^2$ with the following expansion :
\begin{equation}
  \Delta_{c}^{NSV,(i)}(z) = \sum_{k=0}^{2i}\Delta_{c}^{ik}  \ln^k (1-z)
\end{equation}
The following results with the explicit dependence on $\mu_R$ and $\mu_F$ are provided in the ancillary files supplied with the \arXiv \  submission. We also put $\Delta^{30}_c,\cdots,\Delta^{32}_c$ and $\Delta^{40}_c,\cdots,\Delta^{44}_c$ in the ancillary files as they were lengthy. 
\begin{small}
\begin{align}
   \Delta^{10}_c = &
           \ 2 \varphi^{c,(0)}_1 \,,
 \nonumber \\
  \Delta^{11}_c = &
        \     2 \varphi^{c,(1)}_1
          + 4 D^c_1 \,,
 \nonumber \\
  \Delta^{12}_c = &
          \  4 C^c_1\,,
 \nonumber \\
	\Delta_c^{20} = &
  \    2 \varphi_{c,2}^{(0)}   
       + 4 \varphi_{c,1}^{(0)}   \bigg( \gt^{c,(1)}_1 
		+  \g^{c,1}_1 \bigg)
       + 4 f^c_1 \varphi_{c,1}^{(1)}   \zeta_2  
	- 16 f^c_1 C_1^c   \zeta_3 
         + 8 A^c_1 \varphi_{c,1}^{(1)}  \zeta_3
       + 6 A^c_1 \varphi_{c,1}^{(0)}    \zeta_2  \nonumber\\&
        + 2 \big(f^c_1\big)^2
	 - \frac{16}{5} A^c_1 C_1^c     \zeta_2^2 
       - 8 \big(A^c_1\big)^2   \zeta_2 
       + 8 D_1^c f^c_1    \zeta_2 
	+ 16 D_1^c A^c_1    \zeta_3 \,,
	\nonumber \\
	\Delta_c^{21} = &
    \     2 \varphi_{c,2}^{(1)}   
       + 4 \varphi_{c,1}^{(1)}   \bigg(  \gt^{c,(1)}_1 +  \g^{c,1}_1 \bigg)
       - 4 \varphi_{c,1}^{(0)}   \beta_0 
	- 4 f^c_1 \varphi_{c,1}^{(0)}   
       + 16 f^c_1 C_1^c    \zeta_2
       - 2 A^c_1 \varphi_{c,1}^{(1)}   \zeta_2 + 4 D_2^c  \nonumber\\&
       - 8 A^c_1 f^c_1 
       + 64 A^c_1 C_1^c    \zeta_3 
       - 4 D_1^c A^c_1   \zeta_2 
       + 8 D_1^c   \bigg( 
		\gt^{c,(1)}_1 
		+  \g^{c,1}_1 \bigg)
	   \,,
	\nonumber \\
		\Delta_c^{22} =&
    \   2 \varphi_{c,2}^{(2)}  
       - 4 \varphi_{c,1}^{(1)}  \beta_0 
       + 4 C_2^c  
       + C_1^c   \bigg( 8 \gt^{c,(1)}_1 + 8 \g^{c,1}_1 \bigg)
       - 4 f^c_1 \varphi_{c,1}^{(1)}   
       + 4 A^c_1 \varphi_{c,1}^{(0)}   
       - 20 A^c_1 C_1^c    \zeta_2 \nonumber\\&
       + 8 (A^c_1)^2   
       - 8 D_1^c f^c_1   
       - 4 D_1^c    \beta_0 \ \,,
		\nonumber \\
\Delta_c^{23} =&
       - 4 C_1^c   \beta_0 
       - 8 f^c_1 C_1^c  
       + 4 A^c_1 \varphi_{c,1}^{(1)}   
       + 8 D_1^c A^c_1  \,,
	\nonumber\\
	   \Delta_c^{24} = &  \ 8 A_1^c C_1^c \,,
	\nonumber \\
   \Delta_{c}^{33} =&
    \    2 \varphi_{c,3}^{(3)}  
       - 8 \beta_0\Big(\varphi_{c,2}^{(2)}  
       - \varphi_{c,1}^{(1)}    \beta_0
       +  C_2^c\Big)
       -  4 C_1^c   \bigg(  \beta_1 + 6 \beta_0 \gt^{c,(1)}_1 
	+ 2 \beta_0 \g^{c,1}_1 \bigg)
       - 8 \Big(f^c_2 C_1^c   
       + f^c_1 C_2^c\Big)  
\nonumber\\&
       - 4  f^c_1 \varphi_{c,2}^{(2)} 
       + 12 f^c_1 \varphi_{c,1}^{(1)}    \beta_0 
       - 16 f^c_1 C_1^c   \bigg(   \gt^{c,(1)}_1 + \g^{c,1}_1 \bigg)
       + 4 (f^c_1)^2 \varphi_{c,1}^{(1)}   
       + 4 A^c_2 \varphi_{c,1}^{(1)}  
       + 4 A^c_1 \varphi_{c,2}^{(1)}  
       \nonumber\\&
       + 8 A^c_1 \varphi_{c,1}^{(1)}   \bigg( \gt^{c,(1)}_1 +  \g^{c,1}_1 \bigg)
       -\frac{32}{3}  A^c_1 \varphi_{c,1}^{(0)}  \beta_0 
       + 68 A^c_1 C_1^c   \zeta_2 \beta_0 
       - 8 A^c_1 f^c_1 \varphi_{c,1}^{(0)}  
       + 104 A^c_1 f^c_1 C_1^c   \zeta_2 
\nonumber\\&
       - 16 (A^c_1)^2   \beta_0 
       - 20 (A^c_1)^2 \varphi_{c,1}^{(1)}   \zeta_2 
       + 320 (A^c_1)^2 C_1^c   \zeta_3 
       - 32 (A^c_1)^2 f^c_1
       + 16 D_1^c f^c_1   \beta_0 
       + 8 D_1^c (f^c_1)^2   
\nonumber\\&
       + 8 D_1^c A^c_2   
       + 16 D_1^c A^c_1   \bigg(  \gt^{c,(1)}_1 +  \g^{c,1}_1 \bigg)
       - 40 D_1^c (A^c_1)^2   \zeta_2 
       + \frac{16}{3} D_1^c   \beta_0^2 
       + 8 D_2^c A^c_1 \,,
\nonumber \\
\begin{autobreak}
   \Delta_{c}^{34} =
        \frac{16}{3} C_1^c   \beta_0^2 
       + 16 f^c_1 C_1^c   \beta_0
       + 8 C_1^c \Big((f^c_1)^2    
       +  A^c_2 \Big)   + 16 (A^c_1)^3
       + 4 A^c_1 \varphi_{c,2}^{(2)}   
       - \frac{32}{3} A^c_1 \varphi_{c,1}^{(1)}     \beta_0
       + 8 A^c_1 C_2^c  
       + 16 A^c_1 C_1^c   \bigg( \gt^{c,(1)}_1 +  \g^{c,1}_1 \bigg)
       - 8 A^c_1 f^c_1 \varphi_{c,1}^{(1)}  
       + 4 (A^c_1)^2 \varphi_{c,1}^{(0)}   
       - 72  (A^c_1)^2 C_1^c   \zeta_2 
       - \frac{40}{3} D_1^c A^c_1   \beta_0 
       - 16 D_1^c A^c_1 f^c_1  \ \,,
\end{autobreak}
\nonumber \\
\begin{autobreak}
   \Delta_{c}^{35} =
        - \frac{40}{3} A^c_1 C_1^c   \beta_0 
       - 16 A^c_1 f^c_1 C_1^c   
       + 4  (A^c_1)^2 \varphi_{c,1}^{(1)}  
       + 8 D_1^c (A^c_1)^2  \  \,,
\end{autobreak}
\nonumber \\
\begin{autobreak}
   \Delta_{c}^{36} = 
   8 (A_1^c)^2 C_1^c\,,
\end{autobreak}
\nonumber \\
\begin{autobreak}
  \Delta^{45}_c =
      -8 C^c_1 
          \beta_0^3
      - \frac{88}{3} f^c_1 C^c_1  
            \beta_0^2
      -24 (f^c_1)^2 C^c_1   
           \beta_0
      - \frac{16}{3} (f^c_1)^3 C^c_1  
      - \frac{56}{3} A^c_2 C^c_1  
            \beta_0
      -16 A^c_2 f^c_1 C^c_1  
      +4 A^c_1 \varphi_{c,3}^{(3)} 
      -  \frac{56}{3} A^c_1 \varphi_{c,2}^{(2)} 
           \beta_0
      +24  A^c_1 \varphi_{c,1}^{(1)}    \beta_0^2
      + \frac{64}{3} A^c_1 D^c_1  
            \beta_0^2
      - \frac{64}{3} A^c_1 C^c_2  
            \beta_0
      -\frac{1}{3} A^c_1 C^c_1   \bigg(
          40 \beta_1
          +176 \beta_0 \overline {\cal G}^{c,(1)}_1
          +80 \beta_0 g^{c,1}_1
          \bigg)
      -16 A^c_1 f^c_2 C^c_1 
      -8  A^c_1 f^c_1 \varphi_{c,2}^{(2)} 
      + \frac{88}{3} A^c_1 f^c_1 \varphi_{c,1}^{(1)} 
            \beta_0
      + \frac{128}{3} A^c_1 f^c_1 D^c_1 
            \beta_0
      -16 A^c_1 f^c_1 C^c_2 
      - 32 A^c_1 f^c_1 C^c_1   \bigg(
           \overline {\cal G}^{c,(1)}_1
          + g^{c,1}_1
          \bigg)
      + 8 A^c_1 (f^c_1)^2 \varphi_{c,1}^{(1)}   
      + 16 A^c_1 (f^c_1)^2 D^c_1  
      + 8 A^c_1 A^c_2 \varphi_{c,1}^{(1)}   
      + 16 A^c_1 A^c_2 D^c_1   
      +4  (A^c_1)^2 \varphi_{c,2}^{(1)}  
      + 8 (A^c_1)^2 \varphi_{c,1}^{(1)}   \bigg(
           \overline {\cal G}^{c,(1)}_1
          +  g^{c,1}_1
          \bigg)
      - \frac{40}{3}(A^c_1)^2 \varphi_{c,1}^{(0)} 
            \beta_0
      + 8 (A^c_1)^2 D^c_2   
      + 16 (A^c_1)^2 D^c_1   \bigg(
          +  \overline {\cal G}^{c,(1)}_1
          +  g^{c,1}_1
          \bigg)
      + \frac{776}{3}(A^c_1)^2 C^c_1   
            \beta_0 \zeta_2
      -8 (A^c_1)^2 f^c_1 \varphi_{c,1}^{(0)}
      + 240 (A^c_1)^2 f^c_1 C^c_1 \zeta_2  
      - \frac{128}{3}(A^c_1)^3   \beta_0
      -36 (A^c_1)^3 \varphi_{c,1}^{(1)}  \zeta_2
      -72 (A^c_1)^3 D^c_1   \zeta_2
      + \frac{1792}{3} (A^c_1)^3 C^c_1  
           \zeta_3
      -48 (A^c_1)^3 f^c_1 \  \,,  
\end{autobreak}
 \nonumber \\
\begin{autobreak}
  \Delta^{46}_c =
       \frac{64}{3} A^c_1 C^c_1   
           \beta_0^2
      + \frac{128}{3} A^c_1 f^c_1 C^c_1   
           \beta_0
      + 16 A^c_1 (f^c_1)^2 C^c_1   
      + 16 A^c_1 A^c_2 C^c_1   
      + 4 \big(A^c_1\big)^2 \varphi_{c,2}^{(2)}   
      - \frac{40}{3} (A^c_1)^2 \varphi_{c,1}^{(1)}  
           \beta_0
      - \frac{56}{3} (A^c_1)^2 D^c_1 
           \beta_0
      + 8 \big(A^c_1\big)^2 C^c_2   
      + 16 (A^c_1)^2 C^c_1   \bigg(
          \overline {\cal G}^{c,(1)}_1
          + g^{c,1}_1
          \bigg)
      - 8 (A^c_1)^2 f^c_1 \varphi_{c,1}^{(1)}   
      - 16 (A^c_1)^2 f^c_1 D^c_1   
      + \frac{8}{3} (A^c_1)^3 \varphi_{c,1}^{(0)}   
      - 104 (A^c_1)^3 C^c_1  
           \zeta_2
      + 16 \big(A^c_1\big)^4 \  \,,
\end{autobreak}
 \nonumber \\
\begin{autobreak}
  \Delta^{47}_c =
        - \frac{56}{3} (A^c_1)^2 C^c_1   
           \beta_0
      - 16 (A^c_1)^2 f^c_1 C^c_1  
      + \frac{8}{3} \big(A^c_1\big)^3 \varphi_{c,1}^{(1)} 
      + \frac{16}{3} (A^c_1)^3 D^c_1 \ \,,
\end{autobreak}
\nonumber \\
\begin{autobreak}
  \Delta^{48}_c =
        \frac{16}{3} (A^c_1)^3 C^c_1 \ \,.
\end{autobreak}
\end{align}
\end{small}

The symbols $\overline{\mathcal{G}}^{c,(k)}_j$ and $g^{c,k}_j$ are also provided in the 
ancillary files with the \arXiv \ submission.
\section{Expressions of Resummation Constants $h^c_{ij}(\omega)$} \label{ap:hij}
The resummation constants $h^c_{ij}(\omega)$ given in Eq.(\ref{hNSV}) are found to be as following. Here $\bar{L}_{\omega}=\ln(1-\omega)$, $L_{qr} = \ln(\frac{q^2}{\mu_R^2})$, $L_{fr} = \ln(\frac{\mu_F^2}{\mu_R^2})$ and $\omega = 2 \beta_0 as(\mu_R^2) \ln N$.
\begin{small}
\begin{align}
\begin{autobreak}
   h^c_{00}(\omega) =
        \frac{2}{\beta_0} \bar{L}_{\omega} \bigg[   -\gamma^B_{2}   
            C^c_1
    + \gamma^B_1   
            D^c_1
          \bigg] \,,
\end{autobreak}
\nonumber \\
\begin{autobreak}
   h^c_{01}(\omega) =
        \frac{2}{\beta_0} \bar{L}_{\omega}    \bigg[
          - \gamma^B_1 C^c_1
          \bigg] \,,
\end{autobreak}
\nonumber \\
\begin{autobreak}
    h^c_{10}(\omega) =
        \frac{1}{(1-\omega)}\Bigg[
       \frac{2 \beta_1 D_1^c}{\beta_0^2}   \bigg \{
          \gamma^{B}_{1} \omega 
          +  \gamma^{B}_{1} \bar{L}_{\omega} 
          \bigg \}
       -  \frac{2 \beta_1}{\beta_0^2} C_1^c   \bigg \{
           \gamma^{B}_{2} \omega 
           + \gamma^{B}_{2} \bar{L}_{\omega} 
          \bigg\}
       - 2 \frac{D_2^c}{\beta_0} 
          \gamma^{B}_{1} \omega 
       + 2  \frac{C_2^c}{\beta_0}  
          \gamma^{B}_{2} \omega 
       - 2 \varphi_{c,1}^{(1)}  \gamma^{B}_{2} 
       + 2 \varphi_{c,1}^{(0)}   \gamma^{B}_{1} 
       + 2 D_1^c   \bigg\{
           L_{qr} \gamma^{B}_{1} 
          -  L_{fr} \gamma^{B}_{1} 
          + L_{fr} \gamma^{B}_{1} \omega 
          - 2 \gamma^{B}_{2} 
          \bigg\}
       - 2 C_1^c   \bigg\{
            L_{qr} \gamma^{B}_{2} 
          -  L_{fr} \gamma^{B}_{2} 
          +  L_{fr} \gamma^{B}_{2} \omega 
          - 4 \gamma^{B}_{3} 
          \bigg\}\Bigg] \,,
\end{autobreak}
\nonumber \\
\begin{autobreak}
   h^c_{11}(\omega) = 
   \frac{1}{(1-\omega)}\Bigg[
       \frac{\beta_1}{\beta_0^2} C_1^c   \bigg\{
          - 2 \gamma^{B}_{1} \omega 
          - 2 \gamma^{B}_{1} \bar{L}_{\omega}
          \bigg\}
       +  2 \frac{C_2^c}{\beta_0}    \gamma^{B}_{1} \omega 
       - 2 \varphi_{c,1}^{(1)}   \gamma^{B}_{1} 
       + C_1^c   \bigg\{
          - 2 L_{qr} \gamma^{B}_{1} 
          + 2 L_{fr} \gamma^{B}_{1} 
          - 2 L_{fr} \gamma^{B}_{1} \omega 
          + 4 \gamma^{B}_{2}
          \bigg\}  +  \frac{\omega}{(1-\omega)}\bigg\{
        \frac{ \varphi_{c,2}^{(2)}}{ \beta_0}   \gamma^{B}_{1} 
          \bigg\}\Bigg] \,,
\end{autobreak}
\nonumber \\
   h^c_{21}(\omega) =&
         \frac{1}{(1-\omega)^2}\Bigg[
        \frac{\beta_1^2 C_1^c}{ \beta_0^3}   \bigg\{
          - \gamma^{B}_{1} \omega^2 
          + \gamma^{B}_{1} \bar{L}_{\omega}^2 
          \bigg\}
       +  \frac{\beta_2 C_1^c}{ \beta_0^2}  
           \gamma^{B}_{1} \omega^2 
       + \frac{\beta_1 C_2^c}{ \beta_0^2}   \bigg\{
           - 2 \omega 
          +  \omega^2 
          - 2  \bar{L}_{\omega} 
          \bigg\} \gamma^{B}_{1}
          \nonumber\\&
       +  \frac{C_3^c}{\beta_0}   \bigg\{
           2 \gamma^{B}_{1} \omega 
          - \gamma^{B}_{1} \omega^2 
          \bigg\}
       + 2  \frac{\beta_1 \varphi_{c,1}^{(1)}}{\beta_0}  \gamma^{B}_{1} \bar{L}_{\omega} 
       +  \frac{\beta_1 C_1^c}{\beta_0}   \bigg\{
           2 L_{qr} \gamma^{B}_{1} 
          - 4 \gamma^{B}_{2}  
          \bigg\} \bar{L}_{\omega} 
       + 4 \varphi_{c,2}^{(2)}   \gamma^{B}_{2} 
        \nonumber\\&
       - 2 \varphi_{c,2}^{(1)}   \gamma^{B}_{1} 
       + C_2^c   \bigg\{
          - 2 L_{qr} \gamma^{B}_{1} 
          + 2 L_{fr} \gamma^{B}_{1} (1-\omega)^2
          + 4 \gamma^{B}_{2} 
          \bigg\} 
       + 2\beta_0 \varphi_{c,1}^{(1)}   \bigg\{ 
            L_{qr} \gamma^{B}_{1} 
          - 2 \gamma^{B}_{2} 
          \bigg\}
          \nonumber\\&
       + \beta_0 C_1^c   \bigg\{
           L^2_{qr} \gamma^{B}_{1} 
          - 4 L_{qr} \gamma^{B}_{2} 
          - L^2_{fr} \gamma^{B}_{1} 
          + 2 L^2_{fr} \gamma^{B}_{1} \omega
          + 8 \gamma^{B}_{3} 
          - L^2_{fr} \gamma^{B}_{1} \omega^2 
           \bigg\}\Bigg] \,,
\nonumber\\
\begin{autobreak}
   h^c_{22}(\omega) =
         \frac{\omega}{(1-\omega)^3}\Bigg[
        \frac{ -\varphi_{c,3}^{(3)}}{\beta_0}   \gamma^{B}_{1} 
          \Bigg] \,,
\end{autobreak}
\nonumber\\
\begin{autobreak}
   h^c_{32}(\omega) =
        \frac{1}{(1-\omega)^3}\Bigg[
           - 4 \gamma^{B}_{1} \bar{L}_{\omega} \bigg\{
          \frac{\beta_1 \varphi_{c,2}^{(2)}}{\beta_0}
          \bigg\}
       - 6 \varphi_{c,3}^{(3)}    \gamma^{B}_{2} 
       + 2 \varphi_{c,3}^{(2)}    \gamma^{B}_{1} 
       - 4 \beta_0 \varphi_{c,2}^{(2)}   \bigg\{
           L_{qr} \gamma^{B}_{1} 
          - 2 \gamma^{B}_{2} 
          \bigg\}\Bigg] \,,
\end{autobreak}
\nonumber\\
\begin{autobreak}
   h^c_{33}(\omega) =
        \frac{\omega}{(1-\omega)^4}\Bigg[
       \frac{ \varphi_{c,4}^{(4)}}{\beta_0}  
           \gamma^{B}_{1} 
         \Bigg] \,,
\end{autobreak}
\nonumber\\
\begin{autobreak}
   h^c_{42}(\omega) =
\frac{1}{(1-\omega)^4} \bigg[
       \frac{2\beta_1^2}{\beta_0^2} \varphi_{c,2}^{(2)}   \bigg\{
        3  \Lt^2
          - 2  \omega
          - 2  \Lt
          \bigg\}\gamma_{1}^B
       + \frac{ 4\beta_2}{\beta_0} \varphi_{c,2}^{(2)}  \gamma_{1}^B \omega
       + \frac{ 18\beta_1}{\beta_0} \varphi_{c,3}^{(3)}   \gamma_2^B \Lt
       + 24 \varphi_{c,4}^{(4)}    \gamma_{3}^B
       - \frac{6 \beta_1}{\beta_0} \varphi_{c,3}^{(2)}   \gamma_{1}^B \Lt
       -6\varphi_{c,4}^{(3)}   \gamma_2^B
       + 2\varphi_{c,4}^{(2)}   \gamma_{1}^B
       - 4 \beta_1 \varphi_{c,2}^{(2)}   \bigg\{
           L_{qr} \gamma_{1}^B
          - 3 L_{qr} \gamma_{1}^B \Lt
          -2 \gamma_2^B
          +6 \gamma_2^B \Lt
          \bigg\}
       + 18\beta_0 \varphi_{c,3}^{(3)}   \bigg\{
            L_{qr} \gamma_2^B
          - 4 \gamma_{3}^B
          \bigg\}
       - 6\beta_0 \varphi_{c,3}^{(2)}   \bigg\{
           L_{qr} \gamma_{1}^B
          - 2 \gamma_2^B
          \bigg\}
       + 6\beta_0^2 \varphi_{c,2}^{(2)}   \bigg\{
          + L_{qr}^2 \gamma_{1}^B
          - 4L_{qr} \gamma_2^B
          + 8 \gamma_{3}^B
          \bigg\}\bigg] \,,
\end{autobreak}
\nonumber\\




\begin{autobreak}
   h^c_{43}(\omega) =
\frac{2}{(1-\omega)^4} \bigg[
        \frac{ 3\beta_1}{\beta_0} \varphi_{c,3}^{(3)}   \gamma_{1}^B \Lt
       + 4 \varphi_{c,4}^{(4)}  \gamma_2^B
       -\varphi_{c,4}^{(3)}  \gamma_{1}^B
       + 3\beta_0 \varphi_{c,3}^{(3)}   \Big\{
           L_{qr} \gamma_{1}^B
          - 2 \gamma_2^B
          \Big\}\bigg] \,,
\end{autobreak}
\nonumber\\
\begin{autobreak}
   h^c_{44}(\omega) =
         \frac{\omega}{(1-\omega)^5}\Bigg[
       \frac{- \varphi_{c,5}^{(5)}}{\beta_0}  
           \gamma^{B}_{1} 
         \Bigg] \,,
\end{autobreak}
\end{align}
\end{small}
The above results along with the bigger ones $(h^c_{20}(\omega), h^c_{30}(\omega),h^c_{31}(\omega))$ and 
$( h^c_{40}(\omega),h^c_{41}(\omega))$ are all provided in the ancillary files with the \arXiv \  submission.
\section{Expressions of Resummation Constants $\bar g^c_{i}(\omega)$} \label{ap:gbar}
The resummation constants $\bar g^c_{i}(\omega)$ given in Eq.(\ref{PsiNSVN}) are presented below. Here $\bar{L}_{\omega}=\ln(1-\omega)$, $L_{qr} = \ln(\frac{q^2}{\mu_R^2})$, $L_{fr} = \ln(\frac{\mu_F^2}{\mu_R^2})$ and $\omega = 2 \beta_0 a_s(\mu_R^2) \ln N$. Also, $\textbf{D}_i^c$ are the threshold exponent given in \cite{Ravindran:2005vv}. 
 \begin{small}
 \begin{align}
   \bar g^c_1(\omega) = & \
       \frac{A^c_1}{\beta_0}  \Lt \,,
\nonumber \\
   \bar g^c_2(\omega) = & \
        \frac{1}{(1-\omega)}\bigg[ 
          \frac{\bold D^c_1}{2} 
       - \frac{A^c_2 }{\beta_0}   \omega

       +  \frac{ A^c_1 \beta_1 }{\beta_0^2}   \Big(
           \omega
          + \Lt
          \Big)
       - A^c_1   \Big(
           2
          + 2 \gamma_E
          - L_{qr}
          + L_{fr}(1- \omega)
          \Big)\bigg]\,,
\nonumber \\
   \bar g^c_3(\omega) = & \
     \frac{1}{(1-\omega)^2} \bigg[
      \bold D^c_2 \Big\{ \frac{1}{2}\Big\}
     + \bold D_1^c \bigg\{
      - \frac{\Lt}{2}   \frac{\beta_1}{\beta_0}
        + \beta_0   \Big(
          1
          +  \gamma_E
          - \frac{1}{2} L_{qr} 
          \Big) \bigg\}
         
       -  \frac{A^c_3 }{\beta_0} \Big\{1-\frac{\omega}{2}\Big\}\omega
       \nonumber\\&
       + A^c_2 \bigg\{
       +  \frac{\beta_1}{\beta_0^2} \Big(
           \omega
          - \frac{1}{2} \omega^2
          + \Lt
          \Big)
       -  \Big(
           2
          + 2 \gamma_E
          - L_{qr}
          + L_{fr} (1 -\omega)^2
          \Big) \bigg\}
          - \frac{\beta_2 A^c_1}{\beta_0^2}  \frac{ \omega^2}{2}
          \nonumber\\&
       +A^c_1 \bigg\{
       \frac{ \beta_1^2}{2\beta_0^3}   \Big(
            \omega^2
          -  \Lt^2
          \Big)

       + \frac{ \beta_1}{\beta_0}   \Big(
           2 
          + 2 \gamma_E
          - L_{qr} 
          \Big)\Lt
       - 2 \beta_0  \Big(
           2 \gamma_E
          +  \gamma_E^2
          +  \zeta_2
          -  L_{qr}
          \nonumber\\&
          -  L_{qr} \gamma_E
         + \frac{1}{4} L_{qr}^2
          - \frac{1}{4} L_{fr}^2(1- \omega)^2
          \Big) \bigg\} \bigg]\,,
\nonumber\\
  \bar g^c_4(\omega) =& \
  \frac{1}{(1-\omega)^3} \bigg[
      \bold D^c_1 \bigg\{
      \frac{ \beta_1^2 }{2\beta_0^2} \Big(
          -  \omega
          -  \Lt
          +  \Lt^2
          \Big)
      + \frac{\omega \beta_2}{2 \beta_0} 
           
          + \beta_1   \Big(
          1+\gamma_E - \frac{1}{2} L_{qr}\Big)\Big(1- 2  \Lt \Big)
          \nonumber\\&
          + 2\beta_0^2   \Big(
          2\gamma_E
          +   \gamma_E^2
          +  \zeta_2
          -  L_{qr} 
          -  L_{qr}  \gamma_E
          + \frac{1}{4} L_{qr}^2\Big) 
          \bigg\} 
          + \bold D^c_2 \bigg\{
          - \frac{\beta_1}{\beta_0}  \Lt
      + \beta_0    \Big(
            2 
          + 2  \gamma_E
          \nonumber\\&
          - L_{qr}
          \Big) \bigg\}
          + \frac{1}{2} \bold D^c_3

      - \frac{ A^c_4}{\beta_0}\omega \bigg\{
          1
          - \omega
          + \frac{1}{3} \omega^2
          \bigg\}

      + A^c_3 \bigg\{
      \frac{\beta_1}{\beta_0^2 }    \Big(
          \omega
          - \omega^2
          + \frac{1}{3} \omega^3
          + \Lt
          \Big)
          - 2
          \nonumber\\&
          - 2 \gamma_E
          + L_{qr}
          - L_{fr}
          + 3 L_{fr}
          \Big(1 -\omega+\frac{ \omega^2}{3}\Big)\omega
          \bigg\}

      + A^c_2 \bigg\{ 
      \frac{ \beta_1^2}{\beta_0^3}   \Big(
          \omega^2
          - \frac{1}{3} \omega^3
          - \Lt^2
          \Big)

      - \frac{ \beta_2}{\beta_0^2} 
        \nonumber\\&
      \Big(
          1
          - \frac{1}{3} \omega
          \Big)\omega^2 

      + 2\frac{\beta_1}{\beta_0} \Big(
          2
          + 2 \gamma_E 
          -  L_{qr} 
          \Big) \Lt

      - \beta_0  \Big(
          8 \gamma_E
          + 4 \gamma_E^2
          + 4 \zeta_2
          - 4 L_{qr}(1+ \gamma_E)
          \nonumber\\&
          + L_{qr}^2
          - L_{fr}^2
          + 3 L_{fr}^2 
          \Big(1- \omega+\frac{ \omega^2}{3}\Big)\omega
          \Big)\bigg\}

      + A^c_1 \bigg\{
      -\frac{\beta_1^3}{\beta_0^4}  \Big(
          \frac{1}{2} \omega^2
          - \frac{1}{3} \omega^3
          + \Lt \omega
          + \frac{1}{2} \Lt^2
          \nonumber\\&
          - \frac{1}{3} \Lt^3
          \Big)

      + \frac{\beta_1 \beta_2}{\beta_0^3}   \Big(
          \omega
          - \frac{2}{3} \omega^2
          + \Lt 
          \Big) \omega

      -  \frac{ \beta_3}{\beta_0^2} \omega^2   \Big(
          \frac{1}{2}
          - \frac{1}{3} \omega
          \Big)

      + 2  \frac{ \beta_1^2}{\beta_0^2}  
          \Big(\omega
          +  \Lt
          -  \Lt^2\Big)\Big(1
          \nonumber\\&
          +  \gamma_E- \frac{L_{qr}}{2}\Big)

      +  \frac{\beta_2}{\beta_0}  \Big(
          - 2 
          - 2 \gamma_E 
          + L_{qr}  \Big)\omega   

      +  \beta_1 \Big(
          - 4 \gamma_E
          - 2 \gamma_E^2
          - 2 \zeta_2
          + 2 L_{qr}
        - \frac{1}{2} L_{qr}^2
        \nonumber\\&
          + 2 L_{qr} \gamma_E
          
        \Big) \Big(1-2\Lt\Big)
          +  \frac{\beta_1}{2} L_{fr}^2(1-\omega)^3

      -  \beta_0^2   \Big(
          8 \gamma_E^2
          + \frac{8}{3} \gamma_E^3
          + \frac{16}{3} \zeta_3
          + 2(4 \zeta_2 +  L_{qr}^2)
          \nonumber\\&
          (1+ \gamma_E)
          - 4 L_{qr} \gamma_E(2+ \gamma_E)
          - 4 L_{qr} \zeta_2
          - \frac{1}{3} L_{qr}^3
          + \frac{1}{3} L_{fr}^3 (1-\omega)^3 \Big)
          \bigg\}\bigg] .
 \end{align}
\end{small}
As before here also we provide the above results along with $\bar g^c_{5}(\omega)$ in the ancillary files with the \arXiv \ submission.

\bibliographystyle{JHEP}
\bibliography{nsv}
\end{document}